\documentclass[10pt,journal,compsoc]{IEEEtran}
\usepackage{tikz}

\usepackage{framed,graphicx}
\usepackage{booktabs}
\usepackage{multirow}
\usepackage{rotating}
\usepackage{amsmath}
\usepackage{bigstrut}
\usepackage{lettrine}
\usepackage{dblfloatfix} 
\usepackage{color}
\usepackage{listings}
\usepackage{graphics}
\usepackage{eqparbox}
\usepackage{graphics}
\usepackage{amssymb}
\usepackage{colortbl}
\usepackage{mathptmx} 
\usepackage[scaled=.90]{helvet} 
\usepackage{courier}
\usepackage{balance}
\usepackage{picture}
\usepackage{algorithm}
\usepackage{algorithmicx}
\usepackage{algpseudocode}
\usepackage[export]{adjustbox}
\renewcommand{\footnotesize}{\scriptsize}
\definecolor{lightgray}{gray}{0.8}
\definecolor{darkgray}{gray}{0.6}
\definecolor{lavenderpink}{rgb}{0.98, 0.68, 0.82}
\definecolor{celadon}{rgb}{0.67, 0.88, 0.69}

\definecolor{Gray}{gray}{0.95}
\definecolor{LightGray}{gray}{0.975}

\usepackage[lofdepth,lotdepth]{subfig}

\newcommand{\bi}{\begin{itemize}}
  \newcommand{\ei}{\end{itemize}}
\newcommand{\be}{\begin{enumerate}}
  \newcommand{\ee}{\end{enumerate}}
\newcommand{\tion}[1]{\S\ref{sect:#1}}
\newcommand{\fig}[1]{Figure~\ref{fig:#1}}

\newcommand{\eq}[1]{Equation~\ref{eq:#1}}

\usepackage[shortlabels]{enumitem}
\usepackage{url}

\definecolor{steel}{rgb}{.0, .0, .0}
\definecolor{Gray}{rgb}{0.88,1,1}
\definecolor{Gray}{gray}{0.85}
\usepackage[framed]{ntheorem}
\usepackage{framed}
\usepackage{tikz}
\usetikzlibrary{shadows}
\theoremclass{Lesson}
\theoremstyle{break}

\tikzstyle{thmbox} = [rectangle, rounded corners, draw=black,
fill=Gray!40,  drop shadow={fill=black, opacity=1}]

\definecolor{shadecolor}{gray}{0.9}

\newshadedtheorem{lesson}{Result}

\DeclareFixedFont{\ttb}{T1}{txtt}{bx}{n}{12} 
\DeclareFixedFont{\ttm}{T1}{txtt}{m}{n}{12}  

\usepackage{color}
\definecolor{deepblue}{rgb}{0,0,0.5}
\definecolor{deepred}{rgb}{0.6,0,0}
\definecolor{deepgreen}{rgb}{0,0.5,0}

\definecolor{Code}{rgb}{0,0,0}
\definecolor{Decorators}{rgb}{0.5,0.5,0.5}
\definecolor{Numbers}{rgb}{0.5,0,0}
\definecolor{MatchingBrackets}{rgb}{0.25,0.5,0.5}
\definecolor{Keywords}{rgb}{0,0,1}
\definecolor{self}{rgb}{0,0,0}
\definecolor{Strings}{rgb}{0,0.33,0}
\definecolor{Comments}{rgb}{0,0.33,1}
\definecolor{Comments}{rgb}{0.5,0.5,0.5}
\definecolor{Backquotes}{rgb}{0,0,0}
\definecolor{Classname}{rgb}{0,0,0}
\definecolor{FunctionName}{rgb}{0,0,0}
\definecolor{Operators}{rgb}{0,0,0}
\definecolor{Background}{rgb}{1,1,1}
\usepackage{calc}

\usepackage{listings}

\lstnewenvironment{python}[1][]{
 \lstset{
    language=Python,
    basicstyle=\ttfamily\fontsize{3mm}{0.8em}\selectfont,
    breaklines=true,
    prebreak=\raisebox{0ex}[0ex][0ex]{\ensuremath{\hookleftarrow}},
    frame = l,
    showtabs=false,
    columns=fullflexible,
    showspaces=false,
    showstringspaces=false,
    keywordstyle=\bfseries\sffamily,
    emph={def, for, train, predict, score, return, transfer, discover, learner, apply_learner, fails}, 
    emphstyle=\bfseries,
    stringstyle=\color{black!70}\it,
    commentstyle=\color{black!70}\it,
    numbers=right,
    captionpos=t,
    escapeinside={\%*}{*)}
}}{}

\newcommand{\rahul}[1]{{\color{steel}{#1}}}

\newcommand{\respto}[1]{}

\bstctlcite{IEEEexample:BSTcontrol}

\begin{document}
  
  \title{Bellwethers: A Baseline Method\\
        For Transfer Learning}  
  
  \author{Rahul Krishna,~\IEEEmembership{Member,~IEEE,}
    \and Tim~Menzies,~\IEEEmembership{Member,~IEEE}
    \thanks{The authors
      are with the Department of Computer Science, North Carolina State University, USA.
      E-mail: i.m.ralk@gmail.com, tim@menzies.us}
    \thanks{Manuscript received XXXX XX, 20XX; revised XXXX XX, 20XX.}}
  
  \IEEEtitleabstractindextext{%
    \begin{abstract}
    
Software analytics builds quality prediction models
for software projects. Experience shows that (a) the more projects studied, the more
varied are the conclusions; and (b)
project managers lose faith in the results of software analytics if those results keep changing.
To reduce  this  conclusion instability,
we propose the use of ``bellwethers'': given N projects from a community the bellwether
is the project whose data     
yields the best predictions on all  others. The bellwethers offer a way to mitigate conclusion instability because conclusions about a community are stable as long as this bellwether continues as the best oracle. Bellwethers are also simple to discover (just wrap a for-loop around standard data miners). When compared to other  transfer learning methods  (TCA+,
 transfer Naive Bayes, value cognitive boosting), using just the bellwether data to construct a simple transfer learner yields comparable predictions.
Further, bellwethers appear in many SE tasks such as defect prediction, effort estimation, and bad smell detection. We hence recommend using bellwethers as a {\em baseline method} for transfer learning against which  future work should be compared.
      
    \end{abstract}
    
    \pagenumbering{arabic} 
    
    \noindent
    }
    
    \maketitle 
    \IEEEdisplaynontitleabstractindextext

\section{Introduction}

Researchers and industrial practitioners routinely make extensive use of 
software analytics for many diverse tasks such as
\bi
\item
Estimating how long it  takes to integrate  new code~\cite{czer11};
\item
Predicting where bugs are most likely~\cite{ostrand04,Menzies2007a};
\item
Determining how long
it   takes to build new code~\cite{turhan11,koc11b}.
\ei
Large organizations like Microsoft
routinely practice  data-driven policy
development where
organizational policies are learned from an extensive analysis of
large datasets collected from developers~\cite{export:208800,theisen15}.
For more examples of software analytics, see \cite{me13c,bird2015art}.

A premise  of software data analytics is that there exists data from which we can learn models.
When local data is scarce, sometimes it is possible to use data collected from other projects either at the local site, or other sites.
  That is, when building software
quality predictors, it might be best
to look at more than just the local
data. To do this, recent research has been exploring the problem of {\em 
transferring} data from one project to another for the purposes of data 
analytics. These research have focused on two methodological variants 
of transfer learning: (a)~dimensionality transform based techniques by Nam, 
Jing et al.~\cite{Nam2013,Nam2015,Jing2015} and (b)~the similarity based 
approaches of Kocaguneli, Peters and Turhan et  
al.~\cite{kocaguneli2011find,kocaguneli2012,turhan09,peters15}.  One 
problem with transfer learning is  {\em conclusion instability} which may be 
defined as follows:
\begin{quote}
{\em The more
data we inspect from more projects, the more our conclusions change.}
\end{quote}

The problem with conclusion instability is that the assumptions used to 
make prior policy decisions may no longer hold.
Conclusion instability in software engineering, specifically in transfer 
learning, is well documented.
For example, 
Zimmermann
et al.~\cite{zimm09} learned defect predictors from 622 pairs
of projects {\em project1, project2}. In only 4\% of pairs,
   predictors from {\em project1} worked on
{\em project2}. 
Also, Turhan~\cite{me12d} studied defect prediction results from 28 recent  
studies, most of which offered widely differing conclusions about what most 
influences software defects. 

From the perspective of transfer learning, this instability means that learners that rely on the source of data would also become unreliable. Conclusion  instability is  very 
unsettling for software project managers
struggling to find general policies. Hassan~\cite{Hassan17} cautions that 
managers lose faith
 in the results of software analytics if those results keep changing.
 Such instability prevents project managers from 
offering clear guidelines
on  many issues including (a)~when a certain module should be inspected;
(b)~when modules should  be refactored; (c)~where to focus
expensive testing procedures; (d) what return-on-investment
might be expected  after purchasing an
expensive tool; etc.

How to support those managers, who seek stability
in their conclusions, while also allowing new projects
to take full benefit of  the data from recent projects? Perhaps if we cannot {\em generalize} 
from all data, a more achievable goal is to {\em
slow} the pace of conclusion change.
While it may be
a fool's errand  to wait for   globally stable SE
conclusions, one   approach is  to declare one  project as  the {\em ``bellwether''}\footnote{According to the Oxford English Dictionary, the
``bellwether'' is the leading sheep of a flock, with a bell on its neck.} which
should be used to make conclusions about all other projects. Note that conclusions are stable
for as long as this bellwether continues to be the best oracle for that 
community. This ``bellwether'' project would also act as an excellent source to 
perform transfer learning.

In this paper, we first identify a \textit{bellwether effect} and show that it may be used to generate stable conclusions. We offer the following definition:

\bi
\item \textit{The bellwether effect} states that when a community works on 
software, then  there exists one exemplary project, called the bellwether,
which can define   predictors for the others.
\ei
    
\respto{3-2-A} From this bellwether effect we show that we may construct a \textit{baseline 
transfer learner} called the \textit{bellwether method} to benchmark other more complex transfer learners. In other words, 

\bi 
\item In the \textit{bellwether method}, we search for the exemplar bellwether project and construct a transfer learner with it. This transfer learner is then used to predict for effects in future data for that community.
\ei

This work presents a significant extension to our initial findings on bellwethers~\cite{krishna16a}:
\be
\item \textbf{Generalizing Transfer Learners:} Much of the prior
work on transfer learning, including our initial work~\cite{krishna16a} only explored
one domain (defect prediction). Here, we  study the effectiveness of transfer learning
for
\bi
\item Code smells detection (specifically God Class and Feature Envy); 
\item Effort estimation; 
\item Issue lifetime estimation; and 
\item Defect Prediction.
\ei 

 \item \textbf{\respto{3-2-B} Bellwethers as a baseline transfer learner:} Our 
 initial 
 work~\cite{krishna16a}
compared bellwethers against two other transfer learners invented
by ourselves. In this paper, we explore other state-of-the-art 
algorithms such as:
\bi 
\item Transfer Component Analysis (referred to henceforth as 
TCA+)~\cite{Nam2013};
\item Transfer Naive Bayes (hereafter referred to as TNB)~\cite{Ma2012}; and 
\item Value Cognitive Boosting Learner~\cite{vcboost16}.
\ei  
In doing the above, we identified that the transfer learning literature lacks a 
simple baseline to compare and contrast the various transfer learners. To 
address this, we use the bellwether effect to construct a baseline transfer 
learner (using the bellwether method). To  the  best  of  our  knowledge,  
this  is  the  first  report  to offer a baseline for transfer learning and to 
undertake a case study of all the state-of-the-art transfer learners and 
validate their usability in domains other than defect prediction with respect 
to this baseline.

\item \textbf{Addressing Conclusion Instability with Bellwethers:}
We show that depending on the source dataset, there can be large variances in 
the performance of transfer learner. Further, we show that different source 
datasets can lead to different (and often contradicting) conclusions. We show 
that these issues can be potentially addressed using the bellwether dataset.
\item \textbf{Richer Replication Package:} We have made available a updated and 
a much richer replication
package at https://goo.gl/jCQ1Le. The newer replication package
consists of all the datasets used in this paper, in addition
to mechanisms for computation of other statistical
measures.
\ee

Additionally, this paper makes the following  empirical, methodological, and  
pragmatic contributions.
{\em Empirically}, the key contribution of this paper is the discovery that
  simple methods can find general conclusions across multiple SE projects. 
  \respto{3-1-A} 
  While we \textit{cannot} show that this holds for {\em all} SE domains, we can report 
  that it has offer satisfactory results {\em on three out of the four domains 
  that we have studied 
  so far}; i.e. code 
  smell detection, effort estimation, and defect 
  prediction. In one of our domains, issue lifetime estimation, the evidence 
  supporting the usefulness bellwethers was unsatisfactory. \respto{3-1-B} But 
  our results show that seeking bellwethers may be a simple starting point to 
  begin to reason about software projects.

{\em Pragmatically}, we assert that simple methods should always be preferred to more complex ones-- particularly if we hope for those methods to be used widely in the industry. Other researchers agree with our assertion. In a recent paper, Xu et al.~\cite{Xu15a} discuss the cost
of increasing software complexity: as complexity increases users use fewer and fewer of the available configuration options; i.e. they tend to utilize less and less of the power of that software. This is relevant to transfer learning since standard methods, other than bellwethers, come with so many configuration options that even skilled users have trouble exploiting them all.

{\em Methodologically}, the simplicity associated with the discovery and the 
use of bellwethers is encouraging for further research in software engineering.
 Initial experiments with transfer learning in SE built quality predictors from the {\em union} of data taken from  multiple projects. That approach lead to   poor
results, so researchers
turned to {\em relevancy filters} to find what small subset of
the data  was relevant to the current problem~\cite{turhan09}
and then the dimensionality transform methods of Nam, Jing et al were 
developed. In this paper, we demonstrate the use ``bellwethers"  as a baseline 
transfer learning method for software analytics. As described in the next 
section, bellwethers have all the properties desirable for a baseline method 
such as simplicity of implementation and broad applicability. In the case of 
transfer learning, such a baseline would have greatly assisted in justifying 
the need for increasingly complex 
methods~\cite{Nam2013,Nam2015,Jing2015,kocaguneli2011find,kocaguneli2012,turhan09,peters15}.
\respto{3-1-C} 
While we cannot claim that such simple baselines are always better (they fail 
in the case of issue lifetime estimation), the experiments of this paper 
demonstrate that in some cases other cases (code smell detection and effort 
estimation) bellwethers can perform better than more complex algorithms.

The rest of this paper is structured as follows. In \tion{baseline}, we discuss 
the need for baselines and how the bellwether effect can be used for this.
In \tion{rqs}, we present the research questions this paper attempts to answer. Following this, in \tion{instability}, we present an overview of conclusion instability in software engineering. 
This is followed by \tion{bellw}, there we discuss some work on transfer 
learning and our proposed approach (the bellwether method) and their 
implications to software engineering. 
\tion{domains} provides an overview of all the target domains that are studied in this paper. Additionally, for each sub-domain, we discuss our choice of datasets.  
\tion{methods} discusses the research methodology.
In \tion{res}, we answer each of the research questions that we introduced in section \ref{sect:rqs}.
In \tion{discuss}, we discuss the implications of our findings and attempt to answer some other frequently asked questions. In \tion{threats}, some of the threats to validity of our findings are discussed.
Finally, in \tion{conclusion}, we conclude this paper with the following statement: bellwethers may not always be the best
choice for transfer learning. That said, since bellwethers are so simple to discover and use, it is a reasonable
first choice for benchmarking other approaches. To aid in that benchmarking process,
all our scripts and sample  problems
are available on-line in Github\footnote{https://goo.gl/jCQ1Le}.
Also, to simplify all future references to this material,
the same content has been assigned a digital object identifier
in a public-domain repository\footnote{http://doi.org/10.5281/zenodo.891082}.

\section{Baselining with Bellwethers} 
\label{sect:baseline}
Different domains can require different approaches.
According to Wolpert \& Macready~\cite{Wo97}, no single algorithm can ever
be best for all problems. They caution that for every
class of problem where algorithm $A$ performs best, there is some other class 
of problems where $A$  will perform poorly. Hence, when commissioning a transfer learner for a new domain, there is always the need for some experimentation to match the particulars of the domain to particular transfer learning algorithms.

When conducting such commissioning experiments, it is methodologically useful to have a 
{\em baseline} method; i.e. an algorithm which can generate 
{\em floor performance} values. Such baselines let a developer
quickly rule out any method that falls ``below the floor''.
With this, researchers and industrial practitioners
can achieve fast early results, while also gaining some guidance in all their
subsequent experimentation (specifically: ``try to beat the baseline'').

Using baselines for analyzing algorithms has been endorsed by several experienced researchers.
For example, in his textbook on empirical methods for artificial intelligence, Cohen~\cite{cohen95} 
strongly recommends comparing supposedly sophisticated systems against simpler 
alternatives.
In the machine learning community, Holte~\cite{holte93} uses
the OneR baseline algorithm as a {\em scout}
that runs ahead of a more complicated learners as a way to judge
the complexity of up-coming tasks.
In the software engineering community,  Whigham et al.~\cite{Whigham:2015}
recently proposed baseline methods for effort estimation
(for other baseline methods in effort estimation, see Mittas et al.~\cite{mittas13}). 
Shepperd and Macdonnel~\cite{shepperd12z} argue convincingly that measurements are best viewed as ratios compared to measurements taken from some minimal baseline system.
Work on cross-versus within-company cost estimation has also recommended the use of some  very simple  baseline (they recommend regression as their default model)~\cite{Kitchen}.

 In their recent article on baselines in software engineering, Whigham et al.~\cite{Whigham:2015} propose  guidelines for designing a baseline implementation that include:

\be 
\item Be {\em simple} to describe and implement; 
\item Be {\em applicable to a range of models};
\item Be {\em publicly available} via a reference implementation and associated environment for execution;
\ee
In addition to this, we suggest that baselines should also:
\begin{enumerate}
  \setcounter{enumi}{3}
\item Offer {\em comparable performance} to standard methods. While we do not expect a baseline method to
out-perform all state-of-the-art methods, for a baseline to be insightful, it needs to offer a level of performance
that often approaches the state-of-the-art.
\end{enumerate}

We note that the use of \textit{bellwether method} for transfer learning satisfies all the above criteria. The \textit{bellwether method} is very simple in that it just uses the bellwether dataset to construct a prediction model (without any further complex data manipulation).

As to being {\em applicable to a wide range of domains}, in this paper we apply the bellwether method to several sub-domains in SE, i.e., code-smell detection, effort estimation, issue lifetime estimation, and defect prediction.

As to {\em public availability}, a full implementation of bellwethers including all the case studies presented here (including working implementations of other transfer learning algorithms  and our evaluation methods) are available on-line.

In terms of {\em comparative performance}, for each model, we compared the bellwether method's performance against the the established state-of-the-art transfer learners reported in the literature. In  those comparative results, 
bellwethers were usually as good, and sometimes even a little better, than the state-of-the art.

The use of bellwethers benefits practitioners and researchers attempting transfer learning in several ways:
\begin{enumerate}[leftmargin=*]
\item Researchers can use results of bellwethers as the ``sanity checker''. Experiments shows that the use of bellwethers for transfer learning is comparable to, and 
in some cases better than, other complex transfer learners.
Consequently, when designing new transfer learners, researchers can compare their results to bellwether's as a baseline.
\item Practitioners can also use bellwethers as an ``off-the-shelf'' transfer learner.
For example, in three out of the four domains studied here (code-smells, issue lifetimes, effort estimation), there are no established transfer learners. In such cases, we show that practitioners can simply use bellwethers as transfer learners instead of having to develop new transfer learner (or adapt existing ones from other domains). 

\end{enumerate}

\section{Research Questions }
\label{sect:rqs}
\subsection*{RQ1: How prevalent are ``Bellwethers''?}\label{sect:rq1}

\noindent\textbf{Motivation: }
If bellwethers occur infrequently, we cannot rely on them. Hence, this question explores how
common bellwethers are.

\noindent\textbf{Approach: } To answer this question, we  explore four SE domains: defect prediction, effort estimation, issue lifetime estimation, and detection of code smells. Each domain contains multiple ``communities'' of datasets.
For each domain, we ensured that the datasets were as diverse as possible. To this end, data was gathered according to the following rules:
\bi[leftmargin=*]
\item The data has been used in a prior paper. Each of our datasets for defects, code smells, effort estimation, and issue lifetime estimation has been used previously;
\item
The communities are quite diverse; e.g. the NASA projects from the effort estimation datasets
are   proprietary  while the others are open source projects. Similarly, the God Class is a class level smell and 
Feature Envy is a method level design smell.

\item
In addition, where relevant, the 
projects also vary in their granularity of data description (in case of defect 
prediction, we have defects at file, class, or at a function level 
granularity). 
\ei

\noindent\textbf{Results:} \respto{1-5-A}In a result consistent with 
bellwethers 
being prevalent, we find that three out of these four domains have a 
bellwether dataset; i.e. a single dataset from which a superior quality 
predictor 
can be generated for the rest of that community.

\subsection*{RQ2: How does the bellwether dataset fare against within-project dataset?}

\noindent\textbf{Motivation: } One premise of transfer learning is that using data from
other projects are as useful, or better, than using data
from within the same project. This research questions tests 
that this premise holds for bellwethers.

\noindent\textbf{Approach: } 
To answer this question, we reflect on datasets with temporal within-project 
data. One of our communities in defect prediction (APACHE) comes in 
multiple versions. Here, each version is a historical release where version 
$i$ was written before version $j$ where $j > i$. For this community, RQ2 
was explored as follows:

\bi[leftmargin=*]
\item The last version (version $N$) of each project was set aside as a hold-out.
\item Using an older version ($N-1$) we find the bellwether dataset. 
\item A defect predictor was then constructed on the bellwether dataset.
\item The predictor was applied
to the latest version (version $N$).
\ei
We compare the above to using the {\em within-project} data; i.e. for each project:
\bi[leftmargin=*]
\item The last version ($N$) of that project was set aside as a hold-out;
\item The older version ($N-1$) of that project was then used to train a
defect predictor.
\item The predictor was then applied to the latest data ($N$).
\ei

\noindent\textbf{Results: } In our experiments, the  bellwether predictions proved to be   as good or better than, those generated from the local data. Note that, as of now, this has been verified only in defect prediction.

\subsection*{RQ3: How well do transfer learners perform across different 
domains?}
\label{sect:rq2_stmt}
\noindent\textbf{Motivation: } Our reading of the literature is that for  homogeneous transfer learning, the current state of the art is to use TCA+. However, note that this result has only been reported for defect prediction and only for a limited number of datasets. In our previous work we reported that Bellwether was better than relevancy based filtering methods. Here we ask if this is true given newer transfer learning methods and different datasets.

\noindent\textbf{Approach: } To answer this question, we compare the ``bellwether'' method~\cite{krishna16a} against 3 other standard transfer learners: (1) TCA+~\cite{Nam2013}; (2) Transfer Naive Bayes~\cite{Ma2012}; and (3) Value Cognitive Boosting~\cite{vcboost16}. In addition we modify these learners appropriately for different sub-domains under study.

\noindent\textbf{Results: } Our simple \textit{bellwether method's} predictions were observed to be superior than those of other transfer learners in two domains: effort estimation and code smell detection. \textit{Bellwether method's} predictions were a close second in defect prediction.

\subsection*{RQ4: How much data is required to find the bellwether dataset?}

{\bf Motivation: } Our proposal to find bellwethers is to compare the 
performance of pairs of datasets from different projects in a round robin 
fashion. However, conclusion instability (as presented in the introduction 
and further explored in \tion{instability}) is a major issue in SE and the 
primary cause of such conclusion instability is the constant influx of new 
data~\cite{ekanayake2009tracking}. Given this, a natural question that 
arises from our experimental approach is the amount of data that is 
required to find the bellwether dataset given the influx of new data.

\noindent{\bf Approach: } To answer this research question, we again consider 
datasets with historical versions of data similar to RQ3. To discover how 
much bellwether data is required, we  incrementally increase the size of the 
bellwether dataset. We stop increments when (a) we notice no statistical 
improvement in using additional version data, or (b) we notice that there is 
a deterioration of performance scores using additional version data. 
Specifically, assuming that the bellwether project contains versions ${1, ..., 
N}$, we construct a prediction model with version $N$ and measure the 
performance scores, then we repeat this by including versions $N,~N-1$ 
and so on. With this, we hope to offer some empirical evidence as to how 
much data is required to discover the bellwether. 

\noindent{\bf Results: } Our experiments show that program managers need 
not wait 
very long to find their bellwethers --  when there are multiple versions of the 
bellwether project, project managers need to only use the latest version of 
that project to perform analytics. Another interesting finding is that in cases 
with no historical logs, only a few hundred samples usually are sufficient for 
creating and testing candidate bellwethers.

\def\checkmark{\tikz\fill[scale=0.4](0,.35) -- (.25,0) -- (1,.7) -- (.25,.15) -- cycle;} 
\begin{figure*}[!t]
\begin{minipage}{0.37\linewidth}
\resizebox{\linewidth}{!}
{\footnotesize\begin{tabular}{|r|c|c|c|c|}\hline
\begin{sideways}~~~~~\cite{fowler99} and~\cite{Kerievsky2005}~~~~\end{sideways} &\begin{sideways}~~~ ~\cite{Lanza2006}~~~~\end{sideways} & \begin{sideways}~~~~~\cite{sq15} ~~~~\end{sideways} &  \begin{sideways}~~~~~\cite{Yamashita2013} ~~~~\end{sideways}& \begin{sideways}~~~~~Dev. Surv~~~~\end{sideways}\\\hline
  Alt. Classes with Diff. Interfaces & & & & \\
  Combinatorial Explosion~\cite{Kerievsky2005} & & &  & \\
  Comments & & & 11 & VL\\
  Conditional Complexity~\cite{Kerievsky2005} & & & 14  & ?\\
  Data Class & \checkmark & &  &\\
  Data Clumps &  &  &  &\\
  Divergent Change & & &  & \\
  Duplicated Code & \checkmark & \checkmark & 1  & VH\\
  Feature Envy & \checkmark & & 8  &\\
  
  Inappropriate Intimacy & & \checkmark &  & L\\
  Indecent Exposure~\cite{Kerievsky2005} & & &  & ?\\
  Incomplete Library Class & & &  &\\
  Large Class & \checkmark & \checkmark & 4  & VH\\
  Lazy Class/Freeloader & & \checkmark & 7  &\\
  Long Method & \checkmark& \checkmark & 2  & VH\\
  Long Parameter List &  & \checkmark & 9  & L \\
  
  Message Chains & & &  & H\\
  Middle Man & &  &  &\\
  Oddball Solution~\cite{Kerievsky2005} & & &  & \\
  Parallel Inheritance Hierarchies & & &  &\\
  Primitive Obsession &  & &  &\\
  
  Refused Bequest & \checkmark & \checkmark &  & \\ 
  Shotgun Surgery & \checkmark& &  & \\
  Solution Sprawl~\cite{Kerievsky2005} & & &  &\\
  Speculative Generality & & &  & L\\
  Switch Statements &  & &  & L\\
  Temporary Field & & \checkmark &  & ?\\\hline
  \end{tabular}}
  \caption{ Bad   smells from different sources.  Check marks (\protect\checkmark) denote   a bad smell was mentioned.
Numbers or symbolic labels (e.g. "VH") denote  a priorization comment (and
``?'' indicates lack of consensus). Empty cells
denote some bad smell listed in column one that was not found relevant
in other studies.
Note: there are many blank cells.}\label{fig:smells}
  \end{minipage}~~~
\begin{minipage}{0.63\linewidth}
\resizebox{\linewidth}{!}{
{\footnotesize\begin{tabular}{|r|r|r|r|r|r|r|r|l|}\hline
ref & cbo & 
rfc & 
lcom & 
dit & 
noc & 
wmc & 
\begin{sideways}\#projects\end{sideways} & 
size \\\hline
\cite{ref01} & + & + & + & - & - & + & 6 & 95-201\\
\cite{ref02} & + & + & + & - & - & + & 12 & 86 classess (3-12kloc) \\
\cite{ref03} & + & + & - &  &  &  & 1 & 1700  (110kloc)\\
\cite{ref04} & + & + & - & + & + & + & 8 & 113\\
\cite{ref05} & + & + & - & + & + & + & 8 & 114\\
\cite{ref06} & + & + & + & + & - &  & 1 & 83\\
\cite{ref07} &  &  &  & + & + &  & 1 & 32\\
\cite{ref08} &  &  &  & + & - &  & 1 & 42-69\\
\cite{ref09} & + & - & - & - & - & - & 1 & 85\\
\cite{ref10} & - & + &  & - & - & + & 3 & 92\\
\cite{ref11} & + & + & + & - & + & + & 1 & 123  (34kloc)\\
\cite{ref12} & + &  &  & + &  & + & 1 & 706\\
\cite{ref13} & + & + & + & - & + & + & 1 & 145\\
\cite{ref14} & + & + & + & + & - & + & 1 & 3677\\
\cite{ref15} & + & + & + &  &  & + & 1 & ? \\
\cite{ref16} & + & + & + & + & + & + & 3 & ?\\
\cite{ref17} & - & + & + & - & - & + & 8 & 113  \\
\cite{ref18} &  & + & + & + & + &  & 2 & 64  \\
\cite{ref19} &  & - &  & - & - & - & 1 & 3344 modules (2mloc) \\
\cite{ref20} & + & + & + & - & - & + & 5 & 395  \\
\cite{ref21} & + & + &  & - & - & + & 1 & 1412  \\
\cite{ref22} & + & + &  & - & - & + & 2 & 9713  \\
\cite{ref23} & + & + & - & - & - & + & 1 & 145  \\
\cite{ref24} &  &  &  & + & - &  & 1 & 145  \\
\cite{ref25} & - & - & - & - & - & - & 1 & 174  \\
\cite{ref26} & - &  &  &  &  & - & 0 & 50  \\
\cite{ref27} & + & + & - & - & - & + & 1 & 145  \\
\cite{ref28} &  & + &  & + & + &  & 2 & 294  \\\hline
total +  & 18 & 20 & 11 & 11 & 8 & 17\\\cline{1-7} 
total -  & 4 & 3 & 7 & 14 & 16 & 4&\multicolumn{1}{r}{KEY:}&\multicolumn{1}{l}{\colorbox{green}{Strong consensus (over 2/3rds)}}\\\cline{1-7}
\multicolumn{2}{l}{Total percents:}&\multicolumn{5}{r}{ ``*'' denotes majority conclusion  in each column}&\multicolumn{2}{l}{\colorbox{yellow}{Some consensus (less than 2/3rds)}}\\\cline{1-7}
 + & \cellcolor{yellow}* 64\%	&\cellcolor{green}* 71\%&	\cellcolor{pink}* 39\%&	39\%&	29\%&	\cellcolor{yellow}* 61\%&\multicolumn{2}{l}{\colorbox{lightgray}{Weak consensus (about half)}}\\\cline{1-7}
 - & 14\%&	11\%&	25\%&\cellcolor{lightgray}* 50\%&	\cellcolor{lightgray}* 57\%&	14\%&\multicolumn{2}{l}{\colorbox{pink}{No consensus }}\\\cline{1-7}
\end{tabular}}}
\renewcommand{\arraystretch}{1}
\caption{ Contradictory conclusions from OO-metrics studies for defect
  prediction. Studies report significant (``+'') or irrelevant (``-'')
  metrics verified by univariate prediction models. Blank
  entries indicate that the corresponding metric is not evaluated in
  that particular study. Colors comment on the most frequent conclusion
  of each column. CBO= coupling between objects; RFC= response
  for class (\#methods executed by arriving messages); LCOM= lack of
  cohesion (pairs of methods referencing one instance variable,
  different definitions of LCOM are aggregated); NOC= number of
  children (immediate subclasses); WMC= \#methods per class. }\label{fig:turhan}
\end{minipage}

\end{figure*}

\subsection*{\respto{1-3-A} RQ5: How effectively do bellwethers mitigate for conclusion instability?}

{\bf Motivation: } In the previous research questions, we established the 
prevalence of  bellwethers (RQ1), we showed its efficacy in constructing a 
baseline transfer learner (RQ3), and we also showed empirically that we can 
discover bellwethers early in the project's life-cycle (RQ4). Since one of the 
primary motivation for seeking bellwethers is due to existence of conclusion 
instability, in this final research question, we ask how one might use the 
bellwether effect to mitigate the
two sources of instability we identify in \tion{instability_definition}: (a) 
performance 
instability, and (b) 
source 
instability. 

\noindent {\bf Approach: } To answer this question, we take two steps:

\bi[leftmargin=*]
\item To verify if the bellwether effect can be used to mitigate performance 
variations, we reflect on the results of the comparison of various transfer 
learners (note that, these also includes bellwethers as a baseline approach). 
First, we try to determine if different sources of data to construct the transfer 
learners produces variances in the performance. Then, we determine if the 
use of bellwethers can address these variances.
\item Next, to verify if bellwethers can be used to derive stable lessons in the 
presence of a variety of data sources. We determine if using different sources 
of data can lead to different conclusions. Then, we determine how the 
use of bellwethers can offer stable conclusion.
\ei

\noindent{\bf Results: } Our experiments show that all the datasets we have 
explored in the four domains studied here exhibit both performance 
instability and source instability. Performance instability causes large 
variances in performance scores of transfer learners depending on source of 
the data used. By using the bellwether effect, we may identify the bellwether 
data set which can then be used as a stable source to construct transfer 
learners. Further, we show that transfer learners constructed using the 
bellwether dataset offer statistically and significantly greater performance 
scores compared to other data sources. The existence of source instability 
causes different lessons to be derived from different data sources. Bellwether 
effect can be used to tackle this by identifying a bellwether dataset from the 
available data sources. The bellwether dataset can then be used to learn 
lessons. As long as the bellwether dataset remains unchanged, we will (a) 
obtain 
the same performance scores for a transfer learner, and (b) the same 
conclusions from the bellwether dataset.

\section{Conclusion Instability in SE}
\label{sect:instability}

\subsection{\respto{1-1} What is conclusion instability?}
\label{sect:instability_definition}
As and when new data arrives, there is a sudden and an unpredictable 
change in conclusions that are derived from that data source. This 
uncertainty accompanying a change in data is termed as 
\textit{conclusion 
instability}. It manifests itself as large variances in 
conclusions and these instabilities usually challenges the validity of the 
policy decisions made prior to arrival of new data. In addition to making 
generating general policies very difficult, it also causes practitioners to 
distrust decisions made from software analytics tools~\cite{Hassan17}. In 
this paper, we define and categorize conclusion instability into two forms: (a) 
performance instability, and (b) source instability.

\noindent \textit{(a) Performance Instability:} This can be noticed during 
ranking studies undertaken to 
pick a reliable data miner. For instance, many researchers run ranking studies 
where 	performance scores are collected from many classifiers which are 
ranked for tasks such as defect 
prediction~\cite{lessmann08, hall2012, 
elish2008predicting, menzies2010defect, gondra2008applying, 
radjenovic2013software, jiang2008techniques, wang2013using, 
mende2009revisiting, li2012sample, khoshgoftaar2010attribute, 
jiang2009variance, ghotra2015revisiting, jiang2008can, 
tantithamthavorn2016automated, fu2016tuning}. These rankings
are then 
used to identify the ``best'' defect predictor. However, 
these prediction tasks assume that future events to be predicted will be near 
identical to past 
events. Therefore, given data in the from $\{x_{train},y_{train}\}$, 
prediction algorithms use this for \textit{training} in order to form a joint 
distribution $P(X,Y)=P(Y|X)P(Y)$ and estimate the conditional 
$\hat{P}^(Y|X_{test})$. These predictions will be good as long as the 
data is a close approximation of the underlying distribution. As the source of 
the data changes, the joint distribution $P(X,Y)$ changes to reflect this new 
data. This gradual change in the underlying distribution of training data with 
the arrival of new data is called \textit{data drift}. It is widely accepted that this \textit{drift} is
the leading cause of instability of prediction models~\cite{Ca09, Ha06, St09}. Performance instability can result in large variances in the quality of predictions. Numerous researchers~\cite{fu2016tuning, agarwal17} have shown that changing only 
the data and retaining the same defect predictor can result in statistically 
significant differences.

\noindent (b) \textit{Source Instability:} This arises due to the constant influx of 
potential 
new 
data sources. In methods such as transfer learning,  where we translate 
quality predictors learned in one data set to another, arrival of new data would 
require changing models all the time as the transfer learners continually 
exchange new models to the already existing ones. However, as 
demonstrated in subsequent parts of this section, each new data source can 
produce completely different and often contradicting conclusions. Identifying 
a reliable source of data from all the available options is a 
pressing issue; more so for methods such as transfer learning since they 
place an inherent faith in quality the data source. If a change in data source 
can also change the conclusions, then not being able to identify a reliable 
data source would limit one from leveraging the full benefits of  transfer 
learning. 	

Impact of these instabilities can be observed in several domains within 
software 
engineering. The studies explored in the rest this section sample some 
instances of instability and its prevalence in the domains 
of software engineering studied here\footnote{\respto{1-2}Note: Due to relatively 
recency of the research on estimating lifetime of open issues and 
comparatively fewer papers, we omit it from this survey of conclusion 
instability.}. 
Note 
the vast 
contradictions 
in conclusions in each of these domains.

\subsection{Code Smells}
Research on software refactoring endorses the use of code-smells as a 
guide for improving the quality of code as a preventative maintenance. 
However, as discussed below, a lot of the research on bad-smells suffers from 
conclusion instability. 

There is much contradictory evidence on whether programmers should take heed of these guidelines or ignore them. For instance, a systematic literature review conducted by Tufano et al.~\cite{Tufano2015} lists dozens of papers that recommend tools for repair and detection of code smells. On the other hand, several other researchers cast doubt on the value of code smells and their use as triggers for change~\cite{Mantyla2004, Yamashita2013,Sjoberg2013}. 

Further, this contradiction is also frequently seen among domain experts. Researchers caution that developers' cognitive biases can lead to misleading assertions that some things are important  when they are not. According to Passos et al.~\cite{passos11}, developers often assume that the lessons they learn from a few past projects are general to all their future projects. They comment, ``past experiences were taken into account without much consideration for their context''~\cite{passos11}. This warning is echoed by J{\o}rgensen \& Gruschke~\cite{jorgensen09}. They report that the supposed software engineering experts seldom use lessons from past projects to improve their future reasoning and that such poor past advice can be detrimental to new projects.~\cite{jorgensen09}.

Other studies have shown some widely-held views are now questionable given new evidence. Devanbu et al. examined responses from 564 Microsoft software developers from around the world. They comment programmer beliefs can vary with each project, but do not necessarily
correspond with actual evidence in that project~\cite{prem16}.

The above remarks seem to hold true for bad smells. As shown in \fig{smells},
there is a significant disagreement on which bad smells are important and relevant to a particular project. In that figure, the first column lists  commonly mentioned bad smells and comes from Fowler's 1999 text~\cite{fowler99}. The other columns show conclusions from other studies about which bad smells matter most\footnote{The {\em developer survey} column shows the results of an hour-long whiteboard session with a group of 12 developers from a Washington D.C. web tools development company. Participants worked in a round robin manner to rank the bad smells they thought were important (and any disagreements were discussed with the whole group)}. From this figure, it is easy to note the lack of consensus among developers, text books, and tools. They all disagree on which bad smells are important; just because one developer strongly believes in the importance of a bad smell, it does not mean that the same belief transfers to other developers. 

In summary, we seek methods like bellwethers in order to draw stable conclusions. A particular challenge in each of the study in~\fig{smells} is the lack of consistent data source over the period of time these studies were undertaken. In such cases, bellwether datasets can be particularly useful.

\subsection{Defect Prediction}

In the area of defect prediction too there are several examples of conclusion instability. As motivating examples, consider the following two findings: (a) Zimmermann et al.~\cite{zimmermann09} showed that when they learned defect predictors from 622 pairs of projects, in only 4\% of pairs, the defect predictors learned from one project pair worked in another. These contradictory conclusions extend to OO metrics as well; and (b) In our previous work, we conducted a large scale systematic literature review~\cite{menzies2011local}. We distilled our findings into a list of  28 studies. We noted that they offered contradictory conclusions regarding the effectiveness of OO metrics. These findings are tabulated in~\fig{turhan}. The figure offers a troubling prospect for managers of a software project. The only concrete finding they can derive from this figure is that response for class is often a useful indicator of defects. Each study makes a clear, but usually different, conclusion regarding the usefulness of other metrics. 

In a study of conclusion instability, Turhan~\cite{turhan12} showed that the reason for this inconsistency is due to dataset drift. That work reported different kinds of data drift within software engineering data, such as: (1) Source component shift; (2) Domain Shift; (3) Imbalanced Data, etc. Further, he noted that all contribute significantly to the issue of conclusion instability. In our previous work, we offered further evidence to such a drift by demonstrating that different clusters within the data provided completely different models~\cite{menzies2011local}. Further, the models built from specialized regions within a specific data set sometimes perform better than those learned across all data. However, new data is constantly arriving, and finding these specialized regions with new data turns into an arduous task. In such cases, tools like bellwethers offer a way to draw conclusions from a stable project. As long as the bellwether project remains unchanged so does the conclusions we derive from that project. 


\subsection{Effort Estimation}

As with code smell detection and defect prediction, conclusion instability seems to be an inherent property of the datasets commonly explored in this area~\cite{me05a}. For example, consider stability tests conducted on Boehm's COCOMO software effort estimation model by Menzies et al.~\cite{me05a}. There, it was found that only the coefficient on lines of code (loc) was stable while the variance in  dozens of other coefficients were extremely large. In fact, in the case of  five coefficients, the values even changed from positive to negative across different samples in a cross-validation study.

Other studies on effort estimation also report very similar findings. J{\o}rgensen~\cite{jorg04} compared
model-based to expert-based methods in 15 different studies. That study reported that: five studies favored expert-based methods, five found no difference, and five
favored model-based methods.  Similarly, Kitchenham et
al.~\cite{Mendes2007} reviewed seven studies to check the effect of data imported
from other organizations as compared with local data for building effort models. Of these seven studies,
three found that models from other organizations were not
significantly worse than those based on local data, while four found
that they were significantly worse. MacDonell and Shepperd~\cite{macdonell07} also performed a review on effort estimation models by replicating Kitchenham et al.~\cite{Mendes2007}. From a total of 10 studies, two were found to be inconclusive, three supported global models, and five supported local models. Similarly, Mair and Shepperd~\cite{mair05} compared regression to analogy methods for
effort estimation and found conflicting evidence.  From a
total of 20 empirical studies, (a) seven recommended regression for building effort estimators;
(b) four were
indifferent; and (c) nine favored analogy.

\section{Bellwethers in Software Engineering}
\label{sect:bellw}

Bellwethers offer a simple solution to mitigating conclusion instability. Rather than exploring 
all available data for some eternal conclusions in SE, we seek bellwether datasets that can offer stable solutions
over longer stretches of time. When we notice the dataset failing, we may seek different bellwethers. In addition to
this, the ability of bellwethers to offer stable conclusions over long periods of time also simplifies another widely explored problem in SE; i.e., the problem of transfer learning. In this section, we summarize the standard approaches to transfer learning, then discusses how we may simplify transfer learning by using bellwethers as a baseline transfer learner.

\subsection{Transfer Learning}\label{sect:transfer}
When there is insufficient data to apply data miners
to learn defect predictors, {\em transfer learning}
can be used to transfer lessons learned from
other {\em source} projects $S$ to the {\em target} project $T$. 

Initial experiments with transfer learning offered very
pessimistic results. Zimmermann et al.~\cite{zimm09} tried to port models 
between two web 
browsers 
(Internet Explorer and Firefox) and found that cross-project prediction 
was 
still not consistent: a model built on Firefox was useful for Explorer, 
but 
not vice versa, even though both of them are similar applications. 
Turhan's initial experimental results were also very negative:
given data from 10 projects, training on $S=9$ source projects and testing on  
$T=1$ target projects resulted
in alarmingly high false positive rates (60\% or more).
Subsequent research realized that data had to be carefully sub-sampled
and possibly transformed before quality predictors from one source are applied to a target project. 
That work  can be  divided two ways:

\bi
\item {\em Homogeneous} vs {\em heterogeneous};
\item {\em Similarity} vs {\em dimensionality transform}.
\ei

{\em Homogeneous, heterogeneous } transfer learning operates on
source and target data that contain the {\em same, different} attribute names
(respectively).
This paper focuses on homogeneous transfer learning, for the following reason.
As discussed in the introduction,
we are concerned with an IT manager trying to propose general policies across 
their IT organization.
Organizations are defined by what they do---which is to say that within one 
organization there is some overlap in task, tools, personnel,
and development platforms. This overlap justifies the use of lessons derived 
from transfer learning. 

Hence, all our dataset contain overlapping attributes. In our case these 
attributes are the metrics gathered for each of the projects. As evidence for 
this, the datasets explored in this paper
fall into 4 domains; each domain contains so called ``communities'' of data 
sets. Each dataset within a community share the same attributes (see 
\fig{datasets}).

As to other kinds of transfer learning, {\em
	similarity} approaches transfer some subset of the
rows or columns of data from source to target. For
example, the Burak filter~\cite{turhan09} builds its
training sets by finding the $k=10$ nearest code
modules in $S$ for every $t\in T$. However, the Burak filter suffered from
the all too common instability problem (here, whenever the source or target is
updated, data miners will learn a new  model
since  different code modules will satisfy
the $k=10$ nearest neighbor criteria). Other
researchers~\cite{kocaguneli2011find,kocaguneli2012}
doubted that a fixed value of $k$ was appropriate for all
data. That work recursively bi-clustered the source data, then pruned
the cluster sub-trees with greatest
``variance'' (where the
``variance'' of a sub-tree is the variance of
the conclusions in its leaves).  This method
combined row selection with row pruning (of nearby rows with
large variance).  Other similarity
methods~\cite{zhang15} combine domain knowledge
with automatic processing: e.g.  data is
partitioned using engineering judgment before
automatic tools cluster the data.  To address
variations of software metrics between different
projects, the original metric values were
discretized by rank transformation according to
similar degree of context factors.   

Similarity approaches uses data in its raw form and as highlighted above, it 
suffers from instability issues. This prompted research on 
{\em Dimensionality transform} methods. These methods
manipulate the raw source data until it matches the
target. In the case of defect prediction, a
``dimension'' was one of the static code
attributes of \fig{static_metrics}.  

An initial attempt on performing transfer learning with {\em Dimensionality 
transform} was undertaken by Ma et al.~\cite{Ma2012} with an algorithm called 
transfer naive Bayes (TNB). This algorithm used information from all of the 
suitable attributes in the training data. Based on the estimated distribution 
of the target data, this method transferred the source information to 
weight instances the training data. The defect prediction model was constructed using 
these weighted training data.

Nam et al.~\cite{Nam2013} originally proposed a
transform-based method that used TCA based
dimensionality rotation, 
expansion, and contraction to align the source
dimensions to the
target. They also proposed a new approach called TCA+, which selected suitable 
normalization options for TCA.

\begin{figure}[t!]
\small

  ~\hrule~

  {\bf \fig{bellwether_framework}.A: Discover}

  \textit{Discover the bellwether dataset for a given community.}
\rahul{In a community $C$, for all pairs of data from
  projects $P_i, P_j \in C$, do the following:
Construct a prediction model with data from project $P_i$ and predict for the target variable in $P_j$ using this model. Note: The term target variable refers to defects, code-smells, issue lifetime, or effort, depending on the community under consideration.
  Report a bellwether if one $P_i$  generates the best predictions in a majority of  $P_j \in C$. Note: The quality of prediction is measured using G-Score for defect-prediction, code smell estimation, and issue-lifetime estimation and by SA for effort estimation.}
  \begin{minipage}{\linewidth}
  \begin{python}
   def discover(datasets):
   "Identify Bellwether Datasets"
     for data_1, data_2 in datasets:
       def train(data_1):
        "Construct quality predictor"
        return predictor
       def predict(data_1):
         "Predict for quality"
         return predictions
       def score(data_1, data_2):
         "Return accuracy of Prediction"
         return accuracy(train(data_1),\
         test(data_2))

     "Return data with best prediction score"

  \end{python}
  \end{minipage}
  ~\hrule~

  {\bf \fig{bellwether_framework}.B: Transfer}
  {\em Using the bellwether, construct a transfer learner.}
\rahul{
Construct a transfer learner on the bellwether data. The choice of transfer learners may include any transfer learner used in the literature. For more details on this, see~\tion{transfer}. Now, apply it to future projects.}

  \begin{python}
   def transfer(datasets):
     "Transfer Learning with Bellwether Dataset"
     bellwether = discover(datasets)
     def learner(data):
      """
      Construct Transfer Learner, using:
      1. TCA+; 2. TNB; 3. VCB; 4. Bellwether method
      """
     def apply_learner(datasets, learner):
      "Apply transfer learner"
      model = learner(bellwether)
      for data in datasets:
 	   if data != bellwether:
 	    train(model)
	    test(data)
	    yield score(model, data)

  \end{python}

  ~\hrule~

  {\bf \fig{bellwether_framework}.C: Monitor}
   {\em Keep track of the performance of Bellwethers for transfer learning.}
   \rahul{
 If the transfer learner constructed in TRANSFER starts to fail, go back to DISCOVER and update the bellwether.
}

    \begin{python}
   def transfer(datasets):
     "Transfer Learning with Bellwether Dataset"
     def fails(data):
      "Return True if predictions deteriorate"
    \end{python}

  ~\hrule~
 \caption{The Bellwether Framework}
 \label{fig:bellwether_framework}
\end{figure}

The above researchers failed to address the imbalance of classes in 
datasets they studied. In SE, when a dataset is gathered the samples in them tend to 
be skewed toward one of the classes. A systematic literature review on software defect 
prediction carried out by Hall et al.~\cite{hall2012} indicated that data 
imbalance may be connected to 
poor performance. They also suggested more studies should be aware of the need to deal 
with data imbalance. More importantly, they assert that the performance 
measures chosen can \textit{hide} the impact of imbalanced data on the real 
performance of classifiers.

An approach proposed by Ryu et al.~\cite{vcboost16} showed that using Boosting-SVM 
combined with class imbalance learner can be used to address skewed datasets. 
They showed improved performance compared to TNB. More recently, in our previous work~\cite{krishna16a}, we showed that a very simplistic transfer learner can be 
developed using the ``bellwether" dataset with Random Forest. We reported
highly competitive performance scores.

When there are no overlapping attributes (in heterogeneous transfer learning) 
Nam et al.~\cite{Nam2015} found they could dispense with the optimizer in TCA+ 
by combining feature
selection on the source/target following by a Kolmogorov-Smirnov test to find 
associated
subsets of columns. Other researchers take
a similar approach, they prefer instead a canonical-correlation analysis
(CCA) to find the relationships between variables in the source
and target data~\cite{Jing2015}.

Considering all the attempts at transfer learning sampled above, our reading of these literature suggests a surprising lack of consistency in the 
choice of datasets, learning methods, and statistical measures while reporting 
results of transfer learning. Further, there was no baseline approach to compare the 
algorithms against. This partly motivated our study.

\subsection{Bellwether Method}
\label{sect:bellw_process}
In the above section, we sampled  some of the work on transfer
learning in software engineering. This rest of this paper asks the question ``is
the complexity of \tion{transfer} really necessary?''
We believe the answer is \textit{no}. To assert this, we propose a framework that 
assumes some software manager has a watching
brief over $N$ projects (which we will
call the {\em community} ``$C$''). As part of those duties,  they can
access issue reports and static code attributes of the community.
Using that data, this manager will apply the a framework described in 
~\fig{bellwether_framework}
which comprises of three operators-- DISCOVER, TRANSFER, MONITOR. 

\be
\item DISCOVER: {\em Using cross-project data within a community, check if that community has a bellwether dataset.} 
\bi 
\item
For all pairs of data from
projects in a community $P_i, P_j \in C$;
\item
Predict for defects/smells/issue-lifetime/effort in $P_j$ using prediction model from data taken from  $P_i$;
\item
A bellwether exists if one $P_i$  generates the most accurate predictions in a majority of  $P_j \in C$.
\ei
\item TRANSFER: {\em Using the bellwether, generate prediction models on new project data.} That is, having learned the bellwether on past data, we now apply it to future projects.
\item MONITOR: {\em Go back to step 1 if the performance 
statistics seen for 
new projects during TRANSFER start decreasing.} \respto{1-4-A} Specifically, 
\bi 
\item As new data arrives to the projects in a 
community ... 
\item When we note that the prediction performance of bellwether is 
\textit{statistically poorer} than it was before ...  
\item Then we can declare that the bellwether has 
failed\footnote{\respto{3-3-A} \respto{1-4-B} we refrained from
proposing a numerical threshold because 
this is a subjective measure. Even with a fixed dataset, it is
still subject to vary with several other factors such as the
prediction algorithm, the transfer learner, hyper-parameters of
several algorithms used here, etc. We therefore recommend
a more conservative approach to declaring that the
bellwether has failed.}, that is when we would 
ideally eschew that bellwether and look for a newer bellwether using the 
DISCOVER step.
\ei 
\ee

\respto{3-3-B} \respto{1-4-C} On line 3 in~\fig{bellwether_framework}.A, we 
just wrap a 
for-loop around
some all pairs of datasets in a community, i.e, data we try every dataset in a 
round-robin
fashion and report the best performing dataset as the
bellwether. It is important to note that this will not necessarily lead 
to a bellwether. Consider a case where all the datasets have very
similar performance scores – in such a case it would not be possible
to report any dataset as being the bellwether. To identify such similarities in performance, we may use statistical methods such as Scott-Knott tests. If, according to Scott-Knott tests, all the datasets in a community as ranked the same, then we cannot claim that there is a bellwether dataset in that community.  However, as discussed later on 
in this paper, we note that
this was not the case in any of the four sub-domains we study here.
In all cases there is a clear distinction between the best
dataset and the worst dataset.

In addition to this simplicity of \fig{bellwether_framework}. An additional 
benefit of this 
DISCOVER-TRANSFER-MONITOR methodology is the 
ability to optionally replace the Bellwether Method in the TRANSFER stage
with any other transfer learner (like TCA+, VCB, TNB, etc.).

\section{Target Domains}
\label{sect:domains}

The rest of this paper attempts to discover bellwethers and assesses the performance of bellwethers as baseline transfer learning method. For this, we explore 4 domains in SE: code smells, issue lifetime estimation, effort estimation, and defect prediction.

\subsection{Code Smells}

According to   Fowler~\cite{fowler99}, bad smells (a.k.a. code smells)
are ``a surface indication that usually corresponds to a deeper problem''. 
Studies suggest a relationship between code smells and poor maintainability or 
defect proneness~\cite{yamashita2013code,yama13,zazworka2011investigating} and 
therefore, smell detection has become an established method to discover source 
code (or design) problems to be removed through refactoring steps, with the aim 
to improve software quality and maintenance. Consequently, code smells are 
captured by popular static analysis tools, like 
PMD\footnote{https://github.com/pmd/pmd}, 
CheckStyle\footnote{http://checkstyle.sourceforge.net/}, 
FindBugs\footnote{http://findbugs.sourceforge.net/}, and 
SonarQube\footnote{http://www.sonarqube.org/}. Until recently, most detection tools for code smells make use of detection rules 
based on the computation of a set of 
metrics, e.g., well-known object-oriented metrics. These metrics are then used 
to set some thresholds for the detection of a code smell. But these rules lead to 
far too many false positives making it difficult for 
practitioners to refactor code~\cite{krishna2016b}. 

\begin{figure*}
\begin{minipage}[]{0.475\linewidth}
\textbf{{\small Defect}}\\[0.1cm]
\resizebox{\linewidth}{!}{%
  \begin{tabular}{l|l|l|l|l|l}
    \hline
    \multicolumn{1}{c|}{\multirow{2}{*}{Community}} & \multicolumn{1}{c|}{\multirow{2}{*}{Dataset}} & \multicolumn{2}{c|}{\# of instances} & \multicolumn{1}{c|}{\multirow{2}{*}{\# metrics}} & \multicolumn{1}{c}{\multirow{2}{*}{Nature}} \bigstrut\\ \cline{3-4}
    \multicolumn{1}{c|}{} & \multicolumn{1}{c|}{} & \multicolumn{1}{c|}{Total} & \multicolumn{1}{c|}{Bugs (\%)} & \multicolumn{1}{c|}{} & \multicolumn{1}{c}{} \bigstrut\\ \hline
    \multirow{5}{*}{AEEEM} & EQ & 325 & 129 (39.81) & \multirow{5}{*}{61} & \multirow{5}{*}{Class} \\
    & JDT & 997 & 206 (20.66) &  &  \bigstrut\\
    & LC & 399 & 64 (9.26) &  &  \\
    & ML & 1826 & 245 (13.16) &  &  \\
    & PDE & 1492 & 209 (13.96) &  &  \\ \hline
    \multicolumn{1}{l|}{\multirow{3}{*}{Relink}} & Apache & 194 & 98 (50.52) & \multirow{3}{*}{26} & \multirow{3}{*}{File} \\
    \multicolumn{1}{c|}{} & Safe & 56 & 22 (39.29) &  &  \\
    \multicolumn{1}{c|}{} & ZXing & 399 & 118 (29.57) &  &  \\ \hline
    \multirow{10}{*}{Apache} & Ant & 1692 & 350 (20.69) & \multirow{10}{*}{20} & \multirow{10}{*}{Class} \\
    & Ivy & 704 & 119 (16.90) &  &  \\
    & Camel & 2784 & 562 (20.19) &  &  \\
    & Poi & 1378 & 707 (51.31) &  &  \\
    & Jedit & 1749 & 303 (17.32) &  &  \\
    & Log4j & 449 & 260 (57.91) &  &  \\
    & Lucene & 782 & 438 (56.01) &  &  \\
    & Velocity & 639 & 367 (57.43) &  &  \\
    & Xalan & 3320 & 1806 (54.40) &  &  \\
    & Xerces & 1643 & 654 (39.81) &  &  \\ \hline
  \end{tabular}}\\[0.1cm]
 \textbf{{\small Code Smells}}\\[0.1cm]
  \resizebox{\linewidth}{!}{%
    \begin{tabular}{l|l|l|l|l|l}
      \hline
      \multirow{2}{*}{Community}    & \multirow{2}{*}{Dataset} & \multicolumn{2}{c|}{\# of instances} & \multirow{2}{*}{\# metrics} & \multirow{2}{*}{Nature}  \\ \cline{3-4}
      &                          & Samples         & Smelly (\%)        &                             &                          \\ \hline
      \multirow{11}{*}{Feature Envy} & wct                      & 25              & 18 (72.0)          & \multirow{11}{*}{83}        & \multirow{11}{*}{Method} \\
      & itext                    & 15              & 7 (47.0)           &                             &                          \\
      & hsqldb                   & 12              & 8 (67.0)           &                             &                          \\
      & nekohtml                 & 10              & 3 (30.0)           &                             &                          \\
      & galleon                  & 10              & 3 (30.0)           &                             &                          \\
      & sunflow                  & 9               & 1 (11.0)           &                             &                          \\
      & emma                     & 9               & 3 (33.0)           &                             &                          \\
      & mvnforum                 & 9               & 6 (67.0)           &                             &                          \\
      & jasml                    & 8               & 4 (50.0)           &                             &                          \\
      & xmojo                    & 8               & 2 (25.0)           &                             &                          \\
      & jhotdraw                 & 8               & 2 (25.0)           &                             &                          \\ \hline
      \multirow{11}{*}{God Class}    & fitjava                  & 27              & 2 (7.0)            & \multirow{11}{*}{62}        & \multirow{11}{*}{Class}  \\
      & wct                      & 24              & 15 (63.0)          &                             &                          \\
      & xerces                   & 17              & 11 (65.0)          &                             &                          \\
      & hsqldb                   & 15              & 13 (87.0)          &                             &                          \\
      & galleon                  & 14              & 6 (43.0)           &                             &                          \\
      & xalan                    & 12              & 6 (50.0)           &                             &                          \\
      & itext                    & 12              & 6 (50.0)           &                             &                          \\
      & drjava                   & 9               & 4 (44.0)           &                             &                          \\
      & mvnforum                 & 9               & 2 (22.0)           &                             &                          \\
      & jpf                      & 8               & 2 (25.0)           &                             &                          \\
      & freecol                  & 8               & 7 (88.0)           &                             &                          \\ \hline
    \end{tabular}} 
\end{minipage}~~~~~~~~~~~~\begin{minipage}[]{0.4\linewidth}
\textbf{{\small Effort Estimation}}\\[0.1cm]
\resizebox{\linewidth}{!}{%
  \begin{tabular}{l|l|l|l|l}
    \hline
    Community               & Dataset  & Samples & Range (min-max) & \# metrics          \bigstrut\\\hline
    \multirow{5}{*}{Effort} & coc10 & 95      & 3.5 - 2673    & \multirow{5}{*}{24} \bigstrut\\ \cline{2-4}
    & nasa93   & 93      & 8.4 - 8211    &\bigstrut\\ \cline{2-4}
    & coc81    & 63      & 5.9 - 11400   &\bigstrut\\ \cline{2-4}
    & nasa10 & 17      & 320 - 3291.8  &\bigstrut\\ \cline{2-4}
    & cocomo   & 12      & 1 - 22        &\bigstrut\\ \hline
  \end{tabular}}\\[0.1cm]
\textbf{{\small Issue Lifetime}}\\[0.1cm]
\resizebox{\linewidth}{!}{%
  \centering
  \begin{tabular}{l|l|l|l|l}
    \hline
    \multirow{2}{*}{Community}      & \multirow{2}{*}{Dataset} & \multicolumn{2}{c|}{\# of instances} & \multirow{2}{*}{\# metrics} \\ \cline{3-4}
    &                       & Total                  & Closed (\%) &                             \\ \hline
    \multirow{5}{*}{camel}        & 1 day   & \multirow{5}{*}{5056} & 698 (14.0) & \multirow{5}{*}{18} \\
    & 7 days  &                       & 437 (9.0)  &                     \\
    & 14 days &                       & 148 (3.0)  &                     \\
    & 30 days &                       & 167 (3.0)  &                     
    \bigstrut\\
     \hline
    \multirow{5}{*}{cloudstack}   & 1 day   & \multirow{5}{*}{1551} & 658 (42.0) & \multirow{5}{*}{18} \\
    & 7 days  &                       & 457 (29.0) &                     \\
    & 14 days &                       & 101 (7.0)  &                     \\
    & 30 days &                       & 107 (7.0)  &                     
    \bigstrut\\
     \hline
    \multirow{5}{*}{cocoon}       & 1 day   & \multirow{5}{*}{2045} & 125 (6.0)  & \multirow{5}{*}{18} \\
    & 7 days  &                       & 92 (4.0)   &                     \\
    & 14 days &                       & 32 (2.0)   &                     \\
    & 30 days &                       & 45 (2.0)   &                     
    \bigstrut\\
     \hline
    \multirow{5}{*}{node}       & 1 day   & \multirow{5}{*}{2045} & 125 (6.0)  
    & \multirow{5}{*}{18} \\
& 7 days  &                       & 92 (4.0)   &                     \\
& 14 days &                       & 32 (2.0)   &                     \\
& 30 days &                       & 45 (2.0)   &                     \bigstrut\\
\hline
    \multirow{5}{*}{deeplearning} & 1 day   & \multirow{5}{*}{1434} & 931 
    (65.0) & \multirow{5}{*}{18} \\
    & 7 days  &                       & 214 (15.0) &                     \\
    & 14 days &                       & 76 (5.0)   &                     \\
    & 30 days &                       & 72 (5.0)   &                     
    \bigstrut\\
     \hline
    \multirow{5}{*}{hadoop} & 1 day   & \multirow{5}{*}{12191} & 40 (0.0)    & \multirow{5}{*}{18} \\
    & 7 days  &                        & 65 (1.0)    &                     \\
    & 14 days &                        & 107 (1.0)   &                     \\
    & 30 days &                        & 396 (3.0)   &                     
    \bigstrut\\
     \hline
    \multirow{5}{*}{hive}   & 1 day   & \multirow{5}{*}{5648}  & 18 (0.0)    & \multirow{5}{*}{18} \\
    & 7 days  &                        & 22 (0.0)    &                     \\
    & 14 days &                        & 58 (1.0)    &                     \\
    & 30 days &                        & 178 (3.0)   &                     
    \bigstrut\\
     \hline
    \multirow{5}{*}{ofbiz}  & 1 day   & \multirow{5}{*}{6177}  & 1515 (25.0) & \multirow{5}{*}{18} \\
    & 7 days  &                        & 1169 (19.0) &                     \\
    & 14 days &                        & 467 (8.0)   &                     \\
    & 30 days &                        & 477 (8.0)   &                     
    \bigstrut\\
     \hline
    \multirow{5}{*}{qpid}   & 1 day   & \multirow{5}{*}{5475}  & 203 (4.0)   & \multirow{5}{*}{18} \\
    & 7 days  &                        & 188 (3.0)   &                     \\
    & 14 days &                        & 84 (2.0)    &                     \\
    & 30 days &                        & 178 (3.0)   &                     
    \bigstrut\\
     \hline
  \end{tabular}}
\end{minipage}
\caption{Datasets from 4 chosen domains.}
\label{fig:datasets}
\end{figure*}
\begin{figure*}[t!]
\begin{minipage}[]{\linewidth}
\centering
\resizebox{\linewidth}{!}{%
\centering
\begin{tabular}{l|p{2cm}|l|p{2cm}|l|p{2cm}|l|p{2cm}|l|p{2cm}|l|p{2cm}}

	\multicolumn{2}{c|}{Size}                          & 
	\multicolumn{2}{c|}{Complexity}                                        & 
	\multicolumn{2}{c|}{Cohesion}  & 
	\multicolumn{2}{c|}{Coupling}                              & 
	\multicolumn{2}{c|}{Encapsulation}     & 
	\multicolumn{2}{c}{Inheritance}         \bigstrut\\ \hline
	Label   & Description                               & Label           & 
	Description                                          & Label & 
	Description            & Label     & 
	Description                                    & Label & 
	Description                    & Label & Description                      
	\bigstrut\\ \hline
	LOC     & Lines Of Code                             & CYCLO           & 
	Cyclomatic Complexity                                & LCOM  & Lack of 
	Cohesion       & FANOUT/IN & Fan Out/In                                     
	& LAA   & Locality of Attribute Accesses & DIT   & Depth of Inheritance 
	Tree        \bigstrut\\ \hline
	LOCNAMM & LOC (without accessor or mutator)         & WMC             & 
	Weighted Methods Count                               & TCC   & Tight Class 
	Cohesion   & ATFD      & Access to Foreign Data                         & 
	NOAM  & Number of Accessor Methods     & NOI   & Number of 
	Interfaces             \bigstrut\\ \hline
	NOM     & No. of Methods                            & WMCNAMM         & 
	Weighted Methods Count (without accessor or mutator) & CAM   & Cohesion 
	Among classes & FDP       & Foreign Data Providers                         
	& NOPA  & Number of Public Attribute     & NOC   & Number of 
	Children               \bigstrut\\ \hline
	NOPK    & No. of Packages                           & AMW             & 
	Average MethodsWeight                                &       
	&                        & RFC       & Response for a 
	Class                           &       &                                & 
	NMO   & Number of Methods Overridden     \bigstrut\\ \hline
	NOCS    & No. of Classes                            & AMWNAMM         & 
	Average Methods Weight (without accessor or mutator) &       
	&                        & CBO       & Coupling Between 
	Objects                       &       &                                & 
	NIM   & Number of Inherited Methods      \bigstrut\\ \hline
	NOMNAMM & Number of Not Accessor or Mutator Methods & MAXNESTING      & Max 
	Nesting                                          &       
	&                        & CFNAMM    & Called Foreign Not Accessor or 
	Mutator Methods &       &                                & NOII  & Number 
	of Implemented Interfaces \bigstrut\\ \hline
	NOA     & Number of Attributes                      & CLNAMM          & 
	Called Local Not Accessor or Mutator Methods         &       
	&                        & CINT      & Coupling 
	Intensity                             &       
	&                                &       &                                  
	\bigstrut\\ \hline
	&                                           & NOP             & Number of 
	Parameters                                 &       &                        
	& MaMCL     & Maximum Message Chain Length                   &       
	&                                &       &                                  
	\bigstrut\\ \hline
	&                                           & NOAV            & Number of 
	Accessed Variables                         &       &                        
	& MeMCL     & Mean Message Chain Length                      &       
	&                                &       &                                  
	\bigstrut\\ \hline
	&                                           & ATLD            & Access to 
	Local Data                                 &       &                        
	& CA/CE/IC  & Afferent/ Efferent/ Inheritance coupling         &       
	&                                &       &                                  
	\bigstrut\\ \hline
	&                                           & NOLV            & Number of 
	Local Variable                             &       &                        
	& CM        & Changing Methods                               &       
	&                                &       &                                  
	\bigstrut\\ \hline
	&                                           & WOC             & Weight Of 
	Class                                      &       &                        
	& CBM       & Coupling between Methods                       &       
	&                                &       &                                  
	\bigstrut\\ \hline
	&                                           & MAX\_CC/AVG\_CC & 
	Maximum/ Average McCabe                              &       
	&                        &           
	&                                                &       
	&                                &       &                                  
	\bigstrut\\ 
\end{tabular}}
\captionof{figure}{Static code metrics used in defects and code smells data 
sets.}
\label{fig:static_metrics}
\end{minipage}\\[0.2cm]
\begin{minipage}[]{0.5\linewidth}
\centering
\resizebox{\linewidth}{!}{%
  \centering
\begin{tabular}{lll}
 
  \multicolumn{1}{l|}{Commit}                  & 
\multicolumn{1}{l|}{Comment}          & \multicolumn{1}{l}{Issue}    
\bigstrut\\ \hline
  \multicolumn{1}{l|}{nCommitsByActorsT}       & 
\multicolumn{1}{l|}{meanCommentSizeT} & issueCleanedBodyLen           
\bigstrut\\
  \multicolumn{1}{l|}{nCommitsByCreator}       & 
\multicolumn{1}{l|}{nComments}        & nIssuesByCreator              
\bigstrut\\
  \multicolumn{1}{l|}{nCommitsByUniqueActorsT} & 
\multicolumn{1}{l|}{}                 & nIssuesByCreatorClosed        
\bigstrut\\
  \multicolumn{1}{l|}{nCommitsInProject}       & 
\multicolumn{1}{l|}{}                 & nIssuesCreatedInProject       
\bigstrut\\
  \multicolumn{1}{l|}{nCommitsProjectT}        & 
\multicolumn{1}{l|}{}                 & nIssuesCreatedInProjectClosed 
\bigstrut\\
  \multicolumn{1}{l|}{}                        & 
\multicolumn{1}{l|}{}                 & nIssuesCreatedProjectClosedT  
\bigstrut\\
  \multicolumn{1}{l|}{}                        & 
\multicolumn{1}{l|}{}                 & nIssuesCreatedProjectT        
\bigstrut\\ \hline
  \multicolumn{1}{r|}{Misc.}                                        & 
\multicolumn{2}{l}{nActors, nLabels, nSubscribedByT}                  
\bigstrut\\ \hline
\end{tabular}}
  \captionof{figure}{Metrics used in issue lifetimes data.}
  \label{fig:issue_metrics}
  \vfill
\end{minipage}~~\begin{minipage}[]{0.5\linewidth}
\centering
\resizebox{\linewidth}{!}{%
	\centering
\begin{tabular}{l|p{1.5cm}|l|p{2cm}|l|p{1.5cm}|l|p{1.5cm}}
	\multicolumn{2}{c|}{Personnel} & \multicolumn{2}{c|}{Product} & 
	\multicolumn{2}{c|}{System}  & \multicolumn{2}{c}{Other}    \bigstrut\\ 
	\hline
	Label  & Description            & Label  & Description         & Label & 
	Description          & Label & Description           \bigstrut\\ \hline
	ACAP   & Analyst Capability     & CPLX   & Prod. Complexity    & DATA  & 
	Database size        & DOCU  & Documentation         \bigstrut\\ \hline
	APEX   & Applications Exp.      & SCED   & Dedicated Schedule  & PVOL  & 
	Platform volatility  & TOOL  & Use of software tools \bigstrut\\ \hline
	LEXP   & Language Exp.          & SITE   & Multi-side dev.     & RELY  & 
	Required Reliability &       &                       \bigstrut\\ \hline
	MODP   & Modern Prog. Practices & TURN   & turnaround time     & RUSE  & 
	Required Reuse       &       &                       \bigstrut\\ \hline
	PCAP   & Programmer Capability  &        &                     & STOR  & \% 
	RAM               &       &                       \bigstrut\\ \hline
	PLEX   & Platform Exp.          &        &                     & TIME  & \% 
	CPU time          &       &                       \bigstrut\\ \hline
	VEXP   & Virtual Machine Exp.   &        &                     & VIRT  & 
	Machie volatility    &       &                       \bigstrut\\ \hline
	PCON   & Personnel Continuity   &        &                     &       
	&                      &       &                       \bigstrut\\ 
\end{tabular}}
\captionof{figure}{Metrics used in effort estimation dataset.}
\label{fig:effort_metrics}
\end{minipage}
\end{figure*} 

Recently, the research community is changing rapidly in terms of defining 
novel methodologies that incorporate additional information to detect 
code-smells. Much progress has been 
made in towards adopting machine 
learning tools to classify code smells from examples, easing the build of 
automatic code smell detectors, thereby providing a better-targeted detection. 
Kreimer~\cite{kreimer05} proposes an adaptive detection to combine known 
methods for finding
design flaws Large Class and Long Method on the basis of metrics. Khomh et al.~\cite{khomh09} proposed a Bayesian approach to detect 
occurrences of the Blob antipattern on open-source programs. 
Khomh et al.~\cite{khomh11} also presented BDTEX, a GQM approach to build 
Bayesian Belief Networks from the 
definitions of antipatterns.  
Yang et al.~\cite{maiga12} study the judgment of individual users by applying 
machine learning algorithms on code clones. These studies were not included in our comparison as the data was not readily available for us to reuse.

More recently, Fontana et al.~\cite{font16} in their study of several code 
smells, considered 74 systems for their analysis and validation. They 
experimented with 16 different machine learning algorithms. They made available 
their dataset, which we have adapted for our applications in this study. These datasets were generated using the Qualitas 
Corpus (QC) of systems~\cite{tempero}. The Qualitas corpus is composed of 111 
systems written in Java, characterized by different sizes and belonging to 
different application domains. Fontana et al.~\cite{font16} selected a subset of 
74 systems for their analysis. The authors computed a large set of object-oriented 
metrics belonging to class, method, package, and project 
level. A detailed list of metrics and 
their definitions are available in appendices of~\cite{font16}. The code smells repository we use comprises of 22 datasets for two different code 
smells: Feature envy and God Class. The God Class code smell class level code smell that refers to classes 
that tend to centralize the intelligence of the system. Feature Envy is a method level smell that tends to use many attributes of other classes (considering also attributes accessed through accessor methods). 

The number of samples in these datasets are particularly small. For our analysis, we retained 
only datasets with at least 8 samples so that the transfer learners used here function reliably. This lead 
us to a total of 22 datasets shown in \fig{datasets}.

\subsection{Issue Lifetime Estimation}

Open source projects use issue tracking systems to enable effective 
development  and  maintenance  of  their software  systems. Typically, issue  
tracking  systems collect information about system failures, feature requests, 
and system  improvements. Based  on  this  information  and  actual project 
planing, developers select the issues to be fixed. Predicting the time it may 
take to close an issue has multiple benefits for the developers, managers, and  
stakeholders  involved  in  a  software  project.  Predicting
issue lifetime helps software developers better prioritize work;
helps  managers  effectively  allocate  resources  and  improve
consistency  of  release  cycles;  and  helps  project  stakeholders
understand changes in project timelines and budgets. It is also
useful to be able to predict issue lifetime specifically when
the issue is created. An immediate prediction can be used, for
example, to auto-categorize the issue or send a notification if
it is predicted to be an easy fix.

As an initial attempt, Panjer~\cite{panjer} used logistic regression models to 
classify bugs as closing in 1.4, 3.4, 7.5, 19.5, 52.5, and 156 days, and 
greater than 156 days. He was able to achieve an accuracy of 34.9\%. Giger et 
al.~\cite{giger} used models constructed with decision trees to predict for 
issue lifetimes in Eclipse, Gnome, and Mozilla. They were able obtain a peak 
precision of 65\% by dividing time into 1, 3, 7, 14, 30 days. Zhang et al.~\cite{zhang13issue} developed a 
comprehensive system to predict lifetime of issues. They used a 
Markov model with a kNN-based classifier to perform their prediction. More 
recently, Rees-Jones et al~\cite{rees} showed that using Hall's CFS 
feature selector and C4.5 decision tree learner a very reliable prediction of 
issue lifetime could be made. 

\fig{datasets} shows a list of 8 projects used to study issue 
lifetimes. These projects were selected by our industrial partners 
since they use, or extend, software from these projects. It forms a part of 
an ongoing study on prediction of issue lifetime by Rees-Jones et 
al.~\cite{rees}. The authors note that one issue in preparing their data was a small 
number of \emph{sticky} issues. They define sticky issues as one 
which was not yet closed at the 
time of data collection. As recommended by Rees-Jones et 
al.~\cite{rees}, we removed these 
sticky issues from our datasets. 

In raw form, the data consisted of sets of JSON files for each  
repository, each file contained one type of data regarding the 
software repository (issues, commits, code contributors, changes to 
specific files). In order to extract data specific to issue lifetime, 
we did similar preprocessing and feature extraction on the raw 
datasets as suggested by ~\cite{rees}.

\subsection{Effort Estimation}

The nature of effort estimation and the corresponding data is unlike that of other domains. 
Firstly, while domains like defect prediction datasets often store several thousand samples of 
defective and non-defective samples, effort data is usually smaller with only a 
few dozen samples at most. Secondly, unlike defect dataset or code smells, 
effort is measured using, say \textit{man-hours}, which is a continuous 
variable. These differences requires us to significantly modify existing 
transfer learning techniques to accommodate this kind of data. 

Transfer learning attempts have been made in defect prediction before albeit with limited success. 
Kitchenham et al.~\cite{Kitchen} reviewed 7 published transfer studies in 
effort estimation. They found that in most cases, transferred data 
generated worse predictors than using within-project information. Similarly, Ye 
et al.~\cite{yang11} report that the tunings to Boehm’s COCOMO model have 
changed radically for new data collected in the period 2000 to 2009. Kocaguneli et al.~\cite{kocaguneli2012} used 
analogy-based effort estimation with relevancy filtering using a method called 
TEAK for studying transfer learning in effort estimation. He found that it 
outperforms other approaches such as linear regression, neural networks, and 
traditional analogy-based reasoners. Since then, however, newer more 
sophisticated transfer learners have been introduced. Krishna et 
al.~\cite{krishna16a} suggest that relevancy filtering (for defect prediction 
tasks) would never have been necessary in the first place if researchers had 
instead hunted for bellwethers. Therefore, in this paper, we revisit transfer 
learning in effort estimation keeping in mind these changing trends.

For our experiments, we consider effort estimation data expressed in terms of the 
COCOMO ontology: 23 attributes describing a software project, as well as aspects of its
personnel, platform, and system features (see~\fig{effort_metrics} for details). The data is gathered using Boehm's 2000 COCOMO model. The data was made available by Menzies et al.~\cite{Menzies2016} who show that this model works better than (or just as well as) other models they've previously studied. We use 5 datasets shown in \fig{datasets}. Here, COC81 is the original
data from 1981 COCOMO book~\cite{cocomo}. This comes from projects dated from 1970 to 1980. NASA93 is NASA
data collected in the early 1990s about software that supported the planning activities for the International
Space Station. The other datasets are NASA10 and COC05 (the latter is proprietary and cannot be released
to the research community). The non-proprietary data (COC81 and NASA93 and NASA10) are available at http://tiny.cc/07wvjy.

\subsection{Defect Prediction}


Human  programmers  are  clever,  but  flawed.   Coding  adds functionality,  
but  also  defects, so software  will crash (perhaps at the most awkward or 
dangerous time) or  deliver     wrong  functionality. Since  programming    introduces  defects  into  programs, it is important to
test them before they are used.  Testing is expensive. 
According to Lowry et al. software assessment budgets are finite while 
assessment effectiveness increases exponentially with assessment 
effort~\cite{LowryBK98}.   Exponential  costs  quickly  exhaust  finite  
resources, so  standard  practice  is  to  apply  the  best  available methods 
only on code sections that seem most critical. One such approach is to use defect predictors learned from static 
code attributes. Given software described in the attributes of Figures \ref{fig:static_metrics}, 
\ref{fig:issue_metrics}, and \ref{fig:effort_metrics}, data miners can
learn where the probability of software defects is highest. These static code attributes can be automatically collected, even 
for very large systems~\cite{nagappan05}. Although other methods like  manual code reviews are much more accurate in identifying defects, they take much higher effort to find a defect and also are relatively slower. For example, depending on the review methods, 8 to 20 LOC/minute can be inspected and this effort repeats for all members of the review team, which can be as large as four or six people~\cite{me02f}. This is complementary to defect prediction techniques. These techniques 
enable developers to target defect-prone areas faster, but do not guide 
developers toward a particular fix. The defect prediction models are easier 
to use in that sense that they prioritize \textit{both} code review and testing 
resources (these areas complement each other). 

Moreover, defect predictors often  find the location of  70\% (or more)
of the defects in code~\cite{me07b}.
Defect predictors have some level of generality:
predictors learned at NASA~\cite{me07b} have also been found useful elsewhere
(e.g. in Turkey~\cite{tosun10,tosun09}).
The success of this method in  predictors in finding bugs is   markedly
higher than other currently-used
industrial
methods such as manual code reviews. For example, 
a  panel at {\em IEEE Metrics
	2002}~\cite{shu02} concluded that manual software  reviews can find 
	${\approx}60\%$ 
of defects.
In another work, 
Raffo documents the typical    defect detection capability of
industrial review methods:   around 50\%
for full Fagan inspections~\cite{fagan76} to
21\% for less-structured inspections.

Not only do static code defect predictors perform well compared to manual 
methods,
they also are competitive with certain automatic methods.
A recent study at ICSE'14, Rahman et al.~\cite{Rahman:2014} compared
(a) static code analysis tools FindBugs, Jlint, and Pmd and (b)
static code defect predictors
(which they called ``statistical defect prediction'') built using logistic 
regression.
They found  no significant differences in the cost-effectiveness
of these  approaches. Given this equivalence, it is significant to note that 
static code defect prediction can be quickly adapted to new languages by 
building lightweight
parsers that find   information like \fig{static_metrics}. The same is not true 
for   static code analyzers-- these need  extensive modification before they 
can be used on new
languages.

For the above reasons, researchers and industrial practitioners use static 
attributes to guide software  quality predictions. Defect prediction has been favored by 
most transfer learning researchers. Further, 
defect prediction models have been reported
at Google~\cite{lewis13}.
Verification and validation (V\&V) textbooks
\cite{rakitin01} advise using static code complexity attributes
to decide which modules are worth manual inspections.

The defect dataset we have used come from 18 projects grouped into 3 communities 
taken
from previous transfer learning studies. The projects measure defects at various 
levels of granularity 
ranging from function-level to file-level. \fig{datasets} summarizes all 
the communities of datasets used in our experiments. 

For the reasons  discussed in \tion{transfer}, we explore homogeneous
transfer learning using the attributes shared by a community.
That is, this study explores intra-community transfer learning
and not cross-community heterogeneous transfer learning.

The first dataset, AEEEM, was used by \cite{Nam2015}. This dataset was 
gathered by D'Amborse et al. \cite{DAmbros2012}, it contains 61 
metrics: 17 object-oriented metrics, 5 previous-defect metrics, 5 entropy 
metrics 
measuring code change, and 17 churn-of-source code metrics. 

The RELINK community data was obtained from work by Wu et al. \cite{Wu2011} who 
used the Understand tool 
\footnote{$\text{http://www.scitools.com/products/}$}, to measure 26 
metrics that calculate code complexity in order to improve the quality of 
defect prediction. This data is particularly interesting because the 
defect information in it has been manually verified and corrected. It has 
been widely used in defect prediction 
\cite{Nam2015}\cite{Wu2011}\cite{basili1996validation}\cite{ohlsson1996predicting}\cite{kim2011dealing}.

In addition to this, we explored two other communities of datasets from the 
SEACRAFT repository\footnote{$\text{https://zenodo.org/communities/seacraft/}$}. The 
 group of data contains defect measures from several Apache projects. 
It was gathered by Jureczko et al. \cite{Jureczko2010}. This dataset 
contains records the number of known defects for each class using a 
post-release bug tracking system. The classes are described in terms of 20 
OO metrics, including CK metrics and McCabes complexity metrics. Each 
dataset in the Apache community has several versions. There are a total of 38 
different datasets. For more information on this dataset 
see~\cite{krishna2016b}.

\section{Methodology}
\label{sect:methods}

\subsection{Learning Methods}\label{sect:met}
In our datasets, the class variable (defects, code-smells, closed issues, and effort) belong to two categories:
\be
    \item \textit{Discrete classes:} The classes in the case of defect prediction, detection of code-smells, and close time of issues have \textit{two} discrete classes. We therefore use learners as \textit{binary classifiers}.
    \item \textit{Continuous classes:} The class variable in the case of effort estimation takes on continuous values. Here we use learners as \textit{regression} algorithms.
\ee
There are many binary classifiers to predict defects (smells or issue lifetime). A comprehensive study on defect prediction was conducted by Lessmann et al.~\cite{lessmann08}. They endorsed the 
use of Random Forests~\cite{Breiman2001} for defect prediction over 
several other methods. This was also true in detecting code smells~\cite{font16}. When a specific transfer learner did not endorse the use of any classification/regression scheme, we used Random Forests (Note: If an explicit reference was made regarding using a specific prediction algorithm by the authors of other transfer learners used in this paper, we use those predictors instead of random forest. e.g., VCB endorses the use of SVMs).

Random Forests is an ensemble learning method that 
builds several decision trees on randomly chosen subsets of data. The 
final reported prediction is the mode of predictions by the trees. 

When the class variable if discrete (as in binary classification), it is known that the fraction of ``positive'' class samples in the training data affects 
the performance of predictors. \fig{datasets}  shows that in 
most datasets, the percentage of ``positive samples'' (i.e., samples that are defective, smelly, or closed) vary between 10\% to 40\% (except in a few, projects like log4j for instance where it is 58\%). Handling this class imbalance has been shown to improve the quality of 
prediction. 

Pelayo and Dick~\cite{Pelayo2007} report that the defect 
prediction is improved by SMOTE~\cite{Chawla2002}. SMOTE works by 
under-sampling majority-class examples and over-sampling minority class 
examples to balance the training data prior to applying prediction models.

After an extensive experimentation, in this study, we randomly sub-sampled examples until the training data had only positive and negative classes in a ratio of 1:2.

Important methodological notes: 
\be
\item sub-sampling was only applied to 
{\em training} data (so the test data remains unchanged).
\item Authors of several transfer learners studied here recommend 
using different predictors. When replicating their studies, we 
adhere to their recommendations.
\item SMOTE is only applicable for classification problems (defect prediction, code smell detection, and issue lifetime prediction). When performing regression for estimation of effort, we don't apply SMOTE.
\ee

\subsection{Evaluation Strategy}
\label{sect:eval}
\subsubsection{Evaluation for Continuous Classes}

For the effort estimation data in \fig{datasets}, the dependent attribute is development effort, measured in terms of calendar hours (at 152 hours per month, including development and management hours). For this, we use the same learning methods as in \tion{met} used as a regressor instead of a classifier.

To evaluate the quality of the learners used for regression, we make use of 
Standardized Accuracy (SA). The use of SA has been endorsed by several 
researchers in SE~\cite{shepperd2012evaluating, LANGDON201616} 
Standard Accuracy is computed as below:
\begin{equation}
SA=1 - \frac{MAR}{\frac{2}{n^2}\sum_{i=1}^{n}\sum_{j=1}^{j<i}|y_i-y_j|}\times 100
\label{eq:MRE}
\end{equation}

Where, MAR  is the mean of the absolute error for the predictor of interest. E.g. for software project estimation, the average of the absolute difference between the effort predicted and the actual effort the project took.

Higher values of SA are considered to be \textit{better}. Note: Some 
researchers have endorsed the use other metrics such as MMRE to 
measure the quality of regressor in effort estimation. We have made 
available a replication package\footnote{https://goo.gl/jCQ1Le} with this 
and other metrics. Interested readers are encouraged to use these.

\subsubsection{Evaluation for Discrete Classes}

In the context of discrete classes, we define positive and negative classes. 
With defects, instances with one or more defects are considered to belong 
to the ``positive class'' and non-defective instances are considered to 
belong to the ``negative class''. Similarly in code smell detection (smelly 
samples belong to ``positive class'') and in issue lifetime estimation (closed 
issues belong to ``positive class''). Prediction models are not ideal, they  
therefore need to be evaluated in terms of statistical performance 
measures. 

For classification problems we construct a confusion matrix, with this we 
can obtain several performance measures such as: (1) \textit{Accuracy}: 
Percentage of correctly classified classes (both positive and negative); 
(2) \textit{Recall or pd}: percentage of the target classes (defective 
instances) predicted. The higher the pd the better ; (3) \textit{False alarm or pf}: percentage of non-defective 
instances wrongly identified as defective. Unlike pd, lower the pf implies better quality; (4) \textit{Precision}: probability of predicted defects 
being actually defective. Either a smaller number of correctly predicted 
faulty modules or a larger number of erroneously predicted defect-free 
modules would result in a low precision.

\rahul{
There are several trade-offs between the metrics described above. There is 
a trade-off between recall rate and false alarm rate where attempts to increase recall leads to larger false alarm, which is undesirable. There is also a 
trade-off between precision and recall where increasing precision lowers recall and vice-versa. These measures alone do not paint 
a complete picture of the quality of the predictor. Therefore, it is very 
common to apply performance metrics that incorporate a combination of 
these metrics. As a result, some authors generally resort to using metrics such as F1 score to 
assess learners~\cite{fu16, kim2008classifying, me07b, 
wang2016automatically}. However, there exists a peculiar challenge with 
using F-measure that is specific to some software engineering problem -- 
the large imbalance between class variables in the datasets commonly 
studied here. For instance, consider the datasets studied in this paper 
shown in \fig{datasets}. There, a number of datasets have highly skewed 
samples. In these cases, several researchers caution against use of 
common performance metrics such as precision or F-measure. Menzies et 
al.~\cite{Menzies2007a} in their 2007 paper showed the negative impact of 
using these metrics. They caution researchers against the use precision 
when assessing their detectors. They recommend other more stable 
measures especially for highly skewed data sets. This concern is echoed 
by several other researchers in SE~\cite{chawla03, kubat97, shatnawi10}. 
Kubat \& Matwin found that the effect of the negative classes (in our 
context this refers to bug-free/smell-free/closed issues) has a profound 
impact on the outcome of these metrics. As a remedy, these authors 
recommend a new evaluation scheme that combines reliable metrics such 
as recall ($pd$) and false-alarm ($pf$).

One such approach that can combine these metrics is to build a \textit{Receiver Operating 
Characteristic (ROC)} curve. ROC curve is a plot of Recall versus False 
Alarm pairing for various predictor cut-off values ranging from 0 to 1. 
The best possible predictor is the one with an ROC curve that rises as steeply as possible and 
plateaus at pd=1. Ideally, for each curve, we can measure the \textit{Area Under Curve 
(AUC)}, to identify the best training dataset. Unfortunately, building an 
ROC is not straight forward in our case. We have used Random Forest for 
predicting defects owing to it's superior performance over several other 
predictors~\cite{lessmann08}. Note that Random Forest lacks a threshold 
parameter, since this threshold parameter is required in order to generate a 
set of points to plot the ROC curve, Random Forest is not capable of 
producing an ROC curve, instead we produce just one point on the ROC 
curve. It is therefore not possible to compute AUC. 

In a previous work, Ma and Cukic~\cite{ma07} have shown that other 
metrics that measure the distance from perfect classification can be 
substituted for AUC in cases where a ROC curve cannot be generated. 
Accordingly, we use the 
the "G-Score" for combining Pd and Pf. Several authors~\cite{Menzies2007a, shatnawi10} have previously shown that such a measure is justifiably better than other measures when the test samples have imbalanced distribution in terms of classes. G-Score can by computed by measuring the mean (geometric/harmonic) between the Probability of True Positives (Pd) and Probability of true negatives (1-Pf). The choice of using geometric mean or harmonic mean depends on the variance in Pd/Pf values. Mathematically, it is known that in cases where samples tend to take extreme values (such as Pd=0 or Pf=1) harmonic mean provides estimates that are much more stable and also more conservative in it's estimate compared to geometric mean~\cite{xia1999proof}. Therefore, we propose the use of G-Score, measured as follows: 

\begin{equation}
G = \frac{2 \times Pd \times(1-Pf)}{1+Pd-Pf}
\label{eq:G}
\end{equation}
}

In this work, for the sake of consistency with other SE literature, we report the measures of Pd and Pf reported in terms of the G-Score. Also, note that with the formulation in \eq{G}, \textit{larger} G-scores are better.

\subsection{Statistics}
\label{sect:stats}
To overcome the inherent randomness introduced by Random Forests and 
SMOTE, we use 30 repeated runs, each time with a different random number 
seed (we use 30 since that is the minimum needed samples to satisfy 
the central limit theorem). Researchers have endorsed the use of repeated 
runs to gather reliable evidence~\cite{vaux2012replicates}. Thus, we 
repeat the whole experiment independently several times to provide 
evidence that the results are reproducible. The repeated runs provide us 
with a sufficiently large sample size (of size 30) to statistically compare all 
the datasets. 
Each repeated run collects the values of Pd and Pf which are then used to estimate the G-Score using \eq{G}. (Note: We refrain from 
performing a cross validation because the process tends to mix the samples 
from training data (the source) and the test data (other target 
projects), which defeats the purpose of this study.)

To rank these 30 numbers collected as above, we use the Scott-Knott test 
recommended by Mittas and Angelis~\cite{mittas13}. Scott-Knott is a 
top-down clustering approach used to rank different treatments. If that 
clustering finds an statistically significant splits in data, then some 
statistical test is applied to the two divisions to check if they are 
statistically significant different. If so, Scott-Knott recurses into both 
halves.

To  apply Scott-Knott, we sorted a list of  $l=40$ values of \eq{G} 
values found in $ls=4$ different methods. Then, we split $l$ into 
sub-lists $m,n$ in order to maximize the expected value of differences in 
the observed performances before and after divisions. E.g. for lists 
$l,m,n$ of size $ls,ms,ns$ where $l=m\cup n$: 
\[E(\Delta)=\frac{ms}{ls}abs(m.\mu - l.\mu)^2 + \frac{ns}{ls}abs(n.\mu - 
l.\mu)^2\]

We then apply a statistical hypothesis test $H$ to check
if $m,n$ are significantly different. In our case, the conjunction of 
bootstrapping and A12 test. Both the techniques are non-parametric in 
nature, i.e., they do not make gaussian assumption about the data. As for 
hypothesis test, we use a non-parametric bootstrapping test as endorsed 
by Efron \& Tibshirani~\cite[p220-223]{efron93}. Even with statistical 
significance, it is possible that the difference
can be so small as to be of no practical value. This is known as a ``small effect''. To ensure that the statistical significance is not due to ``small effect'' we use effect-size tests in conjunction with hypothesis tests. A popular effect size test used in SE literature is the A12 test. It has been endorsed by several SE researchers~\cite{leech2002call, poulding10, arcuri11, shepperd12a, kampenes07, Kocaguneli2013:ep}. It was first proposed by Vargha and Delany~\cite{vargha2000}. In our context,
given the performance measure 
G, the A12 statistics measures the
probability that one treatment yields higher $G$ values than another. If the two algorithms are equivalent, then A12 = 0.5. Likewise if $A12 \ge 0.6$, then 60\% of the times, values of one treatment are significantly greater that the other. In such a case, it can be claimed that there is \textit{significant effect} to justify the hypothesis test $H$. In our case, we divide the data if \textit{both} bootstrap sampling and effect size test agree that a division is statistically significant (with a
confidence of 99\%) and not a small effect ($A12 \ge 0.6$). Then, we 
recurse on each of these splits to rank G-scores from best to worst.

\subsection{\respto{1-9} Experimental Setup}
\label{sect:expt_setup}
\bi[leftmargin=*]
\item \textbf{Discovering the bellwether:}
\be
\item For each community in every sub-domain, we pick a project $P_i$. We 
use this as the training set to construct a quality prediction model according 
to the learning method described in \tion{met}.
\item Next, we pick another project $P_j \notin P_i$ and retain this as a 
holdout dataset.
\item Then, for every other project $P_k$ where $k \in {1,\dots,n};~k\notin 
\{i,j\}$, that belong to the same community as $\{P_i, P_j\}$, we evaluate the 
performance of $P_i$ for $P_k$ according to the evaluation strategy 
discussed in \tion{eval}. 
\item We repeat steps 1,2, and 3 for all pairs of projects in a community. 
\ee
This whole process is repeated 30 times, with different random number seeds. 
Then, we use the statistical test described in \tion{stats} to rank each 
project 
$P_i$. For every holdout dataset in step 2 above, if there exists one project 
that returns consistently high performance scores, we label that as the 
bellwether.

\item \textbf{Discovering the best transfer learner:}
\be
\item For each community in every sub-domain, we pick a project $P_i$ as in 
\tion{expt_setup}.A. We then use this as the training data to construct the 
transfer learners (TCA+, TNB, VCB, and Bellwether Method).
\item For every other project $P_j$ where $j \in {1,\dots,n};~j\notin i$, that 
belong to the same community as $P_i$, we evaluate the performance of 
each of the transfer learners and use the evaluation strategy discussed in 
\tion{eval} to evaluate their performance. 
\ee
Similar to above, the above steps are repeated 30 times, with different random 
number seeds. Then, we use the statistical test  from \tion{stats} to rank each 
transfer learner.
\ei
\begin{figure*}
\begin{minipage}[]{0.475\linewidth}
\textbf{{\small Defect}}\\[0.16cm]
\resizebox{\linewidth}{!}{%
\begin{tabular}{l|l|l|l|r|r}\hline
\multicolumn{1}{c|}{\multirow{2}{*}{Community}} & \multicolumn{1}{c|}{\multirow{2}{*}{Holdout}} & 
\multicolumn{1}{c|}{\multirow{2}{*}{Test}} & 
\multicolumn{1}{c|}{\multirow{2}{*}{Bellwether(s)}} & 
\multicolumn{2}{c}{G-Score(s)} \\\cline{5-6}
& & & & med & iqr                                \\ \hline
\multirow{5}{*}{AEEEM}       & EQ    & \begin{tabular}[c]{@{}c@{}}$\forall p  \ne EQ$\end{tabular} & LC             & 74 & 4                               \\ %
& JDT   & \begin{tabular}[c]{@{}c@{}}$\forall p  \ne JDT$\end{tabular}  & LC             & 75 & 3                                \\ %
& ML    & \begin{tabular}[c]{@{}c@{}}$\forall p  \ne ML$\end{tabular} & LC             & 75 & 3                                \\ %
& PDE   & \begin{tabular}[c]{@{}c@{}}$\forall p  \ne PDE$\end{tabular}  & LC             & 75 & 4                               \\ \hline
\multirow{3}{*}{Relink}      & Apache               & \begin{tabular}[c]{@{}c@{}}$\forall p  \ne Apache$\end{tabular}      & Zxing          & 67 &  5                              \\ 
& Safe  & \begin{tabular}[c]{@{}c@{}}$\forall p  \ne Safe$\end{tabular}    & Zxing          & 66 &  5                              \\ \hline
\multirow{10}{*}{Apache}     & Ant   & \begin{tabular}[c]{@{}c@{}}$\forall p  \ne Ant$\end{tabular}                  & Lucene         & 66 &   5                             \\ 
& Ivy   &   \begin{tabular}[c]{@{}c@{}}$\forall p  \ne Ivy$\end{tabular}         & Lucene, Poi    & 64 & 5                               \\ 
& Camel &    \begin{tabular}[c]{@{}c@{}}$\forall p  \ne Camel$\end{tabular}        & Lucene, Poi    & 69 &   7                             \\ 
& Poi   &   \begin{tabular}[c]{@{}c@{}}$\forall p  \ne Poi$\end{tabular}         & Lucene, Poi    & 59 &    6                            \\ 
& Jedit &       \begin{tabular}[c]{@{}c@{}}$\forall p  \ne Jedit$\end{tabular}     & Lucene         & 66 &    4                            \\ 
& Log4j &   \begin{tabular}[c]{@{}c@{}}$\forall p  \ne Log4j$\end{tabular}         & Lucene, Poi    & 65 &    5                            \\ 
& Velocity             &  \begin{tabular}[c]{@{}c@{}}$\forall p  \ne Velocity$\end{tabular}          & Lucene         & 67 & 7                               \\ 
& Xalan &    \begin{tabular}[c]{@{}c@{}}$\forall p  \ne Xalan$\end{tabular}        & Lucene, Poi    & 68 & 8                                \\ 
& Xerces               & \begin{tabular}[c]{@{}c@{}}$\forall p  \notin Xerces$\end{tabular}           & Lucene         & 68 & 5                                \\ \hline
\end{tabular}}\\[0.22cm]
\textbf{{\small Code Smells}}\\[0.16cm]
\resizebox{\linewidth}{!}{%
\begin{tabular}{l|l|l|l|l|l}\hline
\multicolumn{1}{c|}{\multirow{2}{*}{Community}} & \multicolumn{1}{c|}{\multirow{2}{*}{Holdout}} & 
\multicolumn{1}{c|}{\multirow{2}{*}{Test}} & 
\multicolumn{1}{c|}{\multirow{2}{*}{Bellwether(s)}} & 
\multicolumn{2}{c}{G-Score(s)} \\\cline{5-6}
& & & & med & iqr                                \\ \hline
\multirow{10}{*}{\begin{tabular}[c]{@{}l@{}}Feature\\ Envy\end{tabular}} & wct  & \begin{tabular}[c]{@{}c@{}}$\forall p \ne wct$\end{tabular} & mvnforum & 92 & 3 \\
 & itext  & \begin{tabular}[c]{@{}c@{}}$\forall p \ne itext$\end{tabular} & mvnforum & 92 & 2 \\
 & hsqldb  & \begin{tabular}[c]{@{}c@{}}$\forall p \ne hsqldb$\end{tabular} & mvnforum & 91 &  4 \\
 & nekohtml  & \begin{tabular}[c]{@{}c@{}}$\forall p \ne nekohtml$\end{tabular} & mvnforum & 89 & 4 \\
 & galleon  & \begin{tabular}[c]{@{}c@{}}$\forall p \ne galleon$\end{tabular} & mvnforum & 90  & 2 \\
 & sunflow  & \begin{tabular}[c]{@{}c@{}}$\forall p \ne sunflow$\end{tabular} & mvnforum & 90  & 3 \\
 & emma  & \begin{tabular}[c]{@{}c@{}}$\forall p \ne emma$\end{tabular} & mvnforum & 92 &1  \\
 & jasml  & \begin{tabular}[c]{@{}c@{}}$\forall p \ne jasml$\end{tabular} & mvnforum & 92 & 2 \\
 & xmojo  & \begin{tabular}[c]{@{}c@{}}$\forall p \ne xmojo$\end{tabular} & mvnforum & 92 & 1 \\
 & jhotdraw  & \begin{tabular}[c]{@{}c@{}}$\forall p \ne jhotdraw$\end{tabular} & mvnforum & 92 & 1 \\\hline
\multirow{10}{*}{\begin{tabular}[c]{@{}l@{}}God\\ Class\end{tabular}} & fitjava  & \begin{tabular}[c]{@{}c@{}}$\forall p \ne fitjava$\end{tabular} & xerces, xalan & 88 & 3 \\
 & wct  & \begin{tabular}[c]{@{}c@{}}$\forall p \ne wct$\end{tabular} & xerces, xalan & 88 & 3 \\
 & hsqldb  & \begin{tabular}[c]{@{}c@{}}$\forall p \ne hsqldb$\end{tabular} & xerces & 87 & 2 \\
 & galleon  & \begin{tabular}[c]{@{}c@{}}$\forall p \ne galleon$\end{tabular} & xerces, xalan & 90 & 2 \\
 & xalan  & \begin{tabular}[c]{@{}c@{}}$\forall p \ne xalan$\end{tabular} & xerces & 91 &2  \\
 & itext  & \begin{tabular}[c]{@{}c@{}}$\forall p \ne itext$\end{tabular} & xerces & 90 &3  \\
 & drjava  & \begin{tabular}[c]{@{}c@{}}$\forall p \ne drjava$\end{tabular} & xerces, xalan & 88 & 2 \\
 & mvnforum  & \begin{tabular}[c]{@{}c@{}}$\forall p \ne mvnforum$\end{tabular} & xerces, xalan & 90 & 3 \\
 & jpf  & \begin{tabular}[c]{@{}c@{}}$\forall p \ne jpf$\end{tabular} & xerces, xalan & 90 & 3 \\
 & freecol  & \begin{tabular}[c]{@{}c@{}}$\forall p \ne freecol$\end{tabular} & xerces & 90 & 4 \\\hline
\end{tabular}} 
\end{minipage}~~~~~~~~~~~~\begin{minipage}[]{0.5\linewidth}
\textbf{{\small Effort Estimation}}\\[0.1cm]
\resizebox{0.9\linewidth}{!}{%
\begin{tabular}{l|l|l|l|l|l}\hline
\multicolumn{1}{c|}{\multirow{2}{*}{Community}} & \multicolumn{1}{c|}{\multirow{2}{*}{Holdout}} & 
\multicolumn{1}{c|}{\multirow{2}{*}{Test}} & 
\multicolumn{1}{c|}{\multirow{2}{*}{Bellwether(s)}} & 
\multicolumn{2}{c}{G-Score(s)} \\\cline{5-6}
& & & & med & iqr                                \\ \hline
\multirow{5}{*}{Effort} & coc10 & $\forall p \ne coc10$ & cocomo & 98 & 2 \\ 
 & nasa93 & $\forall p \ne nasa93$ & cocomo & 99 & 1 \\ 
 & coc81 & $\forall p \ne coc81$ & cocomo & 98 & 2 \\ 
 & nasa10 & $\forall p \ne nasa10$ & cocomo & 98 & 3 \\ \hline
\end{tabular}}\\[0.1cm]
\textbf{{\small Issue Lifetime}}\\[0.1cm]
\resizebox{0.92\linewidth}{!}{%
\centering
\begin{tabular}{l|l|l|l|r|r}\hline
\multicolumn{1}{c|}{\multirow{2}{*}{Community}} & \multicolumn{1}{c|}{\multirow{2}{*}{Holdout}} & 
\multicolumn{1}{c|}{\multirow{2}{*}{Test}} & 
\multicolumn{1}{c|}{\multirow{2}{*}{Bellwether(s)}} & 
\multicolumn{2}{c}{G-Score(s)} \\\cline{5-6}
& & & & med & iqr                                \\ \hline
\multirow{8}{*}{1 Day}
 & cloudstack & \begin{tabular}[c]{@{}l@{}}$\forall p \ne cloudstack$\end{tabular} & camel & 55 & 6 \\
 & cocoon & \begin{tabular}[c]{@{}l@{}}$\forall p \ne cocoon$\end{tabular} & camel & 54 & 8 \\
 & node & \begin{tabular}[c]{@{}l@{}}$\forall p \ne node$\end{tabular} & camel & 49 & 11 \\
 & dl4j & \begin{tabular}[c]{@{}l@{}}$\forall p \ne dl4j$\end{tabular} & camel, qpid & 55 & 5 \\
 & hadoop & \begin{tabular}[c]{@{}l@{}}$\forall p \ne hadoop$\end{tabular} & camel & 57 & 5 \\
 & hive & \begin{tabular}[c]{@{}l@{}}$\forall p \ne hive$\end{tabular} & camel & 55 & 7 \\
 & ofbiz & \begin{tabular}[c]{@{}l@{}}$\forall p \ne ofbiz$\end{tabular} & camel & 54 & 4 \\
 & qpid & \begin{tabular}[c]{@{}l@{}}$\forall p \ne qpid$\end{tabular} & camel, node & 55 & 7 \\\hline
\multirow{8}{*}{7 Days} & camel & \begin{tabular}[c]{@{}l@{}}$\forall p \ne camel$\end{tabular} & ofbiz & 47 & 7 \\ 
 & cloudstack & \begin{tabular}[c]{@{}l@{}}$\forall p \ne cloudstack$\end{tabular} & ofbiz & 47 & 8 \\ 
 & cocoon & \begin{tabular}[c]{@{}l@{}}$\forall p \ne cocoon$\end{tabular} & ofbiz & 48 & 7 \\ 
 & node & \begin{tabular}[c]{@{}l@{}}$\forall p \ne node$\end{tabular} & ofbiz & 48 & 8 \\ 
 & dl4j & \begin{tabular}[c]{@{}l@{}}$\forall p \ne dl4j$\end{tabular} & ofbiz & 47 & 8 \\ 
 & hadoop & \begin{tabular}[c]{@{}l@{}}$\forall p \ne hadoop$\end{tabular} & ofbiz & 46 & 9 \\ 
 & hive & \begin{tabular}[c]{@{}l@{}}$\forall p \ne hive$\end{tabular} & ofbiz & 46 & 9 \\ 
 & qpid & \begin{tabular}[c]{@{}l@{}}$\forall p \ne qpid$\end{tabular} & ofbiz & 47 & 8 \\ \hline
\multirow{8}{*}{14 Days} & camel & $\forall p \ne camel$ & qpid & 38 & 5 \\
 & cloudstk & $\forall p \ne cloudstk$ & qpid & 38 & 5 \\ 
 & cocoon & $\forall p \ne cocoon$ & qpid & 39 & 6 \\
 & node & $\forall p \ne node$ & qpid & 37 & 4 \\
 & dl4j & $\forall p \ne dl4j$ & qpid & 37 & 4 \\
 & hadoop & $\forall p \ne hadoop$ & qpid & 36 & 6 \\
 & hive & $\forall p \ne hive$ & qpid & 38 & 6 \\
 & ofbiz & $\forall p \ne ofbiz$ & qpid & 38 & 4 \\
 & qpid & $\forall p \ne qpid$ & qpid & 39 & 5 \\\hline
\multirow{8}{*}{30 Days} & camel & $\forall p \ne camel$ & qpid & 46 & 6 \\
 & cloudstk & $\forall p \ne cloudstk$ & qpid & 48 & 5 \\
 & cocoon & $\forall p \ne cocoon$ & qpid & 47 & 5 \\
 & node & $\forall p \ne node$ & qpid & 46 & 6 \\
 & dl4j & $\forall p \ne dl4j$ & qpid & 46 & 7 \\
 & hadoop & $\forall p \ne hadoop$ & qpid & 47 & 4 \\
 & hive & $\forall p \ne hive$ & qpid & 48 & 4 \\
 & ofbiz & $\forall p \ne ofbiz$ & qpid & 47 & 5 \\
 & qpid & $\forall p \ne qpid$ & qpid & 46 & 6 \\\hline
\end{tabular}}
\end{minipage}
\caption{Discovering Bellwether datasets with a holdout data. We use the 
experimental setup mentioned in \tion{expt_setup} to discover these 
bellwethers.}
\label{fig:loo}
\end{figure*}

\subsection{Understanding These Results}
\rahul{
	In presenting our results for experiments in~\tion{expt_setup}, we adopted a 
	convention that includes tabulated results. The following remarks need to be 
	made regarding our tables:
	\bi
	\item In~\fig{loo}, we list the results of performing the experiment in 
	\tion{expt_setup}. The column labeled ``Holdout'' represents the holdout 
	dataset. The column labeled ``Test'' represents the test data, i.e., all the 
	remaining data in the community except the holdout. The column 
	``Bellwether(s)'' shows the dataset that was ranked the best from among the 
	test data (and therefore it is the bellwether dataset). Finally, the column 
	``G-score(s)'' is the G-score of training on the bellwether and testing on the 
	holdout dataset.
	\item In Figures~\ref{fig:smells_result}, \ref{fig:issues_result}, 
	\ref{fig:defects_results}, and~\ref{fig:effort_result}, we list the results of 
	performing the experiment in \tion{expt_setup} where we compare the 
	bellwether method with other transfer learners. In these figures, the column 
	labeled ``source'' (the second column) indicates the source from which a 
	transfer learner is built. The remaining datasets within the community are 
	then used as target datasets. The numeric values indicate the \textit{median} 
	performance scores (Standardized Accuracy in case of effort estimation, 
	G-score in the rest), when model is constructed with a ``target'' dataset and 
	tested against all the ``source'' datasets, and this processes repeated 30 
	times for reasons discussed in \S5.5.
	\ei
}

\section{Results}
\label{sect:res}
\begin{figure}
    \centering
    \resizebox{0.66\linewidth}{!}{%
\begin{tabular}{l|l|l|l|l}
& Bellwether                                 & \multicolumn{3}{c}{Local}                                                      \\ \cline{2-5} 
& \multicolumn{1}{c|}{(Lucene)}                                            & 
Train              & Test 
&                                                     \bigstrut\\
& \multicolumn{1}{c|}{(G-score)}                                            
&                    &      & 
G-Score                                             \bigstrut\\ \hline
Xalan    & \cellcolor[HTML]{EFEFEF}{\color[HTML]{000000} 82} & 2.6      & 2.7  & 
56                                                \bigstrut\\
Ant      & \cellcolor[HTML]{EFEFEF}{\color[HTML]{000000} 68}  & 1.6 & 1.7  & 
54                                                \bigstrut\\
Ivy      & \cellcolor[HTML]{EFEFEF}{\color[HTML]{000000} 67} & 
1.4           & 2    & {\color[HTML]{000000} 
63} \bigstrut\\
Camel    & \cellcolor[HTML]{EFEFEF}{\color[HTML]{000000} 62} & 1.4      & 1.6  & 
51                                                \bigstrut\\
Velocity & \cellcolor[HTML]{EFEFEF}{\color[HTML]{000000} 57} &
1.5           & 1.6  & {\color[HTML]{000000} 
32} \bigstrut\\
Jedit    & 61                                                & 4.2 & 4.3  & \cellcolor[HTML]{EFEFEF}{\color[HTML]{000000} 
77} \bigstrut\\
Log4j    & {\color[HTML]{000000} 56} &
1.1           & 1.2  & \cellcolor[HTML]{EFEFEF}{\color[HTML]{000000} 
75} \bigstrut\\
Xerces   & {\color[HTML]{000000} 58} & 1.3      & 1.4  & 
\cellcolor[HTML]{EFEFEF}66                                                \bigstrut\\

\hline
\end{tabular}}
\caption{Bellwether dataset (Lucene) vs. Local Data.  Performance scores are 
G-scores so
{\em higher} values are {\em better}. 
Cells highlighted in \colorbox[HTML]{E8E8E8}{gray} indicate datasets with superior prediction capability. Out of the eight datasets studied here, we note that in five cases the prediction performance of bellwether dataset was superior to within-project dataset.
}
\label{fig:local_result}
\end{figure}
\begin{figure*}[t]
\arrayrulecolor{darkgray}
\centering
\begin{minipage}[c]{0.5\linewidth}
\resizebox{0.67\linewidth}{!}{%
\begin{tabular}{l|l|l|l|l}
\cline{2-5}
   &    Source      & \textbf{Baseline} & TCA & TNB \\\hline
\multicolumn{1}{l|}{\multirow{12}{*}{God Class}}    & 
\textbf{xerces}    & \cellcolor[HTML]{EFEFEF}\textbf{90}         & 
75  & 48  \\
\multicolumn{1}{l|}{}     & xalan   & \cellcolor[HTML]{EFEFEF}89         & 73  
& 39  \\
\multicolumn{1}{l|}{}     & hsqldb   & \cellcolor[HTML]{EFEFEF}88         & 0   
& 0   \\
\multicolumn{1}{l|}{}     & galleon  & \cellcolor[HTML]{EFEFEF}87         & 61  
& 55  \\
\multicolumn{1}{l|}{}     & wct      & \cellcolor[HTML]{EFEFEF}81         & 58  
& 67  \\
\multicolumn{1}{l|}{}     & drjava   & \cellcolor[HTML]{EFEFEF}80         & 58  
& 56  \\
\multicolumn{1}{l|}{}     & jpf      & \cellcolor[HTML]{EFEFEF}79         & 59  
& 65  \\
\multicolumn{1}{l|}{}     & mvnforum & \cellcolor[HTML]{EFEFEF}74         & 43  
& 57  \\
\multicolumn{1}{l|}{}     & freecol  & \cellcolor[HTML]{EFEFEF}69         & 0   
& 0   \\
\multicolumn{1}{l|}{}     & fitjava  & \cellcolor[HTML]{EFEFEF}68         & 40  
& 0   \\
\multicolumn{1}{l|}{}     & itext    & 62         & \cellcolor[HTML]{EFEFEF}72  
& 30  \\ \cline{2-5}
\multicolumn{1}{l|}{}     & W/T/L    &   \cellcolor[HTML]{EFEFEF}\textbf{10/0/1}       & 
1/0/10  & 0/0/11  \bigstrut\\\hline 
\end{tabular}}
\end{minipage}\begin{minipage}[c]{0.5\linewidth}
\resizebox{0.67\linewidth}{!}{%
\begin{tabular}{l|l|l|l|l}
\cline{2-5}
   &    Source      & \textbf{Baseline} & TCA & TNB \\\hline
\multicolumn{1}{l|}{\multirow{12}{*}{Feature Envy}} & \textbf{mvnforum} & 
\cellcolor[HTML]{EFEFEF}\textbf{92}         & 
57  & 61  \\
\multicolumn{1}{l|}{}     & galleon  & \cellcolor[HTML]{EFEFEF}84         & 59  
& 0   \\
\multicolumn{1}{l|}{}     & hsqldb   & \cellcolor[HTML]{EFEFEF}81         & 57  
& 0   \\
\multicolumn{1}{l|}{}     & jhotdraw & \cellcolor[HTML]{EFEFEF}81         & 35  
& 64  \\
\multicolumn{1}{l|}{}     & nekohtml & \cellcolor[HTML]{EFEFEF}81         & 52  
& 57  \\
\multicolumn{1}{l|}{}     & wct      & \cellcolor[HTML]{EFEFEF}81         & 47  
& 0   \\
\multicolumn{1}{l|}{}     & itext    & \cellcolor[HTML]{EFEFEF}74         & 66  
& 0   \\
\multicolumn{1}{l|}{}     & xmojo    & \cellcolor[HTML]{EFEFEF}74         & 0   
& 0   \\
\multicolumn{1}{l|}{}     & emma     & \cellcolor[HTML]{EFEFEF}70         & 74  
& 37  \\
\multicolumn{1}{l|}{}     & jasml    & 66         & \cellcolor[HTML]{EFEFEF}79  
& 0   \\
\multicolumn{1}{l|}{}     & sunflow  & \cellcolor[HTML]{EFEFEF}47         & 0   
& 0   \\ \cline{2-5}
\multicolumn{1}{l|}{}     & W/T/L    &   \cellcolor[HTML]{EFEFEF}\textbf{10/0/1}       & 
1/0/10  & 0/0/11  \bigstrut\\\hline
\end{tabular}}
\end{minipage}
\caption{Code Smells: This figure compares the prediction performance of the bellwether 
dataset (xalan,mvnforum) against other datasets (other rows). \textit{Bellwether Method} against Transfer Learners (columns) for 
detecting code smells. The numerical value seen here are the median 
G-scores from Equation 2 over 30 repeats where one dataset is used as a 
source and others are used as targets in a round-robin fashion.  Higher values 
are better and cells highlighted in gray produce the best Scott-Knott ranks. 
The last row in each community indicate Win/Tie/Loss(W/T/L). The 
\textit{bellwether Method} is the overall best.\\[0.1cm]}
\label{fig:smells_result}
\end{figure*}

\subsection*{RQ1: How prevalent is the ``Bellwether Effect''?}
\label{sect:rq1_1}

The bellwether effect points to an exemplar dataset to construct quality 
predictors from. Ideally, given an adequate transfer learner, such a 
dataset should produce reasonably high  performance scores. \fig{loo} 
documents our findings. We use the setup described in \tion{expt_setup}
to discover bellwethers.
It is immediately noticeable that for each community there is at least one 
dataset that provides 
consistently better predictions when compared to other datasets. For 
example:

\begin{figure*}[t!]
\arrayrulecolor{darkgray}
\centering
\begin{minipage}[c]{0.5\linewidth}
\resizebox{0.75\linewidth}{!}{%
\begin{tabular}{l|l|l|l|l|l}
\cline{2-6}

&       Source       &  Baseline  &  TCA+  &  \textbf{TNB}  &  VCB \bigstrut[t]\\\hline
\multicolumn{1}{l|}{\multirow{10}{*}{1 Day}}   & \textbf{camel}         & 17 & 
2  & 
\cellcolor[HTML]{EFEFEF}55 & 
10 \\ 
\multicolumn{1}{l|}{}& node         & 44 & 24 & \cellcolor[HTML]{EFEFEF}55 & 17 
\\ 
\multicolumn{1}{l|}{}& ofbiz        & 29 & 14 & \cellcolor[HTML]{EFEFEF}53 & 8  
\\ 
\multicolumn{1}{l|}{}& qpid        & 44 & 34 & \cellcolor[HTML]{EFEFEF}49 & 19 
\\ 
\multicolumn{1}{l|}{}& deeplearning & \cellcolor[HTML]{EFEFEF}51 & 42 & 42 & 15 
\\ 
\multicolumn{1}{l|}{}& cocoon       & 7  & 6  & \cellcolor[HTML]{EFEFEF}34 & 13 
\\ 
\multicolumn{1}{l|}{}& cloudstack   & \cellcolor[HTML]{EFEFEF}55 & 32 & 32 & 11 
\\ 
\multicolumn{1}{l|}{}& hive         & 11 & 1  & 22 & \cellcolor[HTML]{EFEFEF}23 
\\ 
\multicolumn{1}{l|}{}& hadoop       & 17 & 0  & 10 & \cellcolor[HTML]{EFEFEF}19 
\\ \cline{2-6}
\multicolumn{1}{l|}{}& W/T/L        &  2/0/7  &  0/0/9  &  
\cellcolor[HTML]{EFEFEF}\textbf{7/0/2}  & 2/0/7   
\bigstrut\\ 
\hline\hline
\multicolumn{1}{l|}{\multirow{9}{*}{7 Days}}  & \textbf{ofbiz}         & 17 & 3  
& \cellcolor[HTML]{EFEFEF}49 & 
11 
\\ 
\multicolumn{1}{l|}{}& camel        & 34 & 6  & \cellcolor[HTML]{EFEFEF}47 & 20 
\\ 
\multicolumn{1}{l|}{}& cloudstack   & 8  & 27 & \cellcolor[HTML]{EFEFEF}38 & 7  
\\ 
\multicolumn{1}{l|}{}& qpid        & 7  & 16 & \cellcolor[HTML]{EFEFEF}38 & 20 
\\ 
\multicolumn{1}{l|}{}& node         & 15 & 33 & \cellcolor[HTML]{EFEFEF}36 & 13 
\\ 
\multicolumn{1}{l|}{}& deeplearning & 15 & 20 & \cellcolor[HTML]{EFEFEF}29 & 10 
\\ 
\multicolumn{1}{l|}{}& cocoon       & 0  & 3  & \cellcolor[HTML]{EFEFEF}22 & 16 
\\ 
\multicolumn{1}{l|}{}& hadoop       & \cellcolor[HTML]{EFEFEF}23 & 0  & 18 & 19 
\\ 
\multicolumn{1}{l|}{}& hive         & 3  & 0  & 7  & \cellcolor[HTML]{EFEFEF}14 
\\ \cline{2-6}
\multicolumn{1}{l|}{}& W/T/L        &  2/0/7  &  0/0/9  &  
\cellcolor[HTML]{EFEFEF}\textbf{7/0/2}  & 2/0/7   
\bigstrut\\ 
\hline
\end{tabular}}
\end{minipage}~~\begin{minipage}[c]{0.5\linewidth}
\resizebox{0.75\linewidth}{!}{%
\begin{tabular}{l|l|l|l|l|l}
\cline{2-6}

&       Source       &  Baseline  &  TCA+  &  \textbf{TNB}  &  VCB \bigstrut\\\hline

\multicolumn{1}{l|}{\multirow{10}{*}{14 Days}} & \textbf{qpid}         & 0  & 
0  & \cellcolor[HTML]{EFEFEF}39 & 
6  \\ 
\multicolumn{1}{l|}{}& cloudstack   & 8  & 8  & \cellcolor[HTML]{EFEFEF}36 & 8  
\\ 
\multicolumn{1}{l|}{}& hadoop       & 0  & 0  & \cellcolor[HTML]{EFEFEF}31 & 22 
\\ 
\multicolumn{1}{l|}{}& deeplearning & 4  & 6  & \cellcolor[HTML]{EFEFEF}30 & 17 
\\ 
\multicolumn{1}{l|}{}& camel        & 1  & 6  & \cellcolor[HTML]{EFEFEF}29 & 18 
\\ 
\multicolumn{1}{l|}{}& cocoon       & 0  & 0  & \cellcolor[HTML]{EFEFEF}19 & 8  
\\ 
\multicolumn{1}{l|}{}& node         & 5  & 4  & \cellcolor[HTML]{EFEFEF}16 & 4  
\\ 
\multicolumn{1}{l|}{}& ofbiz        & 2  & 2  & 7  & \cellcolor[HTML]{EFEFEF}12 
\\ 
\multicolumn{1}{l|}{}& hive         & 0  & 0  & 0  & \cellcolor[HTML]{EFEFEF}14 
\\ \cline{2-6}
\multicolumn{1}{l|}{}& W/T/L        &  0/0/9  &  0/0/9  &  
\cellcolor[HTML]{EFEFEF}\textbf{7/0/2}  & 2/0/7   
\bigstrut\\ 
\hline\hline
\multicolumn{1}{l|}{\multirow{10}{*}{30 Days}} & \textbf{qpid}         & 1  & 
5  & \cellcolor[HTML]{EFEFEF}47 & 
17 \\ 
\multicolumn{1}{l|}{}& cloudstack   & 1  & 13 & \cellcolor[HTML]{EFEFEF}38 & 19 
\\ 
\multicolumn{1}{l|}{}& node         & 2  & 10 & \cellcolor[HTML]{EFEFEF}32 & 16 
\\ 
\multicolumn{1}{l|}{}& camel        & 1  & 1  & \cellcolor[HTML]{EFEFEF}30 & 17 
\\ 
\multicolumn{1}{l|}{}& deeplearning & 1  & 2  & \cellcolor[HTML]{EFEFEF}29 & 14 
\\ 
\multicolumn{1}{l|}{}& cocoon       & 2  & 1  & \cellcolor[HTML]{EFEFEF}24 & 12 
\\ 
\multicolumn{1}{l|}{}& ofbiz        & 4  & 5  & \cellcolor[HTML]{EFEFEF}13 & 5  
\\ 
\multicolumn{1}{l|}{}& hadoop       & 0  & 0  & \cellcolor[HTML]{EFEFEF}7  & 2  
\\ 
\multicolumn{1}{l|}{}& hive         & \cellcolor[HTML]{EFEFEF}16 & 2  & 6  & 9  
\\ \cline{2-6}
\multicolumn{1}{l|}{}& W/T/L        &  1/0/8  &  0/0/9  &  
\cellcolor[HTML]{EFEFEF}\textbf{8/0/1}  & 0/0/9   
\bigstrut\\ 
\hline
\end{tabular}}
\end{minipage}
\caption{Issue Lifetime: This figure compares the prediction performance of 
the bellwether dataset (qpid) against other datasets (rows) and various transfer 
learners (columns) for estimating issue lifetime. The numerical value seen 
here are the median G-scores from Equation 2 over 30 repeats where one 
dataset is used as a source and others are used as targets in a round-robin 
fashion.  Higher  values  are  better  and  cells  highlighted  in  gray  produce  
the  best  Scott-Knott  ranks. The last row in each community indicate 
Win/Tie/Loss(W/T/L).   TNB has the overall best Win/Tie/Loss ratio.}
\label{fig:issues_result}
\end{figure*}
\begin{figure}[t!]
	\centering
\arrayrulecolor{darkgray}
\resizebox{0.75\linewidth}{!}{%
\begin{tabular}{l|l|l|l|l|l}
\cline{2-6}

&     Source     & Baseline & \textbf{TCA+}   & TNB   & VCB    \\\hline
\multicolumn{1}{l|}{\multirow{10}{*}{Apache}} & \textbf{Lucene}   & 
63         & 
\cellcolor[HTML]{EFEFEF}\textbf{69}    
& 57    & 64     \\
\multicolumn{1}{l|}{}                           & Xalan    & 57         & 
\cellcolor[HTML]{EFEFEF}64    
& 59    & 62     \\
\multicolumn{1}{l|}{}                           & Camel    & 60         & 
\cellcolor[HTML]{EFEFEF}63    
& 59    & 44     \\
\multicolumn{1}{l|}{}                           & Velocity & 58         & 
\cellcolor[HTML]{EFEFEF}63    
& 51    & 63     \\
\multicolumn{1}{l|}{}                           & Ivy      & 60         & 
\cellcolor[HTML]{EFEFEF}62    
& 61    & 48     \\
\multicolumn{1}{l|}{}                           & Log4j    & 60         & 
\cellcolor[HTML]{EFEFEF}62    
& 58    & 62     \\
\multicolumn{1}{l|}{}                           & Xerces   & 
57         & 54    
& 58    & \cellcolor[HTML]{EFEFEF}65     \\
\multicolumn{1}{l|}{}                           & Ant      & 
\cellcolor[HTML]{EFEFEF}61         & 52    
& 45    & 55     \\
\multicolumn{1}{l|}{}                           & Jedit    & 
\cellcolor[HTML]{EFEFEF}58         & 43    
& 57    & 49     \\ \cline{2-6} 
\multicolumn{1}{l|}{}                           & W/T/L    & 2/0/7      & 
\cellcolor[HTML]{EFEFEF}\textbf{6/0/3} 
& 0/0/9 & 1/2/06 \\ \hline\hline
\multicolumn{1}{l|}{\multirow{4}{*}{ReLink}}    & \textbf{Zxing}    & 
\cellcolor[HTML]{EFEFEF}\textbf{68}         & \cellcolor[HTML]{EFEFEF}\textbf{67}    
& 53    & 64     \\
\multicolumn{1}{l|}{}                           & Safe     & 
\cellcolor[HTML]{EFEFEF}38         & 34    
& 36    & 31     \\
\multicolumn{1}{l|}{}                           & Apache   & 31         & 31    
& \cellcolor[HTML]{EFEFEF}32    & 31     \\ \cline{2-6} 
\multicolumn{1}{l|}{}                           & W/T/L    & 
\cellcolor[HTML]{EFEFEF}\textbf{0/1/1}      & 0/1/1 
& 1/0/2 & 0/1/2  \\ \hline\hline
\multicolumn{1}{l|}{}                        & \textbf{LC}    & 
\cellcolor[HTML]{EFEFEF}\textbf{75} & \cellcolor[HTML]{EFEFEF}\textbf{75}    & 
73                         & 61                    \\
\multicolumn{1}{l|}{}                        & ML    & 
\cellcolor[HTML]{EFEFEF}73 & \cellcolor[HTML]{EFEFEF}73    & 
67                         & 51                    \\
\multicolumn{1}{l|}{}                        & PDE   & 
70                         & \cellcolor[HTML]{EFEFEF}71    & 
60                         & 57                    \\
\multicolumn{1}{l|}{}                        & JDT   & 
63                         & 64                            & 
\cellcolor[HTML]{EFEFEF}68 & 53                    \\
\multicolumn{1}{l|}{}                        & EQ    & 
59                         & \cellcolor[HTML]{EFEFEF}61    & 
59                         & 57                    \bigstrut\\ \cline{2-6} 
\multicolumn{1}{l|}{\multirow{-6}{*}{AEEEM}} & W/T/L & 
0/2/3                      & \cellcolor[HTML]{EFEFEF}\textbf{2/2/1} & 
1/0/4                      & 0/0/5                 \\ \hline
 \end{tabular}}

\caption{Defect Datasets: This figure compares the prediction performance of the bellwether 
dataset (Lucene,Zxing,LC) against other datasets (other rows). \textit{Bellwether Method} against Transfer Learners (columns) for 
detecting defects. The numerical value seen here are the median G-scores from Equation 2 over 30 repeats where one dataset is used as a source and others are used as targets in a round-robin fashion.  Higher values are better and cells highlighted in gray produce the best Scott-Knott ranks. The last row in each community indicate Win/Tie/Loss(W/T/L). \textit{TCA+} is the overall best transfer learner.}
\label{fig:defects_results}
\end{figure}
\begin{figure}[ht]
\centering
\begin{minipage}[c!]{\linewidth}
\centering
\resizebox{0.64\linewidth}{!}{%
\hfill
\begin{tabular}{l|l|l|l|l}
\cline{2-5}

& Source & Baseline                   & TCA                          & TNB   \\ \hline
& cocomo & \cellcolor[HTML]{EFEFEF}98 & 90                         & 90  \\
& nasa93 & \cellcolor[HTML]{EFEFEF}93 & 85                         & 35  \\
& nasa10 & \cellcolor[HTML]{EFEFEF}90 & 53                         & 65  \\
& coc81  & 83                         & \cellcolor[HTML]{EFEFEF}85 & 60  \\
& coc10  & 55                         & \cellcolor[HTML]{EFEFEF}75 & 73  \\ \cline{2-5} 
\multirow{-6}{*}{FPA} & W/T/L  & 3/0/2                        & 2/0/3                        & 0/0/5 \\ \hline

\end{tabular}}
\end{minipage}
\caption{Effort Estimation: This figure compares the performance of the bellwether dataset (cocomo) against other 
datasets (rows) and Transfer Learners (columns) for estimating effort. The 
numerical value seen are the median Standardized Accuracy scores from Equation 3 over 40 
repeats. \textit{Bellwether Method} has the best Win/Tie/Loss ratio.\\}
\label{fig:effort_result}
\end{figure}
\begin{figure}[b]
\centering
\begin{minipage}[c!]{\linewidth}
\centering
\resizebox{\linewidth}{!}{%
\hfill
\begin{tabular}{lrrrrrr}
\multicolumn{1}{l|}{} & \multicolumn{2}{c|}{Lucene 2.4} & \multicolumn{2}{c|}{Lucene 2.4, 2.2} & \multicolumn{2}{c}{Lucene 2.4, 2.2, 2.0} \\ \cline{2-7} 
\multicolumn{1}{l|}{} & \multicolumn{1}{l|}{G (mean)} & \multicolumn{1}{l|}{G (iqr)} & \multicolumn{1}{l|}{G (mean)} & \multicolumn{1}{l|}{G (iqr)} & \multicolumn{1}{l|}{G (mean)} & G (iqr) \\ \hline
\multicolumn{1}{l|}{Xalan} & \cellcolor[HTML]{EFEFEF}83 & \multicolumn{1}{r|}{\cellcolor[HTML]{EFEFEF}3} & \cellcolor[HTML]{EFEFEF}82 & \multicolumn{1}{r|}{\cellcolor[HTML]{EFEFEF}3} & \cellcolor[HTML]{EFEFEF}84 & \cellcolor[HTML]{EFEFEF}3 \\
\multicolumn{1}{l|}{Poi} & \cellcolor[HTML]{EFEFEF}73 & \multicolumn{1}{r|}{\cellcolor[HTML]{EFEFEF}5} & \cellcolor[HTML]{EFEFEF}71 & \multicolumn{1}{r|}{\cellcolor[HTML]{EFEFEF}4} & \cellcolor[HTML]{EFEFEF}72 & \cellcolor[HTML]{EFEFEF}3 \\
\multicolumn{1}{l|}{Ivy} & \cellcolor[HTML]{EFEFEF}69 & \multicolumn{1}{r|}{\cellcolor[HTML]{EFEFEF}3} & 66 & \multicolumn{1}{r|}{2} & \cellcolor[HTML]{EFEFEF}69 & \cellcolor[HTML]{EFEFEF}2 \\
\multicolumn{1}{l|}{Ant} & 67 & \multicolumn{1}{r|}{2} & 68 & \multicolumn{1}{r|}{1} & \cellcolor[HTML]{EFEFEF}70 & \cellcolor[HTML]{EFEFEF}1 \\
\multicolumn{1}{l|}{Jedit} & \cellcolor[HTML]{EFEFEF}62 & \multicolumn{1}{r|}{\cellcolor[HTML]{EFEFEF}4} & \cellcolor[HTML]{EFEFEF}63 & \multicolumn{1}{r|}{\cellcolor[HTML]{EFEFEF}3} & \cellcolor[HTML]{EFEFEF}62 & \cellcolor[HTML]{EFEFEF}3 \\
\multicolumn{1}{l|}{Xerces} & \cellcolor[HTML]{EFEFEF}56 & \multicolumn{1}{r|}{\cellcolor[HTML]{EFEFEF}9} & 52 & \multicolumn{1}{r|}{5} & 58 & 5 \\
\multicolumn{1}{l|}{Velocity} & \cellcolor[HTML]{EFEFEF}55 & \multicolumn{1}{r|}{\cellcolor[HTML]{EFEFEF}4} & \cellcolor[HTML]{EFEFEF}52 & \multicolumn{1}{r|}{\cellcolor[HTML]{EFEFEF}4} & \cellcolor[HTML]{EFEFEF}55 & \cellcolor[HTML]{EFEFEF}3 \\
\multicolumn{1}{l|}{Camel} & \cellcolor[HTML]{EFEFEF}52 & \multicolumn{1}{r|}{\cellcolor[HTML]{EFEFEF}2} & \cellcolor[HTML]{EFEFEF}54 & \multicolumn{1}{r|}{\cellcolor[HTML]{EFEFEF}2} & \cellcolor[HTML]{EFEFEF}53 & \cellcolor[HTML]{EFEFEF}2 \\
\multicolumn{1}{l|}{Log4j} & \cellcolor[HTML]{EFEFEF}52 & \multicolumn{1}{r|}{\cellcolor[HTML]{EFEFEF}6} & \cellcolor[HTML]{EFEFEF}48 & \multicolumn{1}{r|}{\cellcolor[HTML]{EFEFEF}6} & \cellcolor[HTML]{EFEFEF}50 & \cellcolor[HTML]{EFEFEF}8 \\ \hline
 &  &  &  &  &  &  \\ \cline{2-7}
\multicolumn{1}{l|}{} & \multicolumn{2}{l|}{Lucene 2.4} & \multicolumn{2}{l|}{Lucene 2.4, 2.2} & \multicolumn{2}{l|}{Lucene 2.4, 2.2, 2.0} \\ \cline{1-7}
\multicolumn{1}{l|}{Samples} & \multicolumn{2}{l|}{341} & \multicolumn{2}{l|}{587} & \multicolumn{2}{l|}{782} \\
\multicolumn{1}{l|}{Defect \%} & \multicolumn{2}{l|}{59} & \multicolumn{2}{l|}{59} & \multicolumn{2}{l|}{55} \\ \cline{1-7}
\end{tabular}
}
\end{minipage}
\caption{Experiments with incremental discovery of bellwethers. Note that the latest version of lucene (lucene-2.4) has statistically similar performance to using the other older versions of lucene.}
\label{fig:incr_result}
\end{figure}

\be

\item \textit{Code Smells datasets:} Here we have two datasets which are 
frequently ranked high: $Xerces$ and $Xalan$. But note that $Xerces$ is 
ranked the best in all the cases. Thus, this would be a bellwether dataset for
predicting for the existence of God Classes; this was followed by $hsqldb$ with a G-score of 88\%. Additionally, when $Xalan$ or $Xerces$ were absent in Feature 
Envy, $mvnforum$ was a bellwether with a G-score of {92\%}.

\item \textit{Effort Estimation}: When performing effort estimation, we found that $cocomo$ was the bellwether with 
remarkably high Standardized Accuracy scores of {98\%}. 

\item \textit{Defect datasets:} In the case of defect prediction, Jureczko's bellwether is Lucene (with a G-Score of {69\%}); AEEEM's bellwether is LC (with a G-Score of {75\%}); and Relink's bellwether is ZXing (with a G-Score of {68\%}).

\item \textit{Issue Lifetime:} Finally when predicting for lifetime of issues, 
we discovered the following bellwethers: $camel$ for close time of 1 day with G-Scores of around {55}\%, $ofbiz$ for close time of 7 days with a G-score of around {47}\%, $qpid$ for 14 days and 30 days with G-score of around {38}\%, and {47}\% s respectively.

\ee

Note that in the case of issue lifetime estimation, the G-Scores are particularly low. Here recommend that practitioners monitor the performance of bellwethers and eschew current ones in favor of other better bellwether datasets.

In summary, in three out of the four domains studied here, there was a 
clear bellwether dataset for every community. In the case of issue lifetimes, 
although there was a bellwether, the performances were particular low. 
\respto{1-5-B}Note that this may/may not hold true for other sub-domains of 
SE. The 
study on these other domains are beyond the scope of this work but what 
we can say now is:
\begin{lesson}
\respto{1-5-C} Bellwethers are common in several domains of software 
engineering studied here. ie., 
in defect 
prediction, effort estimation, and code-smell detection.  \\[-0.2cm]
\end{lesson}

\subsection*{RQ2: How does the bellwether dataset fare against within-project dataset?}
\label{sect:rq2}

Having established in RQ1 that bellwethers are prevalent in the 
sub-domains studied here. In
\fig{local_result}, we compare the predictors built on within-project data
against those built with a bellwether. \respto{1-6-A}For this question,
we only used data from the Apache community since it has  
releases ordered historically (which is required to test older data against 
newer data). \respto{1-6-B}Since other sub-domains did not have historically 
data similar to Apache, we were unable to use them for this research question. 
For the Apache community, the bellwether dataset was $Lucene$. 

As seen in \fig{local_result}, the prediction scores with the bellwether is 
very encouraging in case of the Apache datasets.  In 5 out of 8 cases 
($Ant$, $Camel$, $Ivy$, $Xalan$, and $Velocity$), defect prediction models 
constructed with $Lucene$ as the bellwether performed better than 
within-project data. In 3 out of 8 cases ($Jedit$, $Xerces$, and $Log4j$), 
the performance scores of bellwether data were statistically worse than  
within-project data. \respto{1-6-C}Note again that this is true in only one out 
of our the four domains studied, i.e., defect prediction. Therefore, the 
following answer to the this research question is limited this domain.

\begin{lesson}
\respto{1-6-D} For projects in the Apache Community that were evaluated with 
the same quality metrics, training a 
quality prediction model with the Bellwether is better than using 
within-project data  in majority of the cases.\\[-0.2cm]
\end{lesson}

\subsection*{RQ3: How  well  do  transfer  learners  perform  across different domains?}

Figures \ref{fig:smells_result}, 
\ref{fig:issues_result}, \ref{fig:defects_results}, \ref{fig:effort_result} 
show the results of transferring data between different projects in a 
community for code smell detection, issue lifetime estimation, defect 
prediction, and effort estimation. 

\respto{1-7} Note that of the three transfer learners studied here, value 
cognitive boosting (VCB) has some methodological constrains that prevents us 
from translating it to all the domains. VCB was initially designed for defect 
prediction. To enable it to work efficiently, the authors propose the use of 
under-sampling techniques to complement transfer learning. This under-sampling 
required that the datasets have discrete class variables ($\#defects$) and that 
the datasets are sufficiently large. Two of the domains 
considered in this paper do not satisfy these constraints. We could not use VCB 
in code smell detection because our datasets had small sample size (see 
\fig{datasets}) and therefore under-sampling could not be performed. We could 
not use VCB in effort estimation either because the class variable was a 
continuous in nature. Other transfer learners did not have these constraints, 
therefore we were able to translate them to all the domains relatively easily.

These results are expressed in terms 
of win/tie/loss   (W/T/L) ratios:
\be
\item {Code Smells dataset:} From \fig{smells_result} we note that the baseline transfer learner constructed using the bellwether dataset outperforms the other two approaches with a W/T/L of {10/0/1} in both cases.
\item {Issue lifetime dataset:} From \fig{issues_result}, we see that, in this case, TNB outperforms the other three methods. We note a W/T/L ratio for TNB at {7/0/2}. The baseline approach has W/T/L of 2/0/7 (for 1 and 7 days), 1/0/8 (for 14 days), and 0/0/9 (30 days).
\item {Defects dataset:} In the case of \fig{defects_results}, we note that 
TCA+ was generally better than the other three methods with an overall W/T/L 
ratio of {8/3/5}. The was followed by the baseline transfer learner with a 
W/T/L ratio of 2/3/11. Note that this behavior of TCA+ corroborates with 
previous findings by other researchers~\cite{Nam2013}.  
\item {Effort datasets:} In the case of effort estimation, our results are tabulated in~\fig{effort_result}. In this case, the baseline transfer learner once again outperforms the other two methods with a W/T/L ratio of {3/0/2}.
\ee
The key point from the above is that no transfer learning
method is best in all domains (though we would
boast that our bellwether method works best more
often than the other transfer learners).
Hence, when faced with a new community,
software analysts will have to explore multiple
transfer learning methods. In that context, it is
very useful to have an ordering of methods such
that simpler baseline methods are run first
before more complex approaches. Note that:
\bi
\item
When
such an ordering of methods is available
then if the simpler methods achieve acceptable
levels of performance, an analyst might decide to
stop explore more complex methods.
\item
We would argue that bellwethers fall very early
in that ordering; i.e. bellwethers should be the
first simplest transfer learning method
tried before other approaches. 
\ei
That is, although we can't endorse a transfer learner in general, we can offer the bellwether method as a baseline transfer learner which can be used to benchmark other complex transfer learners and seek newer transfer learners that can outperform this baseline. Hence, our answer to this question is:

\begin{lesson}
There is no universal best transfer learner that works across multiple domains. Simpler baseline methods like bellwethers show comparable performances in several domains.\\[-0.2cm]
\end{lesson}

\begin{figure*}[htbp]
  \centering
  \resizebox{0.85\linewidth}{!}{
    \begin{tabular}{|c|l|l|l|l|l|l|}
    	\hline
    \multicolumn{1}{|r|}{} & \multicolumn{1}{r|}{} & 
    \multicolumn{5}{c|}{Feature 
    Ranks} \\\cline{3-7}
    \multicolumn{1}{|r|}{}      & \multicolumn{1}{r|}{Project} & 1st   & 
    2nd   & 
    3rd   & 4th   & 5th \\
    \hline
    \multirow{10}[2]{*}{Apache} & ant   & rfc   & loc   & cam   & ce    & 
    cbo 
    \bigstrut[t]\\
          & \cellcolor[HTML]{EFEFEF}lucene & 
          \cellcolor[HTML]{EFEFEF}loc   & 
          \cellcolor[HTML]{EFEFEF}cbo   & 
          \cellcolor[HTML]{EFEFEF}amc   & 
          \cellcolor[HTML]{EFEFEF}ce    & 
          \cellcolor[HTML]{EFEFEF}rfc \\
          & jedit & loc   & rfc   & amc   & lcom  & avg\_cc \\
          & xerces & cbo   & loc   & cam   & rfc   & ca \\
          & xalan & loc   & amc   & cbo   & lcom3 & rfc \\
          & camel & ca    & mfa   & cbo   & loc   & amc \\
          & velocity & mfa   & cbo   & cam   & loc   & rfc \\
          & poi   & loc   & ce    & lcom  & cbm   & rfc \\
          & log4j & wmc   & cbo   & rfc   & amc   & loc \\
          & ivy   & loc   & rfc   & cam   & ce    & amc \bigstrut[b]\\
    \hline
    \multirow{5}[2]{*}{AEEEM} & JDT   & ce    & wmc   & nbugs & 
    lwmc  & 
    cle \bigstrut[t]\\
          & PDE   & ntb   & cwe   & lloc  & ce    & cle \\
          & EQ    & cee   & loc   & cle   & ce    & cbo \\
          & \cellcolor[HTML]{EFEFEF}LC    & 
          \cellcolor[HTML]{EFEFEF}cwe   & 
          \cellcolor[HTML]{EFEFEF}nbugs & 
          \cellcolor[HTML]{EFEFEF}ce    & 
          \cellcolor[HTML]{EFEFEF}cle   & 
          \cellcolor[HTML]{EFEFEF}lloc \\
          & ML    & fanOut & CvsLinEntropy & loc   & lloc  & npm 
          \bigstrut[b]\\
    \hline
    \multirow{3}[2]{*}{Relink} & Apache & CountLineCodeExe & 
    CountLine & 
    CountLineCode & RatioCommentToCode & AvgEssential 
    \bigstrut[t]\\
          & Safe  & CountStmt & SumCyclomaticStrict & 
          CountLineCode & 
          CountStmtDecl & CountLineCodeExe \\
          & \cellcolor[HTML]{EFEFEF}Zxing & 
          \cellcolor[HTML]{EFEFEF}CountLineCodeDecl & 
          \cellcolor[HTML]{EFEFEF}CountLineCode & 
          \cellcolor[HTML]{EFEFEF}AvgLine 
          & 
          \cellcolor[HTML]{EFEFEF}CountLine & 
          \cellcolor[HTML]{EFEFEF}CountStmtDecl\\
    \hline
    \end{tabular}}%
     \caption{An example of source instability in defect datasets studied 
     here. The rows highlighted in gray indicate the bellwether dataset. Note: Space limitations prohibit showing these for the other communities. Interested readers are encouraged to use our replication package to see more examples of source instability in other communities.}
  \label{fig:instability}%
\end{figure*}%

\subsection*{RQ4: How much data is required to find the bellwether dataset?}

One of our defect dataset allows for a special kind of analysis -- the the 
Apache community (see \fig{datasets}) in the defect datasets has data 
available as historical versions. Using this dataset, we performed an 
empirical study to establish the required amount of bellwether data to make 
reliable predictions. We conducted experiments by incrementally updating 
the versions of the bellwether dataset until we find no significant increase in 
performance, i.e., starting from version $N$ (the latest version) we 
construct a prediction model and measure the performance using G-Score. 
Next, we include an older version $N-1$ to and construct a prediction 
model to measure the performance. This process is repeated by 
incrementally growing the size of the bellwether data by including older 
versions of the bellwether project. With this, the following empirical 
observations can be made: 
\bi
\item \fig{incr_result} documents the results of this experiment. As previously mentioned, we used the defect datasets from the Apache community in \fig{datasets}. In RQ1, it was found the $Lucene$ was the bellwether dataset for that community. In experimenting with different versions of $Lucene$, we found that using only the latest version of $Lucene$ produced statistically similar results to including the older versions of the data. Also, note that we required only 341 samples to achieve good G-scores. 
\item In cases where datasets were not available in the form of past versions, we observed that the size of the bellwether dataset is very small. For instance, consider the code-smells dataset, the bellwether datasets had no more than 12 samples. Similarly, in the case of effort estimation, the bellwether dataset had only 12 samples.
\ei

\begin{lesson}
	\respto{1-8} Not much data is required to find bellwether dataset. In the case 
	of defect 
	prediction, bellwethers can be found by analyzing only the latest version of 
	the project. Even in domains which lack data in the form if historical	
	versions, we were able to discover bellwethers with as few as 25 
	samples.\\[-0.2cm]
\end{lesson}\vspace{-0.5cm}

\subsection*{\respto{1-3-B}RQ5: How effectively do bellwethers mitigate for conclusion instability?}

In \tion{instability_definition},  we discussed two sources of conclusion 
instability, namely performance instability and source instability. We can use 
the bellwether effect to mitigate these two instabilities as follows:
\be
\item Performance 
instability causes data mining tools such as prediction algorithms to offer 
unreliable results (their performance depends on the data 
source). To address this issue, in this 
paper, we propose the use of the bellwether effect. This effect can be used 
to 
discover the bellwether 
data and we can use this data set as a reliable source to construct 
prediction 
models. Figures \ref{fig:smells_result}, 
\ref{fig:issues_result}, \ref{fig:defects_results}, and \ref{fig:effort_result} 
reveal that the bellwether data set can be discovered in three out of the four 
domains we have studied here. Additionally, the performance of an 
appropriate transfer learner (as identified in 
RQ3) with the bellwether dataset is statistically and significantly better than 
using other datasets. As long as the bellwether dataset remains 
unchanged, so will the performance of data mining tools such as transfer 
learners. 

\item Source instability causes  vastly different and often 
contradicting conclusions to be derived from a data source. This sort of 
instability is very prevalent in several domains of software engineering. An example of source instability in the case of defect prediction\footnote{Space limitations do not permit us to show these for the other three domains. As a result, we have made available a replication package with instructions to replicate these for all the other domains.} is shown in~\fig{instability}. This figure shows the rankings of top 5 features that contributed to the 
construction of the transfer learner (TCA+) for defect prediction tasks. It 
can be noted that, with every data 
source, the feature rankings are very different. For instance, if \textit{ant} 
was 
used to construct TCA+, one may conclude that rfc (response for class) 
is the most important feature, but if TCA+ was constructed using 
\textit{lucene}, then we would find that loc is the most important feature (rfc 
is only the $5^{\text{th}}$ most important feature). This sort of instability can 
be addressed by identifying a reliable data source to construct a transfer 
learner. The bellwether dataset is one such 
example of a stable data source. As long as the bellwether data is reliable 
(which can be established using the MONITOR step of 
\fig{bellwether_framework}) and the bellwether data remains unchanged, 
so will the conclusions derived from it. 

\ee

\noindent In summary, we may answer this research question as follows:
\begin{lesson}
The Bellwether Effect can be used to mitigate conclusion instability because as 
long as the bellwether dataset remains unchanged, we 
can (a) 
obtain consistent performance for a transfer learner, and (b) consistent 
conclusions from the bellwether dataset.\\[-0.2cm]
\end{lesson}\vspace{-0.5cm}

\section{Discussion}
\label{sect:discuss}

When reflecting on the findings of this work, there may be
four additional questions that arise. These are discussed below:

\be
\item \textit{Can bellwethers mitigate conclusion instability permanently?}
No- and we should not expect them to.
The aim of
bellwethers is to  {\em slow}, but do not necessarily {\em stop}, the pace of 
new ideas in software
engineering (e.g. as in the paper,  new quality prediction models).
Sometimes, new ideas are essential. Software engineering is a very dynamic 
field
with a high churn in techniques, platforms, developers and tasks.
In such a dynamic environment it is important to change with the times.
That said, changing {\em more} than what is necessary is not desirable-- 
hence this paper. 

\item \textit{How to detect when bellwether datasets need updating?}
The conclusion stability offered by bellwether datasets
only lasts as long as the bellwether dataset
remains useful. Hence, the bellwether dataset's performance must always be 
monitored 
and, if that performance starts to dip,
then seek a new bellwether dataset. 
\item \textit{What happens if a set of data has no useful bellwether 
dataset?}
In that case, there are numerous standard transfer learning methods
that could be used to import lessons learned from other 
data~\cite{kocaguneli2011find,kocaguneli2012,he2013learning,turhan09,peters15,Nam2013,Nam2015,Jing2015}.
That said, the result here is that all the communities of data explored by this 
paper
had useful bellwether datasets. Hence, we would recommend trying the 
bellwether method before moving
on to more complex methods.
\ee

\section{Threats to Validity}
\label{sect:threats}
\subsection{Sampling Bias} 
Sampling bias threatens any classification experiment;
what matters in one case may or may not hold in another case. 
For example, even though we use 100+ open-source datasets in 
this study which come from several sources, they were 
all supplied by individuals.  

That said, this paper shares this sampling bias problem with
every other data mining paper.  As researchers, all we can do is document
our selection procedure for data (as done in \tion{rq1}) and suggest that other researchers
try a broader range of data in future work.

\subsection{Learner Bias} 
\rahul{
For building the quality predictors in this
study, we elected to use random forests. We chose this learner
because past studies shows that, for prediction tasks, the 
results were superior to other more complicated 
algorithms~\cite{lessmann08}. We note that recent studies showed that 
different classifiers are highly complementary, despite obtaining similar 
performances. Thus, the usage of Random Forests is not bulletproof; but it 
can certainly act as a baseline for other algorithms. Exploration of these 
learners is part of our future work.
}
Apart from this choice,
one limitation to our current study is that we have focused here
on homogeneous transfer learning (where the attributes in source
and target are the same).
The implications for heterogeneous transfer learning (where the attributes
in source an target have different names) are not yet clear. We have some
initial results suggesting that a bellwether-like effect occurs when
learning across the communities but those results are
very preliminary. Hence, for the moment, we would conclude:
\bi
\item For the homogeneous case, we recommend using bellwethers rather than similarity-based transfer learning.
\item For the heterogeneous case, we recommend using   dimensionality transforms.
\ei

\subsection{Evaluation Bias}
This paper uses one measure of prediction quality, G (see \eq{G}). 
Other quality measures often used in software engineering to quantify
the effectiveness of prediction~\cite{ma07}~\cite{Menzies2007a}~\cite{fu16}
(discussed in \tion{eval}). A comprehensive analysis using these measures may be performed with our replication package.

\subsection{Random Bias} 

With random forest and SMOTE, there is invariably some degree of randomness that is introduced by both the algorithms. Random Forest, as the name suggests, randomly samples the data and constructs trees which it then uses in an ensemble fashion to make predictions. 

To mitigate these biases, we run
the experiments 30 times (the reruns are equal to 30 in keeping with the central limit theorem). Note that the reported variations over those runs
were very small. Hence, we conclude that parameter bias is theoretically a threat, as researchers we have used the default parameters in all situations. As researchers, all we can do is document our selection procedure for data (as done in \tion{rq1}) and suggest that other researchers try a broader range of data in future work.

\subsection{Parameter Bias}

With all the transfer learners and predictors discussed here, there are a number of internal parameters that have been set by default. The result of changing these parameters may (or may not) have a significant impact on the outcomes of this study. However, it must be noted that the possible number of combinations of these parameters is combinatorial in nature. There do however exist a growing number of literature on parameter optimization in SE. However, a encompassing this beyond the scope of this current paper. 

Hence, we conclude that although parameter bias is a possible threat, as researchers we have used the default parameters in all situations sake of consistency. We recommend that other researchers attempt to toggle these parameters with the use of tuning algorithms in to validate (or possibly refute) our findings.

\section{Conclusion}
\label{sect:conclusion}
\rahul{In this paper, we have undertaken a detailed study of transfer learners. Our results show that regardless of the sub-domain of software 
engineering (code smells, effort, defects or issue lifetimes) or granularity of 
data (file, class, or method), there exists a 
bellwether dataset that can be used to train 
relatively accurate quality prediction models and these bellwethers do not require 
elaborate data mining methods to discover (just a for-loop around the data 
sets) and can be found very early in a project's life cycle. 

We show that bellwether method is a simple baseline for transfer learning. 
The baseline performance offered by the bellwether method would be especially useful for researchers attempting to develop better transfer learners for different domains in software engineering. Further, bellwethers satisfy all the criteria of a baseline method, introduced in \tion{baseline}; i.e., they are simple to code and are applicable to a wide range of domains.

Hence, from a pragmatic engineering perspective there are two main reasons to use bellwethers: (a) they slow down the pace of conclusion change; and (b) they can be use to construct a simple baseline transfer learner with comparable performance to the state-of-the-art.

Finally, we remark that much of the prior work on homogeneous transfer learning, including some of the authors own papers, may have needless complicated the homogeneous transfer learning process.
We strongly recommend that when building increasingly complex automatic methods, researchers should pause and compare their supposedly more sophisticated method against simpler alternatives. Going forward from this
paper, we would recommend that the transfer learning
community uses bellwethers as a baseline method
against which they can test more complex methods.
}

\section*{Acknowledgements}
The work is partially funded by NSF awards \#1506586 
and \#1302169.

\balance
 \bibliographystyle{IEEEtran}
\bibliography{References}

\begin{thebibliography}{100}
\providecommand{\url}[1]{#1}
\csname url@samestyle\endcsname
\providecommand{\newblock}{\relax}
\providecommand{\bibinfo}[2]{#2}
\providecommand{\BIBentrySTDinterwordspacing}{\spaceskip=0pt\relax}
\providecommand{\BIBentryALTinterwordstretchfactor}{4}
\providecommand{\BIBentryALTinterwordspacing}{\spaceskip=\fontdimen2\font plus
\BIBentryALTinterwordstretchfactor\fontdimen3\font minus
  \fontdimen4\font\relax}
\providecommand{\BIBforeignlanguage}[2]{{%
\expandafter\ifx\csname l@#1\endcsname\relax
\typeout{** WARNING: IEEEtran.bst: No hyphenation pattern has been}%
\typeout{** loaded for the language `#1'. Using the pattern for}%
\typeout{** the default language instead.}%
\else
\language=\csname l@#1\endcsname
\fi
#2}}
\providecommand{\BIBdecl}{\relax}
\BIBdecl

\bibitem{czer11}
J.~Czerwonka, R.~Das, N.~Nagappan, A.~Tarvo, and A.~Teterev, ``Crane: Failure
  prediction, change analysis and test prioritization in practice --
  experiences from windows,'' in \emph{Software Testing, Verification and
  Validation (ICST), 2011 IEEE Fourth International Conference on}, march 2011,
  pp. 357 --366.

\bibitem{ostrand04}
T.~J. Ostrand, E.~J. Weyuker, and R.~M. Bell, ``Where the bugs are,'' in
  \emph{ISSTA '04: Proceedings of the 2004 ACM SIGSOFT international symposium
  on Software testing and analysis}.\hskip 1em plus 0.5em minus 0.4em\relax New
  York, NY, USA: ACM, 2004, pp. 86--96.

\bibitem{Menzies2007a}
\BIBentryALTinterwordspacing
T.~Menzies, A.~Dekhtyar, J.~Distefano, and J.~Greenwald, ``{Problems with
  Precision: A Response to "Comments on 'Data Mining Static Code Attributes to
  Learn Defect Predictors'"},'' \emph{IEEE Transactions on Software
  Engineering}, vol.~33, no.~9, pp. 637--640, sep 2007. [Online]. Available:
  \url{http://ieeexplore.ieee.org/lpdocs/epic03/wrapper.htm?arnumber=4288197}
\BIBentrySTDinterwordspacing

\bibitem{turhan11}
B.~Turhan, A.~Tosun, and A.~Bener, ``Empirical evaluation of mixed-project
  defect prediction models,'' in \emph{Software Engineering and Advanced
  Applications (SEAA), 2011 37th EUROMICRO Conference on}.\hskip 1em plus 0.5em
  minus 0.4em\relax IEEE, 2011, pp. 396--403.

\bibitem{koc11b}
E.~Kocaguneli, T.~Menzies, A.~Bener, and J.~Keung, ``Exploiting the essential
  assumptions of analogy-based effort estimation,'' \emph{IEEE Transactions on
  Software Engineering}, vol.~28, pp. 425--438, 2012, available from
  \url{http://menzies.us/pdf/11teak.pdf}.

\bibitem{export:208800}
\BIBentryALTinterwordspacing
A.~Begel and T.~Zimmermann, ``Analyze this! 145 questions for data scientists
  in software engineering,'' in \emph{Proceedings of the 36th International
  Conference on Software Engineering (ICSE 2014)}.\hskip 1em plus 0.5em minus
  0.4em\relax ACM, June 2014. [Online]. Available:
  \url{http://research.microsoft.com/apps/pubs/default.aspx?id=208800}
\BIBentrySTDinterwordspacing

\bibitem{theisen15}
C.~Theisen, K.~Herzig, P.~Morrison, B.~Murphy, and L.~Williams, ``Approximating
  attack surfaces with stack traces,'' in \emph{ICSE'15}, 2015.

\bibitem{me13c}
T.~Zimmermann and T.~Menzies, ``Software analytics: So what?'' \emph{IEEE
  Software}, vol.~30, no.~4, pp. 0031--37, 2013.

\bibitem{bird2015art}
C.~Bird, T.~Menzies, and T.~Zimmermann, \emph{The Art and Science of Analyzing
  Software Data}.\hskip 1em plus 0.5em minus 0.4em\relax Elsevier, 2015.

\bibitem{Nam2013}
J.~Nam, S.~J. Pan, and S.~Kim, ``{Transfer defect learning},'' in
  \emph{Proceedings - International Conference on Software Engineering}, 2013,
  pp. 382--391.

\bibitem{Nam2015}
\BIBentryALTinterwordspacing
J.~Nam and S.~Kim, ``{Heterogeneous defect prediction},'' in \emph{Proc. 2015
  10th Jt. Meet. Found. Softw. Eng. - ESEC/FSE 2015}.\hskip 1em plus 0.5em
  minus 0.4em\relax New York, New York, USA: ACM Press, 2015, pp. 508--519.
  [Online]. Available:
  \url{http://dl.acm.org/citation.cfm?doid=2786805.2786814}
\BIBentrySTDinterwordspacing

\bibitem{Jing2015}
X.~Jing, F.~Wu, X.~Dong, F.~Qi, and B.~Xu, ``{Heterogeneous Cross-Company
  Defect Prediction by Unified Metric Representation and CCA-Based Transfer
  Learning Categories and Subject Descriptors},'' \emph{Proceeding of the 10th
  Joint Meeting of the European Software Engineering Conference and the ACM
  SIGSOFT Symposium on the Foundations of Software Engineering (ESEC/FSE
  2015)}, pp. 496--507, 2015.

\bibitem{kocaguneli2011find}
E.~Kocaguneli and T.~Menzies, ``How to find relevant data for effort
  estimation?'' in \emph{Empirical Software Engineering and Measurement (ESEM),
  2011 International Symposium on}.\hskip 1em plus 0.5em minus 0.4em\relax
  IEEE, 2011, pp. 255--264.

\bibitem{kocaguneli2012}
\BIBentryALTinterwordspacing
E.~Kocaguneli, T.~Menzies, and E.~Mendes, ``{Transfer learning in effort
  estimation},'' \emph{Empirical Software Engineering}, vol.~20, no.~3, pp.
  813--843, jun 2015. [Online]. Available:
  \url{http://link.springer.com/10.1007/s10664-014-9300-5}
\BIBentrySTDinterwordspacing

\bibitem{turhan09}
B.~Turhan, T.~Menzies, A.~B. Bener, and J.~Di~Stefano, ``On the relative value
  of cross-company and within-company data for defect prediction,''
  \emph{Empirical Software Engineering}, vol.~14, no.~5, pp. 540--578, 2009.

\bibitem{peters15}
F.~Peters, T.~Menzies, and L.~Layman, ``{LACE2: Better privacy-preserving data
  sharing for cross project defect prediction},'' in \emph{Proceedings -
  International Conference on Software Engineering}, vol.~1, 2015, pp.
  801--811.

\bibitem{zimm09}
T.~Zimmermann, N.~Nagappan, H.~Gall, E.~Giger, and B.~Murphy, ``Cross-project
  defect prediction: a large scale experiment on data vs. domain vs. process,''
  in \emph{Proceedings of the the 7th joint meeting of the European software
  engineering conference and the ACM SIGSOFT symposium on The foundations of
  software engineering}.\hskip 1em plus 0.5em minus 0.4em\relax ACM, 2009, pp.
  91--100.

\bibitem{me12d}
\BIBentryALTinterwordspacing
T.~Menzies, A.~Butcher, A.~Marcus, T.~Zimmermann, and D.~Cok, ``{Local vs.
  global models for effort estimation and defect prediction},'' in \emph{2011
  26th IEEE/ACM International Conference on Automated Software Engineering (ASE
  2011)}.\hskip 1em plus 0.5em minus 0.4em\relax IEEE, nov 2011, pp. 343--351.
  [Online]. Available:
  \url{http://ieeexplore.ieee.org/lpdocs/epic03/wrapper.htm?arnumber=6100072}
\BIBentrySTDinterwordspacing

\bibitem{Hassan17}
A.~Hassan, ``Remarks made during a presentation to the ucl crest open
  workshop,'' March 2017.

\bibitem{krishna16a}
R.~Krishna, T.~Menzies, and W.~Fu, ``{Too Much Automation? The Bellwether
  Effect and Its Implications for Transfer Learning},'' in \emph{ASE'16}, 2016.

\bibitem{Ma2012}
Y.~Ma, G.~Luo, X.~Zeng, and A.~Chen, ``{Transfer learning for cross-company
  software defect prediction},'' \emph{Information and Software Technology},
  vol.~54, no.~3, pp. 248--256, 2012.

\bibitem{vcboost16}
\BIBentryALTinterwordspacing
D.~Ryu, O.~Choi, and J.~Baik, ``{Value-cognitive boosting with a support vector
  machine for cross-project defect prediction},'' \emph{Empir. Softw. Eng.},
  vol.~21, no.~1, pp. 43--71, feb 2016. [Online]. Available:
  \url{http://link.springer.com/10.1007/s10664-014-9346-4}
\BIBentrySTDinterwordspacing

\bibitem{Xu15a}
\BIBentryALTinterwordspacing
T.~Xu, L.~Jin, X.~Fan, Y.~Zhou, S.~Pasupathy, and R.~Talwadker, ``Hey, you have
  given me too many knobs!: Understanding and dealing with over-designed
  configuration in system software,'' in \emph{Proceedings of the 2015 10th
  Joint Meeting on Foundations of Software Engineering}, ser. ESEC/FSE
  2015.\hskip 1em plus 0.5em minus 0.4em\relax New York, NY, USA: ACM, 2015,
  pp. 307--319. [Online]. Available:
  \url{http://doi.acm.org/10.1145/2786805.2786852}
\BIBentrySTDinterwordspacing

\bibitem{Wo97}
\BIBentryALTinterwordspacing
D.~H. Wolpert and W.~G. Macready, ``No free lunch theorems for optimization,''
  \emph{Trans. Evol. Comp}, vol.~1, no.~1, pp. 67--82, Apr. 1997. [Online].
  Available: \url{http://dx.doi.org/10.1109/4235.585893}
\BIBentrySTDinterwordspacing

\bibitem{cohen95}
P.~R. Cohen, \emph{{Empirical Methods for Artificial Intelligence}}.\hskip 1em
  plus 0.5em minus 0.4em\relax MIT Press, 1995.

\bibitem{holte93}
R.~C. Holte, ``{Very Simple Classification Rules Perform Well on Most Commonly
  Used Datasets},'' \emph{Machine Learning}, vol.~11, p.~63, 1993.

\bibitem{Whigham:2015}
\BIBentryALTinterwordspacing
P.~A. Whigham, C.~A. Owen, and S.~G. Macdonell, ``A baseline model for software
  effort estimation,'' \emph{ACM Trans. Softw. Eng. Methodol.}, vol.~24, no.~3,
  pp. 20:1--20:11, May 2015. [Online]. Available:
  \url{http://doi.acm.org/10.1145/2738037}
\BIBentrySTDinterwordspacing

\bibitem{mittas13}
N.~Mittas and L.~Angelis, ``Ranking and clustering software cost estimation
  models through a multiple comparisons algorithm,'' \emph{IEEE Trans. Software
  Eng.}, vol.~39, no.~4, pp. 537--551, 2013.

\bibitem{shepperd12z}
M.~J. Shepperd and S.~G. MacDonell, ``Evaluating prediction systems in software
  project estimation,'' \emph{Information {\&} Software Technology}, vol.~54,
  no.~8, pp. 820--827, 2012.

\bibitem{Kitchen}
\BIBentryALTinterwordspacing
B.~A. Kitchenham, E.~Mendes, and G.~H. Travassos, ``Cross versus within-company
  cost estimation studies: A systematic review,'' \emph{IEEE Trans. Softw.
  Eng.}, vol.~33, no.~5, pp. 316--329, May 2007. [Online]. Available:
  \url{http://dx.doi.org/10.1109/TSE.2007.1001}
\BIBentrySTDinterwordspacing

\bibitem{ekanayake2009tracking}
J.~Ekanayake, J.~Tappolet, H.~C. Gall, and A.~Bernstein, ``Tracking concept
  drift of software projects using defect prediction quality,'' in \emph{Mining
  Software Repositories, 2009. MSR'09. 6th IEEE International Working
  Conference on}.\hskip 1em plus 0.5em minus 0.4em\relax IEEE, 2009, pp.
  51--60.

\bibitem{fowler99}
M.~Fowler, K.~Beck, J.~Brant, W.~Opdyke, and D.~Roberts, \emph{Refactoring:
  Improving the Design of Existing Code}.\hskip 1em plus 0.5em minus
  0.4em\relax Boston, MA, USA: Addison-Wesley Longman, 1999.

\bibitem{Kerievsky2005}
J.~Kerievsky, \emph{Refactoring to Patterns}.\hskip 1em plus 0.5em minus
  0.4em\relax Addison-Wesly Professional, 2005.

\bibitem{Lanza2006}
M.~Lanza and R.~Marinescu, \emph{{Object-Oriented Metrics in Practice: Using
  Software Metrics to Characterize, Evaluate, and Improve the Design of
  Object-Oriented Systems}}.\hskip 1em plus 0.5em minus 0.4em\relax Springer
  Verlag, 2006.

\bibitem{sq15}
A.~Campbell, ``Sonar{Q}ube: Open source quality management,'' 2015, website:
  tiny.cc/2q4z9x.

\bibitem{Yamashita2013}
A.~Yamashita and L.~Moonen, ``Do developers care about code smells? an
  exploratory survey,'' in \emph{Reverse Engineering (WCRE), 2013 20th Working
  Conference on}, Oct 2013, pp. 242--251.

\bibitem{ref01}
H.~Olague, L.~Etzkorn, S.~Gholston, and S.~Quattlebaum, ``Empirical validation
  of three software metrics suites to predict fault-proneness of
  object-oriented classes developed using highly iterative or agile software
  development processes,'' \emph{Software Engineering, IEEE Transactions},
  vol.~33, no.~6, pp. 402--419, 2007.

\bibitem{ref02}
K.~Aggmakarwal, Y.~Singh, A.~Kaur, and R.~Malhotra, ``Empirical analysis for
  investigating the effect of object-oriented metrics on fault proneness: a
  replicated case study,'' \emph{Software Process: Improvement and Practice},
  vol.~14, no.~1, January 2009.

\bibitem{ref03}
E.~Arisholm and L.~Briand, ``Predicting fault prone components in a {JAVA}
  legacy system,'' \emph{2006 ACM/IEEE international symposium on Empirical
  software engineering}, p.~17, 2006.

\bibitem{ref04}
V.~Basili, L.~Briand, and W.~Melo, ``A validation of object-oriented design
  metrics as quality indicators,'' \emph{Software Engineering, IEEE
  Transactions}, vol.~22, no.~10, pp. 751--761, 1996.

\bibitem{ref05}
L.~Briand, J.~Wust, J.~Daly, and D.~V. Porter, ``Exploring the relationships
  between design measures and software quality in object-oriented systems,''
  \emph{Journal of Systems and Software}, vol.~51, no.~3, pp. 245--273, 2000.

\bibitem{ref06}
L.~Briand, J.~Wust, and H.~Lounis, ``Replicated case studies for investigating
  quality factors in object-oriented designs,'' \emph{Empirical Software
  Engineering}, vol.~6, no.~1, pp. 11--58, 2001.

\bibitem{ref07}
M.~Cartwright and M.~Shepperd, ``An empirical investigation of an
  object-oriented software system,'' \emph{Software Engineering, IEEE
  Transactions}, vol.~26, no.~8, pp. 786--796, 2000.

\bibitem{ref08}
K.~el~Emam, W.~Melo, and J.~Machado, ``The prediction of faulty classes using
  object-oriented design metrics,'' \emph{Journal of Systems and Software},
  vol.~56, no.~1, pp. 63--75, 2001.

\bibitem{ref09}
K.~el~Emam, S.~Benlarbi, N.~Goel, and S.~Rai, ``A validation of object-oriented
  metrics,'' \emph{National Research Council of Canada, NRC/ERB,}, vol. 1063,
  1999.

\bibitem{ref10}
M.~Tang, M.~Kao, and M.~Chen, ``An empirical study on object-oriented
  metrics,'' \emph{Software Metrics Symposium}, vol. Proceedings. Sixth
  International, pp. 242--249, 1999.

\bibitem{ref11}
P.~Yu, T.~Systa, and H.~Muller, ``Predicting fault-proneness using oo metrics
  an industrial case study,'' \emph{Sixth European Conference on Software
  Maintenance and Reengineering}, pp. 99--107, 2002.

\bibitem{ref12}
R.~Subramanyam and M.~S. Krishnan, ``Empirical analysis of ck metrics for
  object-oriented design complexity: implications for software defects,''
  \emph{IEEE Transactions on Software Engineering}, vol.~29, no.~4, pp.
  297--310, April 2003.

\bibitem{ref13}
Y.~Zhou and H.~Leung, ``Empirical analysis of object-oriented design metrics
  for predicting high and low severity faults,'' \emph{Software Engineering,
  IEEE Transactions}, vol.~32, no.~10, pp. 771--789, 2006.

\bibitem{ref14}
T.~Gyimothy, R.~Ferenc, and I.~Siket, ``Empirical validation of object-oriented
  metrics on open source software for fault prediction,'' \emph{Software
  Engineering, IEEE Transactions}, vol.~31, no.~10, pp. 897--910, 2005.

\bibitem{ref15}
T.~Holschuh, M.~Pauser, K.~Herzig, T.~Zimmermann, R.~Premraj, and A.~Zeller,
  ``Predicting defects in {SAP} {J}ava code: An experience report,''
  \emph{Software Engineering - Companion Volume, 2009. ICSE-Companion 2009.
  31st International Conference}, pp. 172--181, 2009.

\bibitem{ref16}
R.~Shatnawi and W.~Li, ``The effectiveness of software metrics in identifying
  error-prone classes in post-release software evolution process,''
  \emph{Journal of Systems and Software}, vol.~81, no.~11, pp. 1868--1882,
  2008.

\bibitem{ref17}
F.~Fioravanti and P.~Nesi, ``A study on fault-proneness detection of
  object-oriented systems,'' \emph{Software Maintenance and Reengineering,
  2001. Fifth European Conference}, pp. 121--130, 2001.

\bibitem{ref18}
M.~Thongmak and P.~Muenchaisri, ``Predicting faulty classes using design
  metrics with discriminant analysis,'' \emph{Software Engineering Research and
  Practice}, pp. 621--627, 2003.

\bibitem{ref19}
G.~Denaro, L.~Lavazza, and M.~Pezze, ``An empirical evaluation of object
  oriented metrics in industrial setting,'' \emph{The 5th CaberNet Plenary
  Workshop, Porto Santo, Madeira Archipelago, Portugal}, 2003.

\bibitem{ref20}
A.~Janes, M.~Scotto, W.~Pedrycz, B.~Russo, M.~Stefanovic, and G.~Succi,
  ``Identification of defect-prone classes in telecommunication software
  systems using design metrics,'' \emph{Information Sciences}, vol. 176,
  no.~24, pp. 3711--3734, 2006.

\bibitem{ref21}
M.~English, C.~Exton, I.~Rigon, and B.~Cleary, ``Fault detection and prediction
  in an open-source software project,'' \emph{Proceedings of the 5th
  International Conference on Predictor Models in Software Engineering}, no.
  1-11, 2009.

\bibitem{ref22}
R.~Shatnawi, ``A quantitative investigation of the acceptable risk levels of
  object-oriented metrics in open-source systems,'' \emph{IEEE Transactions on
  Software Engineering}, vol.~36, no.~2, pp. 216--225, 2010.

\bibitem{ref23}
Y.~Singh, A.~Kaur, and R.~Malhotra, ``Empirical validation of object-oriented
  metrics for predicting fault proneness models,'' \emph{Software Quality
  Journal}, vol.~18, no.~1, pp. 3--35, 2010.

\bibitem{ref24}
D.~Glasberg, K.~el~Emam, W.~Memo, and N.~Madhavji, ``Validating object-oriented
  design metrics on a commercial {JAVA} application,'' \emph{NRC 44146}, 2000.

\bibitem{ref25}
K.~el~Emam, S.~Benlarbi, N.~Goel, and S.~Rai, ``The confounding effect of class
  size on the validity of object-oriented metrics,'' \emph{Software
  Engineering, IEEE Transactions}, vol.~27, no.~7, pp. 630--650, 2001.

\bibitem{ref26}
M.~Thapaliyal and G.~Verma, ``Software defects and object oriented metrics-an
  empirical analysis,'' \emph{International Journal of Computer Applications},
  vol. 9/5, 2010.

\bibitem{ref27}
J.~Xu, D.~Ho, and L.~Capretz, ``An empirical validation of object-oriented
  design metrics for fault prediction,'' \emph{Journal of Computer Science},
  pp. 571--577, July 2008.

\bibitem{ref28}
G.~Succi, ``Practical assessment of the models for identification of
  defect-prone classes in object-oriented commercial systems using design
  metrics,,'' \emph{Journal of Systems and Software}, vol.~65, no.~1, pp.
  1--12, January 2003.

\bibitem{lessmann08}
\BIBentryALTinterwordspacing
S.~Lessmann, B.~Baesens, C.~Mues, and S.~Pietsch, ``{Benchmarking
  Classification Models for Software Defect Prediction: A Proposed Framework
  and Novel Findings},'' \emph{IEEE Trans. Softw. Eng.}, vol.~34, no.~4, pp.
  485--496, jul 2008. [Online]. Available:
  \url{http://ieeexplore.ieee.org/lpdocs/epic03/wrapper.htm?arnumber=4527256}
\BIBentrySTDinterwordspacing

\bibitem{hall2012}
T.~Hall, S.~Beecham, D.~Bowes, D.~Gray, and S.~Counsell, ``A systematic
  literature review on fault prediction performance in software engineering,''
  \emph{IEEE Transactions on Software Engineering}, vol.~38, no.~6, pp.
  1276--1304, Nov 2012.

\bibitem{elish2008predicting}
K.~O. Elish and M.~O. Elish, ``Predicting defect-prone software modules using
  support vector machines,'' \emph{JSS}, vol.~81, no.~5, pp. 649--660, 2008.

\bibitem{menzies2010defect}
T.~Menzies, Z.~Milton, B.~Turhan, B.~Cukic, Y.~Jiang, and A.~Bener, ``Defect
  prediction from static code features: current results, limitations, new
  approaches,'' \emph{Automated Software Engineering}, vol.~17, no.~4, pp.
  375--407, 2010.

\bibitem{gondra2008applying}
I.~Gondra, ``Applying machine learning to software fault-proneness
  prediction,'' \emph{Journal of Systems and Software}, vol.~81, no.~2, pp.
  186--195, 2008.

\bibitem{radjenovic2013software}
D.~Radjenovi{\'c}, M.~Heri{\v{c}}ko, R.~Torkar, and A.~{\v{Z}}ivkovi{\v{c}},
  ``Software fault prediction metrics: A systematic literature review,''
  \emph{Information and Software Technology}, vol.~55, no.~8, pp. 1397--1418,
  2013.

\bibitem{jiang2008techniques}
Y.~Jiang, B.~Cukic, and Y.~Ma, ``Techniques for evaluating fault prediction
  models,'' \emph{Empirical Software Engineering}, vol.~13, no.~5, pp.
  561--595, 2008.

\bibitem{wang2013using}
S.~Wang and X.~Yao, ``Using class imbalance learning for software defect
  prediction,'' \emph{IEEE Transactions on Reliability}, vol.~62, no.~2, pp.
  434--443, 2013.

\bibitem{mende2009revisiting}
T.~Mende and R.~Koschke, ``Revisiting the evaluation of defect prediction
  models,'' in \emph{Proceedings of the 5th International Conference on
  Predictor Models in Software Engineering}.\hskip 1em plus 0.5em minus
  0.4em\relax ACM, 2009, p.~7.

\bibitem{li2012sample}
M.~Li, H.~Zhang, R.~Wu, and Z.-H. Zhou, ``Sample-based software defect
  prediction with active and semi-supervised learning,'' \emph{Automated
  Software Engineering}, vol.~19, no.~2, pp. 201--230, 2012.

\bibitem{khoshgoftaar2010attribute}
T.~M. Khoshgoftaar, K.~Gao, and N.~Seliya, ``Attribute selection and imbalanced
  data: Problems in software defect prediction,'' in \emph{Tools with
  Artificial Intelligence (ICTAI), 2010 22nd IEEE International Conference on},
  vol.~1.\hskip 1em plus 0.5em minus 0.4em\relax IEEE, 2010, pp. 137--144.

\bibitem{jiang2009variance}
Y.~Jiang, J.~Lin, B.~Cukic, and T.~Menzies, ``Variance analysis in software
  fault prediction models,'' in \emph{Software Reliability Engineering, 2009.
  ISSRE'09. 20th International Symposium on}.\hskip 1em plus 0.5em minus
  0.4em\relax IEEE, 2009, pp. 99--108.

\bibitem{ghotra2015revisiting}
B.~Ghotra, S.~McIntosh, and A.~E. Hassan, ``Revisiting the impact of
  classification techniques on the performance of defect prediction models,''
  in \emph{37th ICSE-Volume 1}.\hskip 1em plus 0.5em minus 0.4em\relax IEEE
  Press, 2015, pp. 789--800.

\bibitem{jiang2008can}
Y.~Jiang, B.~Cukic, and T.~Menzies, ``Can data transformation help in the
  detection of fault-prone modules?'' in \emph{Proceedings of the 2008 workshop
  on Defects in large software systems}.\hskip 1em plus 0.5em minus 0.4em\relax
  ACM, 2008, pp. 16--20.

\bibitem{tantithamthavorn2016automated}
C.~Tantithamthavorn, S.~McIntosh, A.~E. Hassan, and K.~Matsumoto, ``Automated
  parameter optimization of classification techniques for defect prediction
  models,'' in \emph{ICSE 2016}.\hskip 1em plus 0.5em minus 0.4em\relax ACM,
  2016, pp. 321--332.

\bibitem{fu2016tuning}
W.~Fu, T.~Menzies, and X.~Shen, ``Tuning for software analytics: Is it really
  necessary?'' \emph{IST}, vol.~76, pp. 135--146, 2016.

\bibitem{Ca09}
J.~Quionero-Candela, M.~Sugiyama, A.~Schwaighofer, and N.~D. Lawrence,
  \emph{Dataset shift in machine learning}.\hskip 1em plus 0.5em minus
  0.4em\relax The MIT Press, 2009.

\bibitem{Ha06}
D.~J. {Hand}, ``{Classifier Technology and the Illusion of Progress},''
  \emph{ArXiv Mathematics e-prints}, Jun. 2006.

\bibitem{St09}
\BIBentryALTinterwordspacing
A.~Storkey, ``When training and test sets are different: Characterizing
  learning transfer,'' in \emph{Dataset Shift in Machine Learning}.\hskip 1em
  plus 0.5em minus 0.4em\relax The {MIT} Press, dec 2008, pp. 2--28. [Online].
  Available: \url{https://doi.org/10.7551/mitpress/9780262170055.003.0001}
\BIBentrySTDinterwordspacing

\bibitem{agarwal17}
\BIBentryALTinterwordspacing
A.~Agrawal and T.~Menzies, ``"better data" is better than "better data miners"
  (benefits of tuning {SMOTE} for defect prediction),'' \emph{CoRR}, vol.
  abs/1705.03697, 2017. [Online]. Available:
  \url{http://arxiv.org/abs/1705.03697}
\BIBentrySTDinterwordspacing

\bibitem{Tufano2015}
\BIBentryALTinterwordspacing
M.~Tufano, F.~Palomba, G.~Bavota, R.~Oliveto, M.~{Di Penta}, A.~{De Lucia}, and
  D.~Poshyvanyk, ``{When and Why Your Code Starts to Smell Bad},'' in
  \emph{2015 IEEE/ACM 37th IEEE Int. Conf. Softw. Eng.}\hskip 1em plus 0.5em
  minus 0.4em\relax IEEE, May 2015, pp. 403--414. [Online]. Available:
  \url{http://ieeexplore.ieee.org/lpdocs/epic03/wrapper.htm?arnumber=7194592}
\BIBentrySTDinterwordspacing

\bibitem{Mantyla2004}
M.~Mantyla, J.~Vanhanen, and C.~Lassenius, ``Bad smells - humans as code
  critics,'' in \emph{Software Maintenance, 2004. Proceedings. 20th IEEE
  International Conference on}, Sept 2004, pp. 399--408.

\bibitem{Sjoberg2013}
D.~Sjoberg, A.~Yamashita, B.~Anda, A.~Mockus, and T.~Dyba, ``Quantifying the
  effect of code smells on maintenance effort,'' \emph{Software Engineering,
  IEEE Transactions on}, vol.~39, no.~8, pp. 1144--1156, Aug 2013.

\bibitem{passos11}
C.~Passos, A.~P. Braun, D.~S. Cruzes, and M.~Mendonca, ``Analyzing the impact
  of beliefs in software project practices,'' in \emph{ESEM'11}, 2011.

\bibitem{jorgensen09}
M.~J{\o}rgensen and T.~M. Gruschke, ``The impact of lessons-learned sessions on
  effort estimation and uncertainty assessments,'' \emph{Software Engineering,
  IEEE Transactions on}, vol.~35, no.~3, pp. 368 --383, May-June 2009.

\bibitem{prem16}
P.~Devanbu, T.~Zimmermann, and C.~Bird, ``Belief \& evidence in empirical
  software engineering,'' in \emph{Proceedings of the 38th International
  Conference on Software Engineering}.\hskip 1em plus 0.5em minus 0.4em\relax
  ACM, 2016, pp. 108--119.

\bibitem{zimmermann09}
T.~Zimmermann, N.~Nagappan, H.~Gall, E.~Giger, and B.~Murphy, ``Cross-project
  defect prediction,'' in \emph{ESEC/FSE'09}, August 2009.

\bibitem{menzies2011local}
T.~Menzies, A.~Butcher, A.~Marcus, T.~Zimmermann, and D.~Cok, ``Local vs.
  global models for effort estimation and defect prediction,'' in
  \emph{Proceedings of the 2011 26th IEEE/ACM International Conference on
  Automated Software Engineering}.\hskip 1em plus 0.5em minus 0.4em\relax IEEE
  Computer Society, 2011, pp. 343--351.

\bibitem{turhan12}
B.~Turhan, ``On the dataset shift problem in software engineering prediction
  models,'' \emph{Empirical Software Engineering}, vol.~17, pp. 62--74, 2012.

\bibitem{me05a}
T.~Menzies, Z.~Chen, D.~Port, and J.~Hihn, ``Simple software cost estimation:
  Safe or unsafe?'' in \emph{Proceedings, PROMISE workshop, ICSE 2005}, 2005,
  available from \url{http://menzies.us/pdf/05safewhen.pdf}.

\bibitem{jorg04}
M.~Jorgensen, ``Realism in assessment of effort estimation uncertainty: It
  matters how you ask,'' \emph{IEEE Trans. Softw. Eng.}, vol.~30, no.~4, pp.
  209--217, 2004.

\bibitem{Mendes2007}
B.~Kitchenham, E.~Mendes, and G.~H. Travassos, ``Cross versus within-company
  cost estimation studies: A systematic review,'' \emph{IEEE Trans. Softw.
  Eng.}, vol.~33, no.~5, pp. 316--329, 2007, member-Kitchenham, Barbara A.

\bibitem{macdonell07}
S.~G. MacDonell and M.~J. Shepperd, ``Comparing local and global software
  effort estimation models -- reflections on a systematic review,'' in
  \emph{Proceedings of the First International Symposium on Empirical Software
  Engineering and Measurement}, ser. ESEM '07.\hskip 1em plus 0.5em minus
  0.4em\relax Washington, DC, USA: IEEE Computer Society, 2007, pp. 401--409.

\bibitem{mair05}
C.~Mair and M.~Shepperd, ``The consistency of empirical comparisons of
  regression and analogy-based software project cost prediction,'' in
  \emph{Empirical Software Engineering, 2005. 2005 International Symposium on},
  nov. 2005, p. 10 pp.

\bibitem{zhang15}
\BIBentryALTinterwordspacing
F.~Zhang, A.~Mockus, I.~Keivanloo, and Y.~Zou,
  ``\BIBforeignlanguage{English}{Towards building a universal defect prediction
  model with rank transformed predictors},''
  \emph{\BIBforeignlanguage{English}{Empirical Software Engineering}}, pp.
  1--39, 2015. [Online]. Available:
  \url{http://dx.doi.org/10.1007/s10664-015-9396-2}
\BIBentrySTDinterwordspacing

\bibitem{yamashita2013code}
A.~Yamashita and S.~Counsell, ``Code smells as system-level indicators of
  maintainability: An empirical study,'' \emph{Journal of Systems and
  Software}, vol.~86, no.~10, pp. 2639--2653, 2013.

\bibitem{yama13}
A.~Yamashita and L.~Moonen, ``Exploring the impact of inter-smell relations on
  software maintainability: An empirical study,'' in \emph{Proceedings of the
  2013 International Conference on Software Engineering}.\hskip 1em plus 0.5em
  minus 0.4em\relax IEEE Press, 2013, pp. 682--691.

\bibitem{zazworka2011investigating}
N.~Zazworka, M.~A. Shaw, F.~Shull, and C.~Seaman, ``Investigating the impact of
  design debt on software quality,'' in \emph{Proceedings of the 2nd Workshop
  on Managing Technical Debt}.\hskip 1em plus 0.5em minus 0.4em\relax ACM,
  2011, pp. 17--23.

\bibitem{krishna2016b}
\BIBentryALTinterwordspacing
R.~Krishna, T.~Menzies, and L.~Layman, ``{Less is More: Minimizing Code
  Reorganization using XTREE},'' \emph{CoRR}, vol. abs/1609.03614, 2016.
  [Online]. Available: \url{http://arxiv.org/abs/1609.03614}
\BIBentrySTDinterwordspacing

\bibitem{kreimer05}
\BIBentryALTinterwordspacing
J.~Kreimer, ``{Adaptive Detection of Design Flaws},'' \emph{Electronic Notes in
  Theoretical Computer Science}, vol. 141, no.~4, pp. 117--136, 2005. [Online].
  Available:
  \url{//www.sciencedirect.com/science/article/pii/S1571066105051844}
\BIBentrySTDinterwordspacing

\bibitem{khomh09}
F.~Khomh, S.~Vaucher, Y.~G. Guéhéneuc, and H.~Sahraoui, ``A bayesian approach
  for the detection of code and design smells,'' in \emph{2009 Ninth
  International Conference on Quality Software}, Aug 2009, pp. 305--314.

\bibitem{khomh11}
\BIBentryALTinterwordspacing
F.~Khomh, S.~Vaucher, Y.-G. Guéhéneuc, and H.~Sahraoui, ``Bdtex: A gqm-based
  bayesian approach for the detection of antipatterns,'' \emph{Journal of
  Systems and Software}, vol.~84, no.~4, pp. 559 -- 572, 2011, the Ninth
  International Conference on Quality Software. [Online]. Available:
  \url{//www.sciencedirect.com/science/article/pii/S0164121210003225}
\BIBentrySTDinterwordspacing

\bibitem{maiga12}
J.~Yang, K.~Hotta, Y.~Higo, H.~Igaki, and S.~Kusumoto, ``Filtering clones for
  individual user based on machine learning analysis,'' in \emph{2012 6th
  International Workshop on Software Clones (IWSC)}, June 2012, pp. 76--77.

\bibitem{font16}
\BIBentryALTinterwordspacing
F.~{Arcelli Fontana}, M.~V. M{\"{a}}ntyl{\"{a}}, M.~Zanoni, and A.~Marino,
  ``{Comparing and experimenting machine learning techniques for code smell
  detection},'' \emph{Empir. Softw. Eng.}, vol.~21, no.~3, pp. 1143--1191, jun
  2016. [Online]. Available: \url{http://dx.doi.org/10.1007/s10664-015-9378-4
  http://link.springer.com/10.1007/s10664-015-9378-4}
\BIBentrySTDinterwordspacing

\bibitem{tempero}
E.~Tempero, C.~Anslow, J.~Dietrich, T.~Han, J.~Li, M.~Lumpe, H.~Melton, and
  J.~Noble, ``Qualitas corpus: A curated collection of java code for empirical
  studies,'' in \emph{2010 Asia Pacific Software Engineering Conference
  (APSEC2010)}, Dec. 2010, pp. 336--345.

\bibitem{panjer}
\BIBentryALTinterwordspacing
L.~D. Panjer, ``Predicting eclipse bug lifetimes,'' in \emph{Proceedings of the
  Fourth International Workshop on Mining Software Repositories}, ser. MSR
  '07.\hskip 1em plus 0.5em minus 0.4em\relax Washington, DC, USA: IEEE
  Computer Society, 2007, pp. 29--. [Online]. Available:
  \url{http://dx.doi.org/10.1109/MSR.2007.25}
\BIBentrySTDinterwordspacing

\bibitem{giger}
\BIBentryALTinterwordspacing
E.~Giger, M.~Pinzger, and H.~Gall, ``Predicting the fix time of bugs,'' in
  \emph{Proceedings of the 2Nd International Workshop on Recommendation Systems
  for Software Engineering}, ser. RSSE '10.\hskip 1em plus 0.5em minus
  0.4em\relax New York, NY, USA: ACM, 2010, pp. 52--56. [Online]. Available:
  \url{http://doi.acm.org/10.1145/1808920.1808933}
\BIBentrySTDinterwordspacing

\bibitem{zhang13issue}
\BIBentryALTinterwordspacing
H.~Zhang, L.~Gong, and S.~Versteeg, ``Predicting bug-fixing time: An empirical
  study of commercial software projects,'' in \emph{Proceedings of the 2013
  International Conference on Software Engineering}, ser. ICSE '13.\hskip 1em
  plus 0.5em minus 0.4em\relax Piscataway, NJ, USA: IEEE Press, 2013, pp.
  1042--1051. [Online]. Available:
  \url{http://dl.acm.org/citation.cfm?id=2486788.2486931}
\BIBentrySTDinterwordspacing

\bibitem{rees}
M.~Rees-jones, M.~Martin, C.~College, and T.~Menzies, ``{Better Predictors for
  Issue Lifetime},'' \emph{{}}, pp. 1--8, 2017.

\bibitem{yang11}
\BIBentryALTinterwordspacing
Y.~Yang, L.~Xie, Z.~He, Q.~Li, V.~Nguyen, B.~Boehm, and R.~Valerdi, ``{Local
  bias and its impacts on the performance of parametric estimation models},''
  in \emph{Proceedings of the 7th International Conference on Predictive Models
  in Software Engineering - Promise '11}.\hskip 1em plus 0.5em minus
  0.4em\relax New York, New York, USA: ACM Press, 2011, pp. 1--10. [Online].
  Available: \url{http://dl.acm.org/citation.cfm?doid=2020390.2020404}
\BIBentrySTDinterwordspacing

\bibitem{Menzies2016}
\BIBentryALTinterwordspacing
T.~Menzies, Y.~Yang, G.~Mathew, B.~Boehm, and J.~Hihn, ``Negative results for
  software effort estimation,'' \emph{Empirical Software Engineering}, pp.
  1--26, 2016. [Online]. Available:
  \url{http://dx.doi.org/10.1007/s10664-016-9472-2}
\BIBentrySTDinterwordspacing

\bibitem{cocomo}
B.~W. Boehm \emph{et~al.}, \emph{Software engineering economics}.\hskip 1em
  plus 0.5em minus 0.4em\relax Prentice-hall Englewood Cliffs (NJ), 1981, vol.
  197.

\bibitem{LowryBK98}
M.~Lowry, M.~Boyd, and D.~Kulkami, ``Towards a theory for integration of
  mathematical verification and empirical testing,'' in \emph{Automated
  Software Engineering, 1998. Proceedings. 13th IEEE International Conference
  on}.\hskip 1em plus 0.5em minus 0.4em\relax IEEE, 1998, pp. 322--331.

\bibitem{nagappan05}
N.~Nagappan and T.~Ball, ``Static analysis tools as early indicators of
  pre-release defect density,'' in \emph{ICSE 2005, St. Louis}, 2005.

\bibitem{me02f}
T.~Menzies, D.~Raffo, S.~on~Setamanit, Y.~Hu, and S.~Tootoonian, ``Model-based
  tests of truisms,'' in \emph{Proceedings of IEEE ASE 2002}, 2002, available
  from \url{http://menzies.us/pdf/02truisms.pdf}.

\bibitem{me07b}
T.~Menzies, J.~Greenwald, and A.~Frank, ``Data mining static code attributes to
  learn defect predictors,'' \emph{IEEE Transactions on Software Engineering},
  January 2007, available from \url{http://menzies.us/pdf/06learnPredict.pdf}.

\bibitem{tosun10}
A.~Tosun, A.~Bener, and R.~Kale, ``{AI}-based software defect predictors:
  Applications and benefits in a case study,'' in \emph{Twenty-Second IAAI
  Conference on Artificial Intelligence}, 2010.

\bibitem{tosun09}
A.~Tosun, A.~Bener, and B.~Turhan, ``Practical considerations of deploying ai
  in defect prediction: A case study within the {T}urkish telecommunication
  industry,'' in \emph{PROMISE'09}, 2009.

\bibitem{shu02}
F.~Shull, V.~B. ad~B.~Boehm, A.~Brown, P.~Costa, M.~Lindvall, D.~Port, I.~Rus,
  R.~Tesoriero, and M.~Zelkowitz, ``What we have learned about fighting
  defects,'' in \emph{Proceedings of 8th International Software Metrics
  Symposium, Ottawa, Canada}, 2002, pp. 249--258.

\bibitem{fagan76}
M.~Fagan, ``Design and code inspections to reduce errors in program
  development,'' \emph{IBM Systems Journal}, vol.~15, no.~3, 1976.

\bibitem{Rahman:2014}
F.~Rahman, S.~Khatri, E.~T. Barr, and P.~Devanbu, ``Comparing static bug
  finders and statistical prediction,'' in \emph{Proceedings of the 36th
  International Conference on Software Engineering}.\hskip 1em plus 0.5em minus
  0.4em\relax ACM, 2014, pp. 424--434.

\bibitem{lewis13}
\BIBentryALTinterwordspacing
C.~Lewis, Z.~Lin, C.~Sadowski, X.~Zhu, R.~Ou, and E.~J. Whitehead~Jr., ``Does
  bug prediction support human developers? findings from a google case study,''
  in \emph{Proceedings of the 2013 International Conference on Software
  Engineering}, ser. ICSE '13.\hskip 1em plus 0.5em minus 0.4em\relax
  Piscataway, NJ, USA: IEEE Press, 2013, pp. 372--381. [Online]. Available:
  \url{http://dl.acm.org/citation.cfm?id=2486788.2486838}
\BIBentrySTDinterwordspacing

\bibitem{rakitin01}
S.~Rakitin, \emph{Software Verification and Validation for Practitioners and
  Managers, Second Edition}.\hskip 1em plus 0.5em minus 0.4em\relax Artech
  House, 2001.

\bibitem{DAmbros2012}
\BIBentryALTinterwordspacing
M.~D'Ambros, M.~Lanza, and R.~Robbes, ``{Evaluating defect prediction
  approaches: a benchmark and an extensive comparison},'' \emph{Empir. Softw.
  Eng.}, vol.~17, no. 4-5, pp. 531--577, aug 2012. [Online]. Available:
  \url{http://link.springer.com/10.1007/s10664-011-9173-9}
\BIBentrySTDinterwordspacing

\bibitem{Wu2011}
R.~Wu, H.~Zhang, S.~Kim, and S.-C. Cheung, ``{ReLink},'' in \emph{Proc. 19th
  ACM SIGSOFT Symp. 13th Eur. Conf. Found. Softw. Eng. - SIGSOFT/FSE
  '11}.\hskip 1em plus 0.5em minus 0.4em\relax New York, New York, USA: ACM
  Press, 2011, p.~15.

\bibitem{basili1996validation}
V.~R. Basili, L.~C. Briand, and W.~L. Melo, ``A validation of object-oriented
  design metrics as quality indicators,'' \emph{Software Engineering, IEEE
  Transactions on}, vol.~22, no.~10, pp. 751--761, 1996.

\bibitem{ohlsson1996predicting}
N.~Ohlsson and H.~Alberg, ``Predicting fault-prone software modules in
  telephone switches,'' \emph{Software Engineering, IEEE Transactions on},
  vol.~22, no.~12, pp. 886--894, 1996.

\bibitem{kim2011dealing}
S.~Kim, H.~Zhang, R.~Wu, and L.~Gong, ``Dealing with noise in defect
  prediction,'' in \emph{Software Engineering (ICSE), 2011 33rd International
  Conference on}.\hskip 1em plus 0.5em minus 0.4em\relax IEEE, 2011, pp.
  481--490.

\bibitem{Jureczko2010}
\BIBentryALTinterwordspacing
M.~Jureczko and L.~Madeyski, ``{Towards identifying software project clusters
  with regard to defect prediction},'' in \emph{Proc. 6th Int. Conf. Predict.
  Model. Softw. Eng. - PROMISE '10}.\hskip 1em plus 0.5em minus 0.4em\relax New
  York, New York, USA: ACM Press, 2010, p.~1. [Online]. Available:
  \url{http://portal.acm.org/citation.cfm?doid=1868328.1868342}
\BIBentrySTDinterwordspacing

\bibitem{Breiman2001}
\BIBentryALTinterwordspacing
L.~Breiman, ``{Random forests},'' \emph{Machine learning}, pp. 5--32, 2001.
  [Online]. Available:
  \url{http://link.springer.com/article/10.1023/A:1010933404324}
\BIBentrySTDinterwordspacing

\bibitem{Pelayo2007}
\BIBentryALTinterwordspacing
L.~Pelayo and S.~Dick, ``{Applying Novel Resampling Strategies To Software
  Defect Prediction},'' in \emph{NAFIPS 2007 - 2007 Annu. Meet. North Am. Fuzzy
  Inf. Process. Soc.}\hskip 1em plus 0.5em minus 0.4em\relax IEEE, jun 2007,
  pp. 69--72. [Online]. Available:
  \url{http://ieeexplore.ieee.org/lpdocs/epic03/wrapper.htm?arnumber=4271036}
\BIBentrySTDinterwordspacing

\bibitem{Chawla2002}
N.~V. Chawla, K.~W. Bowyer, L.~O. Hall, and W.~P. Kegelmeyer, ``{SMOTE:
  Synthetic minority over-sampling technique},'' \emph{J. Artif. Intell. Res.},
  vol.~16, 2002.

\bibitem{shepperd2012evaluating}
M.~Shepperd and S.~MacDonell, ``Evaluating prediction systems in software
  project estimation,'' \emph{Information and Software Technology}, vol.~54,
  no.~8, pp. 820--827, 2012.

\bibitem{LANGDON201616}
\BIBentryALTinterwordspacing
W.~B. Langdon, J.~Dolado, F.~Sarro, and M.~Harman, ``Exact mean absolute error
  of baseline predictor, marp0,'' \emph{Information and Software Technology},
  vol.~73, pp. 16 -- 18, 2016. [Online]. Available:
  \url{http://www.sciencedirect.com/science/article/pii/S0950584916000057}
\BIBentrySTDinterwordspacing

\bibitem{fu16}
\BIBentryALTinterwordspacing
W.~Fu, T.~Menzies, and X.~Shen, ``Tuning for software analytics: Is it really
  necessary?'' \emph{Information and Software Technology}, vol.~76, pp. 135 --
  146, 2016. [Online]. Available:
  \url{http://www.sciencedirect.com/science/article/pii/S0950584916300738}
\BIBentrySTDinterwordspacing

\bibitem{kim2008classifying}
S.~Kim, E.~J. Whitehead~Jr, and Y.~Zhang, ``Classifying software changes: Clean
  or buggy?'' \emph{IEEE Transactions on Software Engineering}, vol.~34, no.~2,
  pp. 181--196, 2008.

\bibitem{wang2016automatically}
S.~Wang, T.~Liu, and L.~Tan, ``Automatically learning semantic features for
  defect prediction,'' in \emph{Proceedings of the 38th International
  Conference on Software Engineering}.\hskip 1em plus 0.5em minus 0.4em\relax
  ACM, 2016, pp. 297--308.

\bibitem{chawla03}
N.~V. Chawla, ``C4. 5 and imbalanced data sets: investigating the effect of
  sampling method, probabilistic estimate, and decision tree structure,'' in
  \emph{Proceedings of the ICML}, vol.~3, 2003.

\bibitem{kubat97}
M.~Kubat, S.~Matwin \emph{et~al.}, ``Addressing the curse of imbalanced
  training sets: one-sided selection,'' in \emph{ICML}, vol.~97.\hskip 1em plus
  0.5em minus 0.4em\relax Nashville, USA, 1997, pp. 179--186.

\bibitem{shatnawi10}
R.~Shatnawi, ``A quantitative investigation of the acceptable risk levels of
  object-oriented metrics in open-source systems,'' \emph{IEEE Transactions on
  software engineering}, vol.~36, no.~2, pp. 216--225, 2010.

\bibitem{ma07}
Y.~Ma and B.~Cukic, ``Adequate and precise evaluation of quality models in
  software engineering studies,'' in \emph{Predictor Models in Software
  Engineering, 2007. PROMISE'07: ICSE Workshops 2007. International Workshop
  on}, May 2007, pp. 1--1.

\bibitem{xia1999proof}
D.-F. Xia, S.-L. Xu, and F.~Qi, ``A proof of the arithmetic mean-geometric
  mean-harmonic mean inequalities,'' \emph{RGMIA research report collection},
  vol.~2, no.~1, 1999.

\bibitem{vaux2012replicates}
D.~L. Vaux, F.~Fidler, and G.~Cumming, ``Replicates and repeats—what is the
  difference and is it significant?'' \emph{EMBO reports}, vol.~13, no.~4, pp.
  291--296, 2012.

\bibitem{efron93}
B.~Efron and R.~J. Tibshirani, \emph{An introduction to the bootstrap}, ser.
  Mono. Stat. Appl. Probab.\hskip 1em plus 0.5em minus 0.4em\relax London:
  Chapman and Hall, 1993.

\bibitem{leech2002call}
N.~L. Leech and A.~J. Onwuegbuzie, ``A call for greater use of nonparametric
  statistics.'' in \emph{Annual Meeting of the Mid-South Educational Research
  Association}.\hskip 1em plus 0.5em minus 0.4em\relax ERIC, 2002.

\bibitem{poulding10}
S.~Poulding and J.~A. Clark, ``Efficient software verification: Statistical
  testing using automated search,'' \emph{IEEE Transactions on Software
  Engineering}, vol.~36, no.~6, pp. 763--777, 2010.

\bibitem{arcuri11}
A.~Arcuri and L.~Briand, ``A practical guide for using statistical tests to
  assess randomized algorithms in software engineering,'' in \emph{ICSE'11},
  2011, pp. 1--10.

\bibitem{shepperd12a}
M.~J. Shepperd and S.~G. MacDonell, ``Evaluating prediction systems in software
  project estimation,'' \emph{Information {\&} Software Technology}, vol.~54,
  no.~8, pp. 820--827, 2012.

\bibitem{kampenes07}
V.~B. Kampenes, T.~Dyb{\aa}, J.~E. Hannay, and D.~I.~K. Sj{\o}berg, ``A
  systematic review of effect size in software engineering experiments,''
  \emph{Information {\&} Software Technology}, vol.~49, no. 11-12, pp.
  1073--1086, 2007.

\bibitem{Kocaguneli2013:ep}
E.~Kocaguneli, T.~Zimmermann, C.~Bird, N.~Nagappan, and T.~Menzies,
  ``{Distributed development considered harmful?}'' in \emph{Proceedings -
  International Conference on Software Engineering}, 2013, pp. 882--890.

\bibitem{vargha2000}
A.~Vargha and H.~D. Delaney, ``A critique and improvement of the cl common
  language effect size statistics of mcgraw and wong,'' \emph{Journal of
  Educational and Behavioral Statistics}, vol.~25, no.~2, pp. 101--132, 2000.

\bibitem{he2013learning}
Z.~He, F.~Peters, T.~Menzies, and Y.~Yang, ``Learning from open-source
  projects: An empirical study on defect prediction,'' in \emph{Empirical
  Software Engineering and Measurement, 2013 ACM/IEEE International Symposium
  on}.\hskip 1em plus 0.5em minus 0.4em\relax IEEE, 2013, pp. 45--54.

\end{thebibliography}
\end{document}